\documentclass{revtex4}
\usepackage{epsfig}
\usepackage{psfrag}
\usepackage{graphicx}
\usepackage{amsfonts}
\usepackage[figuresright]{rotating}
\usepackage{amssymb}
\usepackage{amsmath}
\usepackage{subfigure}
\usepackage{multirow}
\usepackage{tabularx}
\usepackage{bm}
\usepackage{color}

\def\Im {\mbox{Im}}

\newcommand{\ma}[1]{\mathsf{#1}}

\newcommand{\mbb}[1]{\mathbb{#1}}
\newcommand{\Ss}{\ma{S}}
\newcommand{\tSS}{\tilde{\Ss}}

\newcommand{\rv}{{\bm r}}
\newcommand{\bv}{{\bm b}}
\newcommand{\bhv}{\hat{\bv}}
\newcommand{\bh}{\hat{b}}
\newcommand{\Gv}{\bm G}
\newcommand{\Cav}{\bm C}

\newcommand{\uv}{{\bm u}}
\newcommand{\tv}{{\bm t}}
\newcommand{\fv}{{\bm f}}
\newcommand{\pv}{{\bm p}}
\newcommand{\xv}{{\mathbf x}}

\newcommand{\dv}{{\bm d}}
\newcommand{\nv}{{\bm n}}
\newcommand{\ve}{{\bm v}}
\newcommand{\vs}{v}

\newcommand{\g}{g}
\newcommand{\rme}{\rm e}
\newcommand{\rmi}{\rm i}
\newcommand{\rmd}{\rm d}
\newcommand{\qv}{{\bm q}}
\newcommand{\av}{{\bm a}}

\newcommand{\Bv}{{\bm B}}

\newcommand{\Qv}{{\mathbf Q}}
\newcommand{\Cv}{{\mathbf C}}
\newcommand{\Dv}{{\mathbf D}}
\newcommand{\Kv}{{\mathbf K}}
\newcommand{\Rv}{{\bm R}}
\newcommand{\Xv}{{\bm X}}

\newcommand{\ee}{{e}} %This will be the extension variable, which can be changed here
\newcommand{\EE}{E} %This will be the extension variable, which can be changed here
\newcommand{\la}{\ell}

\newcommand{\st}{\varepsilon}
\newcommand{\stm}{\pmb{\st}}

\newcommand{\rank}{\text{rank}}
\newcommand{\nullity}{\text{nullity}}
\newcommand{\btheta}{\bar{\theta}}
\newcommand{\tnu}{\tilde{\nu}}
\newcommand{\Ith}{Index theorem }
\newcommand{\Itha}{Index theorem}
\newcommand{\fref}[1]{figure \ref{#1}}
\newcommand{\eref}[1]{(\ref{#1})}
\newcommand{\sref}[1]{section \ref{#1}}
\newcommand{\Sref}[1]{Section \ref{#1}}
\newcommand{\Fref}[1]{Figure \ref{#1}}
\newcommand{\msh}{G} % This is th shear modulus
\newcommand{\SSS}{SSS } % State of self stress abreviation
\newcommand{\SSSs}{SSSs } % States of self stress abreviation
\newcommand{\SSSa}{SSS} % States of self stress abreviation
\newcommand{\SSSsa}{SSSs} % States of self stress abreviation

\newcommand\thide[1]{\bgroup\color{blue}\bfseries{}\egroup}

\begin{document}
\title{Phonons and elasticity in critically coordinated
lattices}
\author{T C Lubensky$^1$, C L Kane$^1$, Xiaoming Mao$^2$, A Souslov$^3$ and Kai Sun$^2$}
\address{$^1$ Department of Physics and Astronomy, University
of Pennsylvania, Philadelphia, Pennsylvania, 19104, USA;
tom@physics.upenn.edu}
\address{$^2$ Department of Physics, University of Michigan, Ann
Arbor, MI 48109, USA}
\address{$^3$ School of Physics, Georgia Institute of Technology, Atlanta, Georgia 30332, USA}

%\tableofcontents

%\listoffigures
\begin{abstract}
Much of our understanding of vibrational excitations and
elasticity is based upon analysis of frames consisting of sites
connected by bonds occupied by central-force springs, the
stability of which depends on the average number of neighbors
per site $z$. When $z<z_c \approx 2d$, where $d$ is the spatial
dimension, frames are unstable with respect to internal
deformations. This pedagogical review focuses on properties of
frames with $z$ at or near $z_c$, which model systems like
randomly packed spheres near jamming and network glasses. Using
an index theorem, $N_0 - N_S = dN - N_B$ relating the number of
sites, $N$, and number of bonds, $N_B$, to the number, $N_0$,
of modes of zero energy and the number, $N_S$, of states of
self stress, in which springs can be under positive or negative
tension while forces on sites remain zero, it explores the
properties of periodic square, kagome, and related lattices for
which $z=z_c$ and the relation between states of self stress
and zero modes in periodic lattices to the surface zero modes
of finite free lattices (with free boundary conditions). It
shows how modifications to the periodic kagome lattice can
eliminate all but trivial translational zero modes and create
topologically distinct classes, analogous to those of
topological insulators, with protected zero modes at free
boundaries and at interfaces between different topological
classes.

\end{abstract}
%\pacs{}
\date{\today}
\maketitle
%\date

\section{Introduction\label{sec:intro}}

Understanding and controlling mechanical stability is important
to fields ranging from structural engineering to granular
materials and glasses.  We do not want buildings or bridges to
fail, and we want to be able to manage the elastic response of
materials of everyday life.  This review focusses on the
elastic and dynamical properties of periodic ball-and-spring
networks that are at or near mechanical collapse.  This may
seem like a fairly narrow subject, but it is one of
considerable richness and impact.

\begin{figure}
\centering
\includegraphics{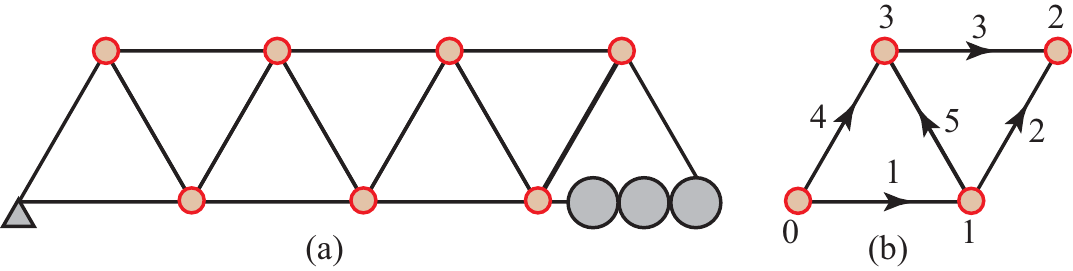}
\caption{(a) The Warren Truss.  This is an isostatic structure composed of equilateral triangles with $N=9$ sites
and $N_B = 15$ connections, so that $2 N - N_B = 3$.  The lower left site (indicated by the small triangle)
is fixed with respect to the earth, and the lower
right  site is constrained to
move only horizontally along a track with wheels.  These constraints reduce the number of
free degrees of freedom of the sites by $3$ to $N_{\text{free}}=2 N-3 =15 = N_B$.  (b) A reduced version of
the isostatic Warren Truss indicating conventions for labeling sites and bonds.
The arrows indicate indicate bond directions following the convention described in section \ref{ssec:equil-comp},
Eqs.~(\ref{eq:bond-vector}) to (\ref{eq:equil2}).}
\label{fig:WarrenTruss}
\end{figure}

In a remarkable 1864 paper \cite{Maxwell1864}, James Clerk
Maxwell undertook the first systematic study of the mechanical
stability of frames consisting of points, which we will refer
to as \emph{sites}, with connections, which we will usually
refer to as \emph{bonds}, between them as a model for such
real-world structures as the Warren Truss (patented in 1848)
shown in \fref{fig:WarrenTruss}. He defined a  ``stiff" frame
as one in which ``the distance between two points cannot be
altered without changing the length of one or more
connections".  He showed that a stiff frame containing $N$
sites in $d$ dimensions requires
\begin{equation}
N_B = dN-f(d)
\label{eq:Maxwell-rule1}
\end{equation}
connections, where $f(d)=d(d+1)/2$ is the number of rigid
translations and rotations under free boundary conditions.
Under periodic boundary conditions, which Maxwell did not
consider, $f(d) = d$. This relation, often referred to as
\emph{Maxwell's rule}, can be reexpressed as a critical
coordination number ($z \equiv 2 N_B/N$),
\begin{equation}
z^N_c = 2 d - 2\frac{f(d)}{N}.
\label{eq:z_c}
\end{equation}
If $z<z^N_c$, the system is not stiff, and if $z>z^N_c$, it is.
As we shall discuss in more detail in the next section,
Maxwell's rule for the stability of frames requires
modification \cite{Calladine1978}.  Nevertheless, it provides a
useful and universally used benchmark for the analysis of the
stability of frames. We will refer to free frames, i.e., ones
under free boundary conditions, satisfying Maxwell's rule as
\emph{Maxwell frames}. It is fairly common practice to use the
term \emph{isostatic} for frames satisfying Maxwell's rule.
Though isostatic frames do satisfy Maxwell's rule, they have
more restrictive properties, which we will discuss in
\sref{sec:Max-Index}.

Our principal interest here is in frames whose sites can be
collected into identical contiguous unit cells whose origins
lie on a Bravais lattice and which fill some region of space.
We will refer to these frames as \emph{lattices} and if they
are subjected to periodic (free) boundary conditions as
periodic (free) lattices.  Any frame, even those that are not
lattices, e.g., whose sites are randomly distributed, can also
be subjected to periodic (free) boundary conditions in which
case we will refer to them as periodic (free) frames. For
reasons that we will justify more fully later, we will use the
term \emph{periodic Maxwell frame (lattice)} for periodic
frames (lattices) with average coordination number $z_c=2d =
z_c^\infty$ rather than $z_c^N$. Free frames can be liberated
from periodic ones by cutting of order $N^{(d-1)/d}$ bonds.

Maxwell's analysis applies to frames with an arbitrary number
of sites and bonds. In addition to being a workhorse of the
structural engineering community
\cite{Heyman1999,Kassimali2005}, it has seen extensive use
(though not always with attribution to Maxwell) in physics,
materials science, and mathematics. It is a critical component
of the theory of structural glasses
\cite{Phillips1981,Thorpe1983,Phillips1985,ThorpePhi2000a,BoolchandTho2005},
rigidity percolation
\cite{Feng1984,JacobsTho1995,JacobsTho1996,GuyonCra1990,ChubynskyTho2007},
framework silicates like $\beta$-cristobalite
\cite{Hammonds1996}, jamming of packed spheres
\cite{Liu1998,Wyart2005a,WyartWit2005b,vanHecke2010,LiuNag2010a},
biopolymer networks
\cite{BroederszMac2014,Head2003,Wilhelm2003,Heussinger2006,HeussingerFre2007a,HuismanLub2011,Broedersz2011,MaoLub2013a},
and of some theories of protein folding \cite{WellsTho2005}.
Rigidity percolation and jamming generally involve central
forces only, and the Maxwell relation and its generalization
can be applied to these problems directly.  Structural glasses
and biopolymer networks have bending forces that require a
modification of the Maxwell rules, and we will not make much
contact with them in what follows.

\begin{figure}
\centering
\includegraphics{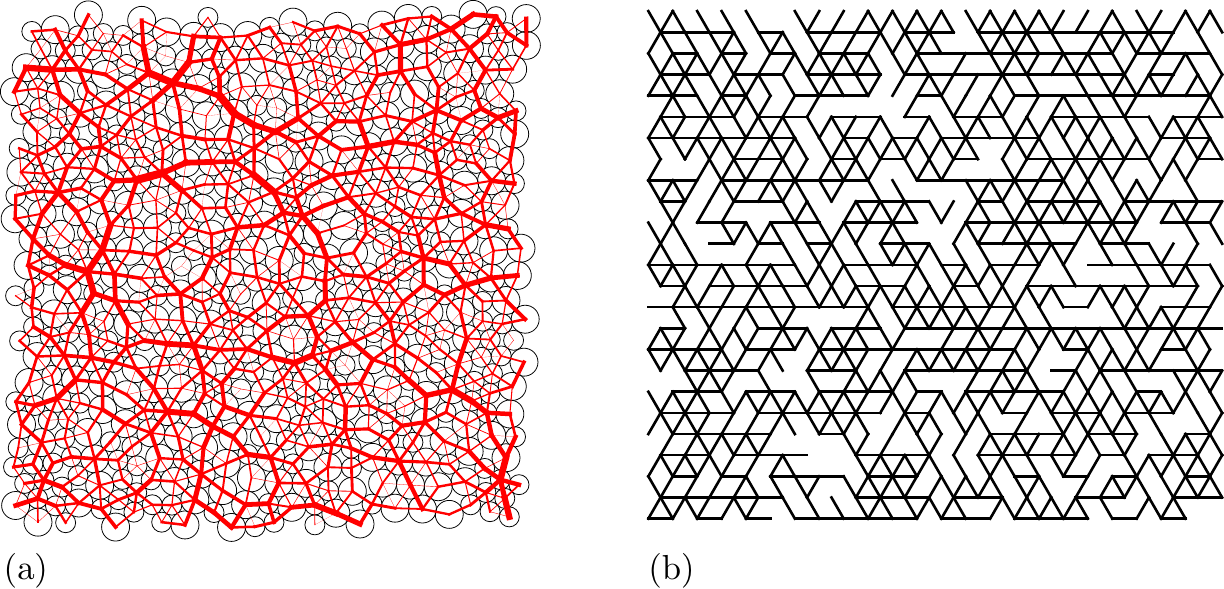}
\caption{(a) packed discs of two sizes just above the
jamming transition. The dark red lines are chains of force that
are a response to the pressure required to pack the particles
at $z>z_c$ (courtesy of Carl Goodrich). (b) A representative
bond-diluted lattice near the rigidity percolation threshold of
$z_c \approx  3.96$.}
\label{fig:spheres-at-jamming}
\end{figure}

Both randomly packed spheres (\fref{fig:spheres-at-jamming}(a))
and diluted elastic networks (\fref{fig:spheres-at-jamming}(b))
pass from a state that does not support stress to one that does
as the number of bonds or contacts increases.  At the critical
point separating these two states, the coordination number is
at or near the Maxwell critical value of $z_c$.  The rigidity
percolation transition is generally continuous
\cite{Feng1984,JacobsTho1995,JacobsTho1996,ChubynskyTho2007} in
two-dimensions, and it is well-described by the language of
critical phenomena, applied so successfully to percolation in
random resistor networks
\cite{StaufferAha1994,Kirkpatrick1973}. Both shear and bulk
elastic moduli increase from zero as a power law in $(z-z_c)$,
and there is a divergent length scale associated with the
probability that two sites are in the same rigid cluster as a
function of their separation. In three dimensions, the rigidity
transition is apparently first order \cite{ChubynskyTho2007}.

Jammed systems exhibit different behavior. They are constructed
by increasing the density of spheres in a fixed volume until
they first resist a further increase.  At this point, they are
jammed, and their bulk modulus $B$  (which resists compression)
is nonzero \cite{OhernNag2002,OHern2003}, but their shear
modulus $G$ is of order $1/N$
\cite{GoodrichNag2012,GoodrichNag2014}. At large $N$, $G$
increases linearly in $\Delta z = (z-z_c)$
\cite{Durian1995,OhernNag2002,OHern2003}. The density of states
of systems with $\Delta z >0$ exhibits a crossover from
Debye-like behavior ($\sim \omega^{d-1}$ in $d$ dimensions as a
function of frequency $\omega$) at low frequency to a
flat-plateau beyond a characteristic frequency $\omega^* \sim
(\Delta z)$ \cite{OHern2003,SilbertNag2005}. There are two
diverging length scales, $l^*\sim(\Delta z)^{-1}$ and $l_T\sim
(\Delta z)^{-1/2}$, which can be extracted
\cite{SilbertNag2005} by, respectively, comparing $\omega^*$ to
the longitudinal and transverse sound frequencies $c_L l^{-1}$
and $c_T l^{-1} \sim (\Delta z)^{1/2} l^{-1}$ at wavenumber $q
\sim l^{-1}$, where $c_L \sim B^{1/2}$ and $c_T \sim G^{1/2}$
are the longitudinal and transverse sound velocities. Other
interpretations of these lengths invoke finite-size effects in
the modes of a finite sample cut from a larger one
\cite{WyartWit2005b,Wyart2005a,Wyart2005} for $l^*$ and
stability of the system to boundary perturbations
\cite{GoodrichLiu2013a,Schoenholz2103}, crossover between
strong and weak scattering at $\omega^*$
\cite{XuNa2009,XuNag2010}, or correlations in systems with
$z<z_c$ \cite{LernerWya2012,DuringWya2013} for $l_T$.

\begin{figure}
\centering
\includegraphics{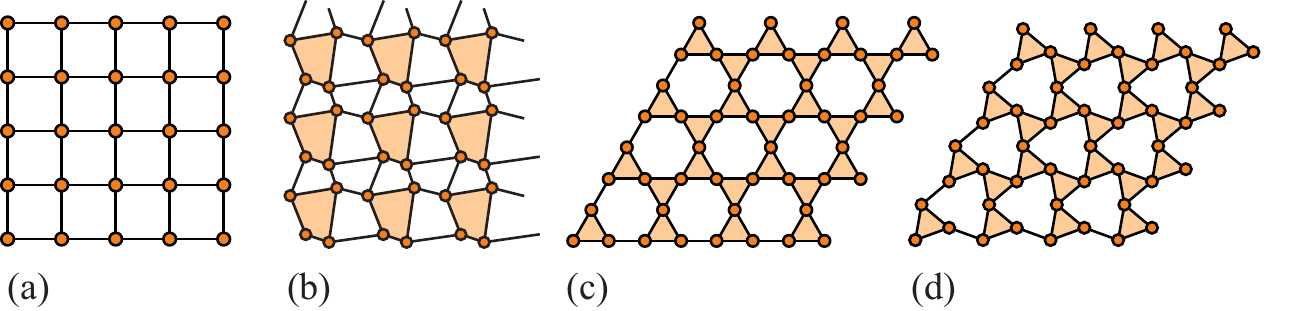}
\caption{(a) Square, (b) distorted square, (c) kagome lattice, and (d) twisted kagome lattice}
\label{fig:square-kagome}
\end{figure}

Though the transition to elastic stability in both rigidity
percolation and jamming occurs at or near the Maxwell critical
point $z=z_c$, properties at and above the two transitions are
quite different.  Presumably, the difference is a reflection of
the different geometrical arrangements of the Maxwell lattices
at or near the two transitions.  An interesting question then
is what precisely are the differences. Can they be quantified,
and if so how? These questions then raise a broader question of
whether there are wider classes of Maxwell frames that lead to
elastic and vibrational structure different from those of the
percolation and jamming lattices. As a first step toward
answering that question, it is useful to study periodic Maxwell
lattices, like the square and kagome lattices shown in
\fref{fig:square-kagome} with NN bonds and $z=2d=4$, and the
reduced coordination lattices they spawn when bonds are cut to
create free surfaces. They can easily be moved away from the
Maxwell limit toward greater rigidity by uniformly
\cite{Souslov2009} or randomly \cite{MaoLub2010} adding further
neighbor bonds, and they have an advantage that they lend
themselves to exact calculations of elastic response both in
the uniform and random systems (via effective medium theory).
In addition, in spite of their simplicity, they present a
surprisingly rich phenomenology that inform us generally about
Maxwell frames.

Much of the language
\cite{Calladine1978,Pellegrino1993,PellegrinoCal1986,GuestHut2003}
for probing the mechanical properties of frames used in this
review was developed by members of the Structural Mechanics
group at the University of Cambridge, which provided an elegant
generalization of Maxwell's relation based on general
principles of linear algebra and used it to deepen our
understanding of the more subtle properties of frames. Though
well-known in the engineering community, it is less so in the
physics and materials science communities.  This review will,
therefore, begin in section \ref{sec:Max-Index}  with a fairly
comprehensive review of this work. It is followed by five more
sections and four appendices. Section \ref{sec:elas-lim} deals
the long-wavelength elastic limit and how it can be calculated
using the developments in references
\cite{Pellegrino1993,PellegrinoCal1986}. Section
\ref{sec:periodic-lattices} tailors the formalism of the
preceding sections specifically to periodic lattices.  Section
\ref{sec:examples} looks in detail at the properties of the
three simple lattices in \fref{fig:square-kagome}, and section
\ref{sec:topological} explores lattices with topologically
protected states at interfaces, which are closely related to
topologically protected electronic states in the quantum Hall
effect \cite{halperin82,haldane88}, polyacetylene \cite{ssh}
and topological insulators
\cite{km05b,bhz06,mb07,fkm07,HasanKane2010,QiZhang2011}.
Section \ref{sec:review} presents some final thoughts and
speculates about future directions. The four appendices provide
mathematical detail and display derivations of various
important relations.

\Sref{sec:examples} contains several subsections.
\Sref{ssec:square-lattice} with its simple analytical treatment
of the square lattice sets the stage for the study in
\sref{ssec:kagome} of the simple kagome lattice. Both lattices
have lines of zero modes in their Brillouin zones arising from
their having straight sample-traversing filaments of colinear
bonds. \Sref{ssec:twisted-kagome} then explores how the simple
geometrical operation of ``twisting" neighboring triangles in
the standard kagome lattice [\fref{fig:square-kagome}(c)] to
convert it to the twisted lattice [\fref{fig:square-kagome}(d)]
gaps the phonon spectrum of the simple kagome lattice at all
wavenumbers except at the origin and leads to zero-frequency
Rayleigh surface modes \cite{Landau1986} not present in either
the square or standard kagome lattices.

This review is intended to be as much a pedagogical
introduction to periodic Maxwell lattices as an overview of the
subject.  Except in section \ref{sec:topological}, which
requires the use of some fairly subtle concepts detailed in
reference \cite{KaneLub2014}, it provides sufficient
calculational detail that even someone totally new to the
subject should be able to follow it.  Though, as the
introductory paragraphs have indicated, this review was
inspired in part by jamming and rigidity percolation, it is not
about these subjects, and they will be considered only when
they have direct overlap with the ideas being presented.

\section{Generalized Maxwell relation as an index theorem
\label{sec:Max-Index}}
\subsection{The Maxwell rule and states of self stress \label{ssec-Max-SSS}}
\begin{figure}
\centering
\includegraphics{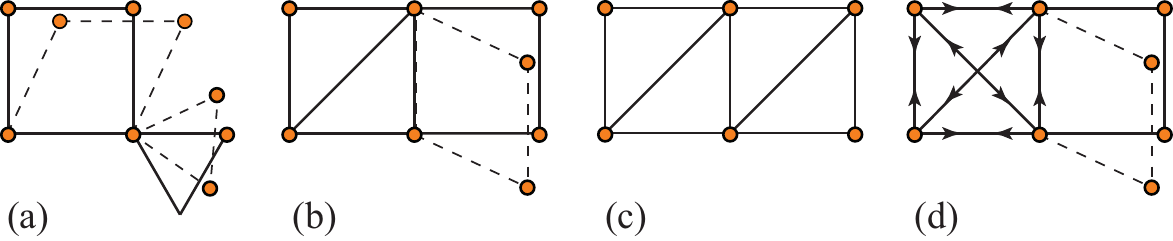}
\caption{(a) to (c) Frames satisfying the Maxwell rule.  (a) has $6$ sites, $7$ bonds,
$5$ zero modes, and two mechanisms indicated by the dotted bonds. (b) has $6$ sites, $8$ bonds, $4$ zero modes,
and one mechanism. (c) and
(d) are constructed from (b) by adding an additional diagonal bond. (c) satisfies the
Maxwell rule with only the three trivial zero modes.  (d) has $4$ zero modes
and one state of self stress indicated by the arrows on the bonds in the left square.}
\label{fig:selftstress1}
\end{figure}

Each site in $d$-dimensions has $d$ independent translational
degrees of freedom, and in the absence of constraints on point
motion, a collection of $N$ points without connections has
$N_{\text{free}}=dN$ zero-frequency displacement modes, which
we will refer to as \emph{zero modes}. In the presence of
constraints, $N_{\text{free}}$ will be less that $dN$. Each
connection reduces the number of zero modes by one. Thus if
there are $N_B$ connections and no constraints, there are
\begin{equation}
N_0 = dN - N_B
\label{eq:simplecount}
\end{equation}
zero modes. Of these, $f(d)$ are the trivial ones associated
with rigid translations and rotations. Any other zero modes
involve internal displacements of the sites and are generally
called \emph{mechanisms} \cite{Calladine1978} in the
engineering literature and \emph{floppy modes} in the physics
literature \cite{Thorpe1983}. Equation (\ref{eq:simplecount})
reexpressed in terms of the number of mechanisms $M$ is
\begin{equation}
M= dN - N_B - f(d) .
\label{eq:Maxwell-mech}
\end{equation}
We will refer to Eqs.~(\ref{eq:simplecount}) and
(\ref{eq:Maxwell-mech}) as the \emph{Maxwell count}. A frame is
stiff if it has no mechanisms. Setting $M=0$ yields the Maxwell
rule [Eqs.~(\ref{eq:Maxwell-rule1})].
\Fref{fig:selftstress1}(a) depicts a simple frame that obeys
Maxwell's count.  It consists $N=6$ sites and $N_B=7$ bonds,
and it has $N_0 = 2 \times 6 - 7 =5$ zero modes and $M = N_B =
5-3=2$ mechanisms.

The simple Maxwell rule does not apply to all frames
\cite{Calladine1978}. Consider the two-square frame with $N=6$
sites and $N_B=8$ bonds shown in \fref{fig:selftstress1}(b). It
has one mechanism as expected from the Maxwell count. If an
extra bond is added, Maxwell's rule would say that the frame is
stiff with no mechanisms. The extra bond, however, can be
placed as a diagonal in the right square
[\fref{fig:selftstress1}(c)] or as an extra diagonal in the
left square [\fref{fig:selftstress1}(d)].  In the first case,
there are no mechanisms, and Maxwell's rule applies.  In the
second case, however, the mechanism present before the extra
bond was added remains, and the Maxwell count is violated.  But
the left square with crossed diagonal bonds has an extra
\emph{redundant} bond not needed for its rigidity. It also has
a new and interesting property: the outer bonds of the square
can be placed under tension (compression) and the inner
diagonal bonds under compression (tension) such that the net
force on all sites is zero. This is a \emph{state of self
stress}, which, because of its repeated use in this review, we
will usually abbreviate as SSS. This theme can clearly be
repeated with each added bond either decreasing the number of
zero modes or increasing the number of states of self stress to
yield the modified Maxwell count \cite{Calladine1978}:
\begin{equation}
N_0 = dN -N_B + N_S \qquad \text{ or } \qquad N_0 - N_S = dN -N_B ,
\label{eq:index1}
\end{equation}
where $N_S$ is the number of SSSs. This is an \emph{index
theorem} \cite{AtoyahSin1963,Nakahara2003} , which we will
derive in section \ref{ssec:equil-comp}, relating mode and
self-stress count to geometric properties of the lattice. We
will refer to it simply as the \Ith.

\begin{figure}
\centering
\includegraphics{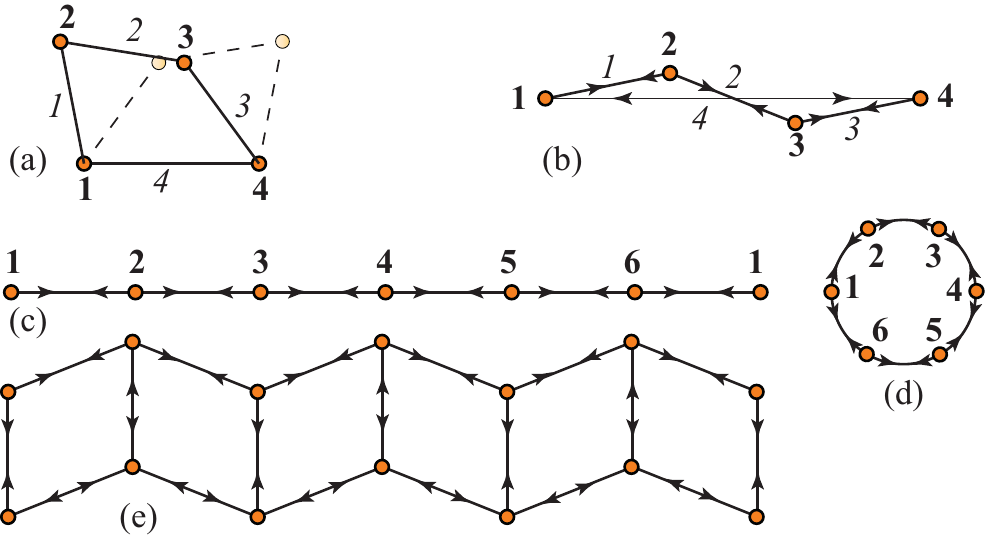}
\caption{States of self stress: (a) A frame with four sites ($\mathbf{1}$-$\mathbf{4}$) and four bonds
(italic $\it{1}$, $\it{2}$, $\it{3}$, and $\it{4}$) with
$4$ zero modes and one finite mechanism (dashed lines); (b) A frame in which the length of
bond $\it{4}$ is equal to the sum of the lengths of bonds $1$ to $3$.  Now there is
a \SSS in which bond $\it{4}$ is under compression (tension) and the other three
are under tension (compression).  Both sites $\mathbf{2}$ and $\mathbf{3}$ can undergo
infinitesimal displacements without changing the length of any bonds, and there are
two infinitesimal mechanisms.
(c) A line of parallel bonds forming a sample-traversing filament under periodic boundary conditions
as depicted in (d).  The forces on all sites are zero if all of the bonds
are under equal tension or compression, and there is one \SSS.
(e) A zigzag ladder under periodic boundary conditions with one \SSS.}
\label{fig:selftstress2}
\end{figure}

Two types of mechanisms can be distinguished: ``finite" ones in
which finite-amplitude displacements of sites stretch no bonds
and ``infinitesimal" ones in which bond lengths do not change
to first order in the magnitude of displacements but do so to
second (or higher) order.  The \Ith\cite{Calladine1978}, as we
shall show below, follows from the assumption of a linear
relation between site displacements and bond lengths, and it
applies only to infinitesimal displacements, i.e., it counts
both finite and infinitesimal mechanisms but does not identify
which is which. Figures \ref{fig:selftstress2}(a)-(b) show how
a finite mechanism can be converted into two infinitesimal
mechanisms and one \SSSa. A configuration of self-stress that
is particularly important for the current study is any straight
line of bonds under periodic boundary conditions, which we will
refer to as straight \emph{filaments}, as shown in figures
\ref{fig:selftstress2}(c)-(d). Changing the straight filament
to a zigzagged one removes this state of self stress. On the
other hand the ``zigzaging" periodic ladder configuration shown
in \fref{fig:selftstress2}(e) has one \SSSa, rather than the
two that a straight ladder would have. Tensions alternate in
sign from bond to bond in this \SSSa, a property, which will be
important in what follows, that prevents it from having any
zero wave-number component.

A system in which there are neither any mechanisms ($M=0$) nor
any states of self-stress ($N_S=0$) is \emph{isostatic} .  A
finite isostatic system necessarily satisfies the Maxwell
relation $z=z^N_c$, but a system with $z=z^N_c$ can have any
number of mechanisms provided it is equal to the number of
\SSSsa. The distinction between satisfying the Maxwell rule and
being isostatic is often lost in the literature, and it is
common practice to refer to any system that satisfies Maxwell's
rule as isostatic.  In this review, we will keep the
distinction, referring to any free frame satisfying Maxwell's
rule as a Maxwell frame, reserving the term isostatic for those
free Maxwell frames satisfying $M=N_S = 0$. As we shall see in
\sref{ssec:Iso-Per}, the extension of this definition to
periodic frames presents some problems \cite{GuestHut2003}.
Since the term isostatic has become so prevalent, we propose in
that section a definition of this term that is in the spirit of
the definition for free frames and consistent with common usage
for periodic frames.

\subsection{Equilibrium and compatibility matrices\label{ssec:equil-comp}}
In the absence of external forces, the equilibrium force at
each site in a frame is determined by the tensions in the bonds
it shares with other sites. This is true whether or not the
site is in mechanical equilibrium; if the force at a site is
nonzero, the mass at that site will accelerate according to
Newton's laws. If forces at sites arising from bond tensions
are nonzero, they can be balanced by external loads to create
an equilibrium situation in which the total force on each site
is zero. Clearly, in mechanical equilibrium, the external loads
are the negative of the forces at each site arising from bond
tensions.

For central forces, the tension in a bond is parallel to the
bond.  Thus its direction is specified by bond orientation, but
its magnitude and sign can vary. Let $\ma{F}$ be a vector in
the $dN$-dimensional space, $V_{\ma{F}}$,  of the $d$
components of force at each site on the lattice exerted by
tensions in the bonds that terminate on it, and let $\ma{T}$ be
a vector in the $N_B$-dimensional space, $V_{\ma{T}}$, of the
of the bond tensions, which are scalars of either sign.
External loads at sites are represented by the $dN$-dimensional
vector $\ma{L}$. In equilibrium when sites do not accelerate
(or move if there is external friction), $\ma{L}=-\ma{F}$.
Since the relation between forces and tensions is linear, there
is a $d N \times N_B$ dimensional matrix $\ma{Q}$ (with $dN$
rows and $N_B$ columns), called the \emph{equilibrium} matrix,
that maps $V_\ma{T}$ to $V_\ma{F}$:
\begin{equation}
\ma{Q} \, \ma{T} = -\ma{F} = \ma{L} ,
\label{eq:QTF}
\end{equation}
where the final relation only applies in static situations.

The null space or kernel of $\ma{Q}$, $\ker(\ma{Q})$ of
dimension $\nullity(\ma{Q})$, is the set of all vectors mapped
to the zero vector. Any vector in the null space of $\ma{Q}$
represents a state of self stress because it corresponds to
tensions on a set of bonds for which the forces on all sites
are zero. Thus $\nullity(\ma{Q})$ is equal to the number of
\SSSs $N_S$. Vectors $\ma{T}$ not mapped into the null space of
$\ma{Q}$ are in the \emph{orthogonal complement},
$\oc(\ma{Q})$, of $\ker(\ma{Q})$. The dimension of
$\oc(\ma{Q})$ is equal the to the rank of $\ma{Q}$. The
rank-nullity theorem of linear algebra \cite{Birkhoff-Mac1998}
relates the rank and nullity of a matrix to its column
dimension:
\begin{equation}
\rank(\ma{Q}) + \nullity(\ma{Q}) = \rank(\ma{Q}) + N_S = N_B.
\label{eq:rank-nullityQ}
\end{equation}
Elongation of bonds are determined by the displacements of the
sites to which they are attached. The elongations of individual
bonds are necessarily parallel to bond vectors for central
forces. Let $\ma{U}$ be a vector in the $dN$-dimensional space,
$V_{\ma{U}}$, of site displacements and $\ma{\EE}$ be a vector
in the $N_B$-dimensional space, $V_\ma{\EE}$, of bond
elongations. The $N_B \times dN$-dimensional compatibility
matrix $\ma{C}$ maps $V_\ma{U}$ to $V_\ma{\EE}$:
\begin{equation}
\ma{C} \, \ma{U} = \ma{\EE} .
\label{eq:CUE}
\end{equation}
The null space of $\ma{C}$ is the set of displacements $\ma{U}$
that do not change the length of bonds, i.e., the set of zero
modes of the system; thus, $\nullity(\ma{C})=N_0$. The
rank-nullity theorem applied to $\ma{C}$ yields
\begin{equation}
\rank( \ma{C}) + \nullity(\ma{C}) = \rank(\ma{C})+ N_0 = d N .
\label{eq:rank-nullityC}
\end{equation}

The equilibrium and compatibility matrices are not independent:
they are matrix transposes of each other. To see this, we can
calculate the work done under infinitesimal distortions of the
system in the presence of sites forces (and thus necessarily
tensions in the bonds) in two ways: First the work $W_L$ done
by external loads in displacing sites and second the work $W_T$
done by bond tensions in stretching bonds.  The two
calculations must yield the same result:
\begin{equation}
W_L = \ma{L}^T \ma{U} = \ma{T}^T \ma{Q}^T \ma{U} =
W_T = \ma{T}^T \ma{E} = \ma{T}^T \ma{C} \ma{U} ,
\label{eq:workQC}
\end{equation}
where the superscript $T$ refers to the transpose of a matrix.
Since this relation is valid for all $\ma{U}$ (even $\ma{U}$ in
the null space of $\ma{C}$) it must be that $\ma{C}= \ma{Q}^T$.
The rank of a matrix is equal to the rank of its transpose,
$\rank(\ma{Q}) = \rank(\ma{C})$, and subtracting
\eref{eq:rank-nullityQ} from \eref{eq:rank-nullityC} yields the
\Ith of \eref{eq:index1}. The equilibrium and compatibility
matrices have proven useful in many contexts, and we mention
only one in which frames with spring lengths that do not
correspond to the length of the bonds they occupy are
constructed from frames in which they do \cite{YanWya2013}.

To construct $\ma{Q}$ and $\ma{C}$ for a particular frame, it
is necessary to have explicit representations for the vectors
$\ma{U}$, $\ma{E}$, $\ma{T}$, and $\ma{F}$. A frame consists of
$N$ sites labeled $s=1, ..., N$ at equilibrium positions
$\Rv(s)$ connected by $N_B$ bonds labeled $\beta = 1, ...,
N_B$. Under frame distortions, site positions change to
\begin{equation}
\Xv(s) = \Rv(s) + \uv(s) ,
\end{equation}
where $\uv(s)$ with components $u_i(s)$, $ i = 1, ..., d$, is
the displacement vector at site $s$. The force at site $s$ is
$\fv(s)= (f_1(s) , ... f_d(s))$, and the $dN$-dimensional
displacement and force vectors are, respectively,
$\ma{U}=(\uv_1, ...,\uv_N )$ and $\ma{F} = (\fv_1, ..., \fv_N
)$.

Each bond $\beta = [s_\beta,s'_\beta]$ connects a pair of sites
$s_\beta$ and $s'_\beta$, whose separation in the equilibrium
frame is the vector
\begin{equation}
\bv(\beta)\equiv \bv([s_\beta,s'_\beta])= \Rv(s'_\beta ) - \Rv(s_\beta) \equiv
b_\beta\bhv_\beta ,
\label{eq:bond-vector}
\end{equation}
from $s_\beta$ to $s'_\beta$, where $b_\beta$ is the length of
bond $\beta$ and $\bhv_\beta$ is the unit vector along bond
$\beta$. The arrows in \fref{fig:WarrenTruss}(b) show an
arbitrarily chosen choice of directions of vectors
$\bhv_{\beta}$ for bonds in a simple frame. Other choices will
lead to different equations relating forces to tensions, but no
the physical tensions in the frame. Let the tension in bond
$\beta$ be $t_\beta$, and associate with that bond a vector
tension,
\begin{equation}
\tv_\beta = t_\beta \bhv_\beta .
\end{equation}
With this convention, the force exerted on sites $s_\beta$ and
$s'_\beta$ by bond $[s_\beta,s'_\beta]$ are, respectively,
$\tv_\beta$ and $-\tv_\beta$, or, equivalently, the force on a
site $s$ from a bond $\beta$ that it shares with another site
is $\tv_\beta$ if $\bhv_\beta$ points away from the site and
$-\tv_\beta$ if it points toward the site.  With these
definitions, we can construct the compatibility matrix from the
the bond elongation relations,
\begin{equation}
e_\beta = \bhv_\beta\cdot (\uv(s'_\beta) -\uv(s_\beta)) ,
\label{eq:compatibility2}
\end{equation}
and the equilibrium matrix from the site force equations,
\begin{equation}
\fv(s) = \sum_{\beta=1}^{z_s} \text{sign} (\beta) \tv_\beta ,
\label{eq:equil2}
\end{equation}
where $z_s$ is the number of bonds site $s$ shares with its
neighbors and $\text{sign} (\beta)= +1 (-1)$ if the arrow of
bond $\beta$ points away from (toward) site $s$.

It is instructive to carry out explicit calculations of
$\ma{Q}$ and $\ma{C}$ for a simple frame.  We consider the
frame shown in \fref{fig:WarrenTruss}(b) with $N=4$, $N_B = 5$
and $2 N - N_B =3$, the number of modes of rigid translation
and rotation. This frame has no floppy modes and no state of
self stress, and it is isostatic. The figure labels sites
($s=0, ... ,3)$ , bonds ($\beta= 1, ... ,5$), and bond
directions. The five displacement-elongation equations are
\begin{align}
e_1 & = \bhv_1\cdot (\uv_1 - \uv_0) \qquad & e_2 = \bhv_2\cdot (\uv_2 - \uv_1)
\qquad & e_3 = \bhv_3 \cdot (\uv_2 - \uv_3) \nonumber\\
e_4 & = \bhv_4 \cdot (\uv_3 - \uv_0 ) \qquad & e_5 = \bhv_5 \cdot (\uv_3 - \uv_1 ) ,
\qquad &
\label{eq:elongations-1}
\end{align}
and the four vector force equations are
\begin{align}
\fv_0 & = \tv_1 +\tv_4 \qquad & \fv_1 = -\tv_1 + \tv_2 + \tv_5\nonumber \\
\fv_2 & = - \tv_2 - \tv_3 \qquad & \fv_3 = \tv_3 - \tv_4 - \tv_5 .
\label{eq:forces-2}
\end{align}
The $8 \times 5$ compatibility and $5\times 8$ equilibrium
matrices are easily constructed from equations
(\ref{eq:elongations-1}) and (\ref{eq:forces-2}). If, however,
we are interested only in internal deformations of the frame
and not its uniform translations and rotations, we can apply
three constraints to the motion, most easily by pinning site
$0$ so that $\uv_0 = 0$ and placing site $1$ on a horizontal
rail as shown in \fref{fig:WarrenTruss}(a) to fix $u_{1,y} = 0$
so that $N_{\text{free}}=5$.  We also allow $\fv_0$ and
$f_{1,x}$ to take on whatever values needed to satisfy the
constraints, so they do not enter into our equations.  This
leaves us with a $5 \times 5$ compatibility matrix,
\begin{equation}
\ma{C} =
    \begin{pmatrix}
    1 & 0 & 0 & 0 & 0 \\
    -\tfrac{1}{2} & \tfrac{1}{2} & \tfrac{\sqrt{3}}{2} & 0 & 0 \\
    0 & 1 & 0 & -1 & 0 \\
    0 & 0 & 0 & \tfrac{1}{2} & \tfrac{\sqrt{3}}{2} \\
    \tfrac{1}{2} & 0 & 0& -\tfrac{1}{2} & \tfrac{\sqrt{3}}{2}
    \end{pmatrix} ,
\end{equation}
mapping $\ma{U} = (u_{1,x}, u_{2,x},u_{2y},u_{3,x},u_{3,y})$ to
$\ma{E} = (e_1,e_2,e_3,e_4,e_5 )$.  The equilibrium matrix,
mapping $\ma{T} = (t_1,t_2,t_3, t_4,t_5)$ to $\ma{L}=-\ma{F} =
-(f_{1,y},f_{2,x},f_{2,y},f_{3,x},f_{3,y})$ constructed from
Eq.~(\ref{eq:forces-2}), is trivially equal to $\ma{C}^T$. Both
$\ma{Q}$ and $\ma{C}$ are square invertible matrices: their
nullspaces are empty, both $N_0=0$ and $M=0$, and the system is
isostatic as required. Thus, the tensions on the bonds are
uniquely determined by the forces on the sites and vice versa:
the frame is \emph{statically determinate}. And the elongations
of the bonds are uniquely determined by the site displacements
and vice versa: the frame is \emph{kinematically determinate}.
Thus, an alternative, and perhaps preferable, definition of an
isostatic frame is that it be both statically and kinematically
determinate. Another way of dealing with the trivial zero modes
is to introduce ``reaction forces" to yield $8 \times 8$
matrices $\ma{Q}$ and $\ma{C}$.

\subsection{The dynamical matrix}
So far, we have only discussed tensions and stretches of bonds
without specifying any relation among them.  In our
``ball-and-spring" frames, each bond $\beta$ is occupied by a
Hooke's law spring whose energy is half its spring constant
$k_b$ times the square of its elongation.  Let $\ma{k}$ be the
$N_B \times N_B$ diagonal matrix of spring constants, then the
elastic energy of the lattice is
\begin{equation}
V_{\text{el}} = \frac{1}{2} \ma{E}^T \ma{k} \ma{E} =
\frac{1}{2}\ma{U}^T \ma{K} \ma{U} ,
\label{eq:Vel}
\end{equation}
where
\begin{equation}
\ma{K} = \ma{Q} \ma{k} \ma{Q}^T =\ma{C}^T \ma{k} \ma{C}
\end{equation}
is the $dN \times dN$ \emph{stiffness} matrix. Normal-mode
frequencies depend on mass as well as the stiffness matrix. The
kinetic energy requires the introduction of a mass matrix
$\ma{M}$.  We will restrict our attention of frames in which
the mass of all mass points is equal to $m$, in which case
$\ma{M} = m \,\ma{I}$, where $\ma{I}$ is the unit matrix, and
the kinetic energy is
\begin{equation}
E_{\text{kin}}=
\tfrac{1}{2}m  \dot{\ma{U}}^T \dot{\ma{U}},
\end{equation}
where $\dot{\ma{U}}$ is the velocity vector. Normal modes are
then eigenvectors of the \emph{dynamical} matrix:
\begin{equation}
\ma{D} =
\tfrac{1}{m} \ma{K} .
\end{equation}
The Lagrangian in the presence of eternal loads is thus
\begin{equation}
L = \tfrac{1}{2}m\dot{\ma{U}}^T \dot{\ma{U}}-V_{\text{el}}-\ma{U}^T \ma{L} ,
\end{equation}
and the equation of motion is
\begin{equation}
m \ddot{\ma{U}} =-\frac{\partial{V}_\text{el}}{\partial \ma{U}^T} -\ma{L}= - \ma{K} \ma{U} -\ma{L}
=\ma{F} - \ma{L} ,
\end{equation}
which vanishes when the external load $\ma{L}$ is equal to the
force $\ma{F} = -\ma{K}\ma{U} = - \ma{Q} \ma{T}$ exerted by
bond stretching. Note that the equilibrium matrix can be used
to calculate $\ma{F}$ whether or not the system is in static
equilibrium or not.  On the other hand $\ma{L}= \ma{Q}\ma{T}$
only in equilibrium when there is no acceleration (assuming no
friction forces).

\section{The elastic limit \label{sec:elas-lim}}
\subsection{Strain and the elastic energy}
Strain is a measure of macroscopic distortions of an elastic
medium. The macroscopic deformation tensor
$\bm{\lambda}=\bm{I}+ \bm{\eta}$, where $\bm{I}$ is the unit
tensor, determines displacements at boundary sites $s_B$ of
either a finite sample or the periodic box under periodic
boundary conditions:
\begin{equation}
\Xv(s_B) = \bm{\lambda}\Rv (s_B) , \text{   or   }
\uv(s_B) = \bm{\eta}\, \Rv(s_B),
\label{eq:deformation1}
\end{equation}
and the macroscopic strain tensor is
\begin{equation}
\stm=\tfrac{1}{2}(\bm{\lambda}^T \bm{\lambda} - \bm{I}) =
\tfrac{1}{2}(\bm{\eta}+\bm{\eta}^T +\bm{\eta}^T \bm{\eta} )
\approx \tfrac{1}{2}( \bm{\eta}+\bm{\eta}^T ),
\end{equation}
where the final form is the linearized limit, which is all that
concerns us here.

The elastic energy density associated with the macroscopic
strain is
$$
f_{\text{el}} = \tfrac{1}{2}K_{ijkl} \st_{ij} \st_{kl} ,
$$
where $K_{ijkl}$ is the elastic constant tensor and the
Einstein convention on repeated indices is understood. The
elastic strain $\st_{ij}$ is symmetric and has $a_d=d(d+1)/2$
independent components in $d$ dimensions. It can be expressed
\cite{Kittel1971,AshcroftMer1976} as an $a_d$-dimensional
vector (Voigt notation), which in two dimensions takes the
form,
\begin{equation}
\stm_V=(\st_{xx},\st_{yy}, \st_{xy} ).
\end{equation}
The elastic tensor is then an $a_d \times a_d$ matrix, which in
two dimension is
\begin{equation}
\mbb{K} =
    \begin{pmatrix}
    K_{xxxx} & K_{xxyy} &2 K_{xxxy} \\
    K_{xxyy} & K_{yyyy} &2 K_{yyxy} \\
   2K_{xxxy} & 2 K_{yyxy} & 4 K_{xyxy}
    \end{pmatrix}
    \xrightarrow{\text{isotropic}}
    \begin{pmatrix}
    B + \msh & B-\msh & 0 \\
    B-\msh & B+ \msh & 0 \\
    0 & 0 & \msh
    \end{pmatrix} ,
\label{eq:elas-matrix}
\end{equation}
where the final form is the isotropic limit, where $B=\lambda +
\mu$ is the bulk modulus and $\msh=\mu$ is the shear modulus
with $\lambda$ and $\mu$ the standard Lam\'{e} coefficients
\cite{Landau1986}. The linearized Cauchy stress tensor is
\begin{equation}
\sigma_{ij} = K_{ijkl} \st_{kl} .
\end{equation}

Mechanical stability requires that all $a_d$ eigenvalues of the
Voigt elastic matrix $\mbb{K}$ be positive. Thus the elastic
energy of an elastically stable system can be expressed as a
sum of squares of $a_d$ independent linear combinations of
strains (the eigenvectors of the elastic matrix) with positive
coefficients (the eigenvalues of the elastic matrix). We shall
see shortly that some or all of the eigenvalues of the elastic
matrix in lattices at or near the Maxwell limit may be zero.

\subsection{Elastic limit and States of Self Stress}

Calculations of the elastic tensor require some method of
maintaining strain.  The usual picture is that boundary sites
of a finite frame are clamped to conform with
\eref{eq:deformation1}.  Since these sites are fixed, their
displacements and the forces associated with them do not enter
the calculation of $\ma{Q}$ and $\ma{C}$.  An alternative
approach, which we implement, is to apply periodic boundary
conditions (PBCs) to the frame, which may or may not be
composed of repeated unit cells.  In this approach, it is the
boundaries of the periodic cell that satisfy
\eref{eq:deformation1}.

If the positions of all sites, rather than just boundary sites,
displace according to Eq.~(\ref{eq:deformation1}), the elastic
distortion is \emph{affine}.  Under such distortions, the
relative displacement of the sites associated with bond $\beta$
is
\begin{equation}
\uv(s'_\beta) - \uv(s_\beta)= \bm{\eta}\,(\Rv(s'_\beta) -
\Rv(s_\beta) ) = \bm{\eta}\, \bv_\beta ,
\end{equation}
and the affine stretch of bond $\beta$ is
\begin{equation}
e_\beta^{\text{aff}} = \bhv_\beta^T \,\bm{\lambda} \, \bv_\beta =
\hat{b}_{\beta,i} \st_{ij} b_{\beta,j} ,
\label{eq:affine-stretch}
\end{equation}
where we used the fact that $\hat{b}_{\beta,i} b_{\beta,j}$ is
symmetric in $i$ and $j$ to convert $\bm{\eta}$ to $\stm$ in
the linearized limit. The affine elastic energy density is then
\begin{equation}
f_{\text{el}}^{\text{aff}} = \frac{1}{2V} \,
\ma{E}_{\text{aff}}^T \,\ma{k} \,\ma{E}_{\text{aff}} ,
\end{equation}
where $V$ is the volume and $\ma{E}_{\text{aff}}$ is the vector
of affine elongations $e_\beta^{\text{aff}}$.

Affine response throughout a sample is the exception rather
than the rule.  It is guaranteed to occur in absolutely
homogeneous systems and in lattices with one site per unit
cell, but it can occur in certain other systems with special
relations among elastic constants or special arrangements of
sites in periodic systems with multi-site unit cells (for
example, as we shall see, in the kagome lattice). Generally,
however, the forces at at least some sites under an affine
strain imposed by macroscopic stain at the boundary are
nonzero, and these sites will relax to positions of zero force
(or equivalently until the energy reaches its minimum value),
in which cases their displacements are nonaffine. In
\ref{app:elas-SSS}, we show that the result of this relaxation
is that the elastic energy density becomes
\cite{PellegrinoCal1986,Pellegrino1993,Goodrich2014}
\begin{equation}
f_{\text{el}}  =  \frac{1}{2V} \ma{E}_{\text{aff},s}^T [(\ma{k}^{-1})_{ss})]^{-1}
\ma{E}_{\text{aff},s}
\xrightarrow{\ma{k} \rightarrow k\,\ma{I}}
\frac{k}{2V} \sum_{\alpha} (\ma{E}_{\text{aff}}\cdot \hat{\ma{t}}_\alpha)^2
\label{eq:elastic-self-stress}
\end{equation}
where $\ma{E}_{\text{aff},s}$ and $(\ma{k}^{-1})_{ss}$ are the
projections of $\ma{E}_{\text{aff}}$ and $\ma{k}^{-1}$ onto
$\ker(\ma{Q})$, and $\hat{\ma{t}}_\alpha$ is the $\alpha$th
orthonormal basis vector of $\ker(\ma{Q})$. Thus, only the
projections of the affine displacement vectors onto states of
self-stress contribute to the elastic energy.

Equation (\ref{eq:elastic-self-stress}) encodes a great deal of
information.
\begin{itemize}
\item First, it shows that lattices cannot be elastically
    stable unless they have \SSSs in the presence of
    conditions that constrain the macroscopic strain - a
    simple reflection of the fact that forces on each site
    must be zero once equilibrium is reached in the
    presence of imposed strains, which necessarily induce
    bond tension.
\item Second, only those states of self-stress with a
    nonzero overlap with the affine bond elongations
    contribute to the elastic energy. These states
    necessarily traverse the sample, and they are
    \emph{load-bearing}. The straight filament of
    \fref{fig:selftstress2}(c) (wound to a circle -
    \fref{fig:selftstress2}(d)), whose bonds all have the
    same sign of tension provides an example of a
    load-bearing state, whereas the zigzag state of
    \fref{fig:selftstress2}(e) and the localized crossed
    square of \fref{fig:selftstress1}(d) does not.
\item Third, because it is a sum of squares of linear
    combinations of strain, it shows that there must be at
    least $a_d =d(d+1)/2$ load-bearing \SSSs to produce an
    elastically stable system with an elastic matrix with
    $a_d$ positive eigenvalues.
\end{itemize}

\subsection{Isostaticity and periodic boundary conditions\label{ssec:Iso-Per}}

In \sref{ssec-Max-SSS}, isostatic lattices were defined as ones
that are both kinematically ($N_0 = f(d)$) and statically
determinate ($N_S=0$).  This definition is unambiguous for
finite free frames.    It would seem natural to define an
isostatic frame under PBCs in the same way, but there is a
problem with this definition \cite{GuestHut2003}. Under PBCs,
the shape and size of the frame boundary is fixed; the
compatibility matrix (and by extension, the equilibrium matrix)
applies to displacements and does not apply to changes in the
shape or size of the periodic boundary \cite{Dagois-Heck2012,
GoodrichNag2012,GoodrichNag2014}, which are described by
macroscopic strain. In order for a lattice to be elastically
stable, it must have at least $d(d+1)/2$ \SSSsa. Reference
\cite{GuestHut2003} defines a lattice under PBCs to be
statically determinate if $N_S=d(d+1)/2$ and kinematically
determinate if $N_0 = d$, but it does not propose applying the
terms \emph{isostatic} to such systems.  We propose calling a
frame under PBCs isostatic if $N_0 =d$ and $N_S$ lies between
$1$ and $d(d+1)/2$. This corresponds more or less to common
usage in the jamming literature.  If greater precision is
required, the value of $N_S$ can be specified via the term
$N_S$-isostatic.

The above discussion applies to any frame, whether it is a
lattice with unit cells on a Bravais lattice or not, and in
particular to finite-frame approximations to infinite random
systems such as those encountered studies of jamming.  These
frames can be subjected to PBCs by wrapping them on a torus to
create a single cell in which sites separated by repeat
distances of the torus are identified.  We will refer to these
as toroidal PBCs. Since there is only one cell with randomly
placed sites, there is no wavenumber $\qv$ to index the $dN$
vibrational states. Alternatively, the single (large) cell can
be periodically repeated in a Bravais lattice, in which case
there are $dN$ vibrational bands with wave-number-dependent
frequencies.  The $\qv=0$ limit of these bands are the
vibrational states under the toroidal PBCs and are generally
the ones of physical interest, though interesting information
about the stability of jammed structures can be obtained from
an examination of spectrum at $\qv \neq 0$
\cite{Schoenholz2103}.

Lattices under periodic boundary conditions are special.  They
consist of $N_c$ periodically repeated unit cells with $n$
sites and $n_b$ bonds, and the \Ith reads $N_0-N_S = (dn -
n_b)N_c$ .  In a periodic Maxwell lattice, $n_b=dn$, and
$N_0=N_S$. Thus if $N_0$ has its minimum value of $d$, $N_S=d$,
and under the definition proposed above, such a lattice would
be isostatic. It is impossible to have $N_0=d$ and for $N_S$ to
have any other value than $d$ in the interval between $1$ and
$d(d+1)/2$ because a change in $N_S$ requires a change in $n$
or $n_b$ and thus of $dn-n_b$, which would lead to a change in
$N_0-N_S$ of order $N_c \gg 1$ rather than of order one. Thus
we can uniquely define an \emph{isostatic periodic lattice
under} PBCs to be one with $N_0=N_S=d$.

Finite frames can be constructed from ones subjected to PBCs by
cutting bonds along the periodic boundaries and ``liberating"
the sites attached to them. Since opposite sides of the
boundary are equivalent, it is only necessary to cut bonds
along half of the boundary to liberate a full free lattice.
Thus, an $N_x \times N_y$ free square lattice is obtained by
cutting $N_x+N_y$ bonds in the lattice under PBCs. The cutting
process thus reduces the number of bonds by of order
$N^{(d-1)/d}$ in a lattice of $N$ sites in $d$ dimensions and
increases $N_0-N_S$ by the same amount.  If the periodic
lattice is isostatic, then there are necessarily of order
$N^{(d-1)/d}$ extra zero modes in the finite lattice. If on the
other hand, the lattice under PBCs has of order $N^{(d-1)/d}$
\SSSs that are removed on cutting, there may be no increase in
zero modes at all upon cutting.

\section{Periodic Lattices\label{sec:periodic-lattices}}

Our primary interest is in periodic lattices, and we review
here notation that we will use to describe them.

\subsection{Notation}
A general lattice has $N_c$ unit cells consisting of $n$ sites
and $n_b$ bonds so that the total number of sites and bonds
are, respectively, $N= N_c n$ an $N_B = N_c n_b$. The Bravais
lattice vectors $\bm{R}_\la$, where $\la=(l_1, \cdots l_d )$
with each $l_i$ an integer, are linear combinations of the
primitive translation vectors $\bm{a}_j$, $j=1, \cdots d$:
\begin{equation}
\bm{R}_{\la} = \sum_{j}^d l_j\bm{a}_j .
\end{equation}
The positions of sites and bonds in cell $\la$ in the
undistorted lattice with a basis are
\begin{eqnarray}
\bm{R}_{\la, \mu} & = & \bm{R}_\la + \bm{r}_{\mu},  \qquad \mu = 1, \cdots , n \nonumber\\
\bm{R}_{\la,\beta} & = & \bm{R}_\la + \bm{r}_\beta, \qquad \beta = 1, \cdots , n_b ,
\end{eqnarray}
where $\bm{r}_\mu$ and $\bm{r}_\beta$ are, respectively the
positions of the sites and bonds relative to the origin of the
unit cell. The positions of lattice sites in a distorted
lattice are
\begin{equation}
\bm{X}_{\la,\mu} = \bm{R}_{\la,\mu} + \bm{u}_{\mu}(\la) ,
\end{equation}
where $\bm{u}_\mu (\la)$ with Cartesian components $u_{\mu,i}
(\la) \equiv u_\sigma (\la)$ is the displacement vector of site
$(\la,\mu)$. The components of $\ma{U}$ are thus the $dN$
displacements $u_\sigma(\la)$. The Cartesian components
$f_\sigma (\la)$ of the force vector  $\bm{f}_\mu (\la)$ are
the $dN$ components of $\ma{F}$.  The $N_B$ bond elongations
$\ee_\beta (\la)$ and bonds tensions $t_\beta (\la)$ are the
components of the of $\ma{E}$, and $\ma{T}$, respectively.
Fourier transforms in periodic lattices are defined in the
usual way in terms of wavenumbers $\qv$ in the first Brillouin
zone:
\begin{align}
\bm{u}_\mu (\la) & = \frac{1}{N_c} \sum_{\bm{q}} e^{i \bm{q}\cdot (\bm{R}_\la + \bm{r}_\mu)}
\bm{u}_\mu (\bm{q}), \qquad
& \bm{u}_\mu (\bm{q}) & = \sum_{\la} e^{- i \bm{q}\cdot (\bm{R}_\la + \bm{r}_\mu)}
\bm{u}_\mu (\la ) \\
t_\beta (\la) &= \frac{1}{N_c} \sum_{\bm{q}} e^{i \bm{q}\cdot (\bm{R}_\la + \bm{r}_\beta)}
t_\beta (\bm{q}) \qquad
& t_\beta (\bm{q}) &=\sum_{\la} e^{-i \bm{q}\cdot (\bm{R}_\la + \bm{r}_\beta)}
t_\beta (\la) ,
\end{align}
and similarly for $\bm{f}_\mu$, $\ee_\beta$ and other site and
bond variables. These quantities can also be defined without
the site and bond basis vectors, $\bm{r}_\mu$ and
$\bm{r}_\beta$ in the exponentials, and we will usually leave
them out.  They will, however, be of use in our discussion of
topological states in \sref{sec:topological}. The components of
the equilibrium matrix $\ma{Q}$ in periodic lattices are
$\Qv_{\sigma,\beta}(\la, \la')$. Its Fourier transform is
\begin{equation}
\Qv_{\sigma \beta} (\bm{q}) = \sum_{\la} e^{-i \bm{q}\cdot (\bm{R}_{\la,\sigma}-\bm{R}_{0,\beta})}
\Qv_{\sigma\beta} (\la,0) .
\end{equation}
Again, the basis vectors in the exponents are not required. The
compatibility matrix is $\Cv (\qv) = \Qv^\dag ( \qv)$.

With these definitions, we can reexpress equations
(\ref{eq:QTF}) and (\ref{eq:CUE}) as
\begin{equation}
\Qv(\bm{q}) \,\mathbf{t}\, (\bm{q}) = -\mathbf{f}(\bm{q}) \qquad
\Cv\, (\bm{q})\, \mathbf{u} (\bm{q}) = \mathbf{e}(\qv) ,
\end{equation}
where the ``math boldface" vectors are defined as
$\mathbf{t}(\qv)=(t_1 (\qv), ...,t_{n_b}(\qv))$ and
$\mathbf{u}(\qv) = (\uv_1 (\qv), ..., \uv_n (\qv))$ and
similarly for $\mathbf{e}(\qv)$ and $\mathbf{f} (\qv)$, and
$\mathbf{Q}(\qv)$ is the $dn \times n_b$ matrix with components
$\bm{Q}_{\sigma \beta}(\qv)$. There is one equation for each of
the $N_c$ values of $\bm{q}$, giving us as many independent
equations as \eref{eq:QTF} and \eref{eq:CUE}. The \Ith applies
to these equations for each $\qv$:
\begin{equation}
n_0(\qv) - n_s(\qv) = dn - n_b ,
\end{equation}
where $n_0(\qv) = \dim\ker(\Cv)$ is the number of zero modes
and $n_s(\qv) = \dim\ker(\Qv)$ is the number of states of self
stress at wavevector $\qv$. Of course, $N_0 = \sum_\qv
n_0(\qv)$ and $N_S = \sum_{\qv} n_s(\qv)$.

In periodic Maxwell lattices, $dn = n_b$, and there are exactly
as many zero modes as states of self stress at each $\qv$.
Under PBCs, there are always $d$ zero modes that arise from
translational invariance, and these are necessarily at $\qv=0$,
implying that there are at least $d$ \SSSs at $\qv=0$.  There
may be more, but each additional \SSS will require an
additional zero mode, which is a $\qv=0$ mechanism. At nonzero
$\qv$, there is no general reason for zero modes to exist, but
if they do, the are necessarily accompanied by \SSSsa.  We will
see that this is a common theme in our study of specific
lattices in \sref{sec:examples}: removing states of self stress
eliminates zero modes and ``gaps" the phonon spectrum.

Following \eref{eq:Vel}, the potential energy in terms of a
periodic harmonic lattice is
\begin{equation}
V_{\text{el}}= \frac{1}{2N_c} \sum_{\bm{q}} \mathbf{e}^\dag (\bm{q})\, \mathbf{k}
\, \mathbf{e}(\bm{q}) =\frac{1}{2N_c}\sum_{\bm{q}}\mathbf{u}^\dag (
\bm{q})
\mathbf{K}(\qv) \mathbf{u}(\bm{q}) ,
\end{equation}
where $\mathbf{k}$ is the $n_b \times n_b$ diagonal matrix of
spring constants, and
\begin{equation}
\mathbf{K} (\qv) = \mathbf{Q}(\bm{q})\, \mathbf{k}\, \mathbf{Q}^\dag(\bm{q}) \equiv
m \mathbf{D} (\qv)
\label{eq:stiffness2}
\end{equation}
is the dynamical matrix. In periodic systems, nearest-neighbor
($NN$), next-nearest neighbor ($NNN$), and further-neighbor
bonds are well defined, and bond vectors can be expressed as
the direct sum of $NN$ and $NNN$ components, e.g. $\mathbf{\ee}
= \mathbf{\ee}_{NN}\oplus\mathbf{\ee}_{NNN}$, and elastic
constant matrix and dynamical matrices can be decomposed into
$NN$ and $NNN$:
\begin{equation}
\mathbf{D} (\qv) = \mathbf{D}_{NN} (\qv)+ \mathbf{D}_{NNN} (\qv) .
\end{equation}

\subsection{The elastic limit}

\subsubsection{The elastic energy\label{ssec:elastic2}}

Under affine strain, the strain of equivalent bonds in
different unit cells are identical, and we can describe affine
strain in terms of the $n_b$ dimensional vector
$\mathbf{e}_{\text{aff},s}$. As a result, the elastic energy
depends only on the projection of the affine strain onto the
$\qv=0$ \SSSsa. Following \eref{eq:elastic-self-stress} and
\ref{app:elas-SSS}, the elastic energy density is thus
\begin{equation}
f_{\text{el}} = \frac{1}{2 v_c} \mathbf{e}_{\text{aff},s}^T (\mathbf{k}_{ss}^{-1})^{-1}
\mathbf{e}_{\text{aff},s} \rightarrow \frac{1}{2v_c}k \sum_\alpha
(\mathbf{e}_{\text{aff}}\cdot \hat{\mathbf{t}}_\alpha )^2 ,
\label{eq:elastic-self-stressP}
\end{equation}
where $v_c = N_c/V$ is the volume of a unit cell and
$\mathbf{k}_{ss}$ is the projection of the $\mathbf{k}$ on to
the nullspace of $\mathbf{Q}(\qv=0)$. The final form with
$\hat{\mathbf{t}}_{\alpha}$ ($\alpha = 1, ..., n_s(0)$) basis
vectors for the null space of $\mathbf{Q}(\qv=0)$ applies when
there is a single spring constant $k$.

Equation (\ref{eq:elastic-self-stressP}) constrains the elastic
energy of \emph{Maxwell} lattices, which depend on the number,
$m_0$, of $\qv=0$ mechanisms and on the overlap of $\qv=0$ SSSs
with the $\qv=0$ affine bond elongation
$\mathbf{e}_{\text{aff}}$. Consider periodic Maxwell lattices,
for which with $N_0 = N_S$.
\begin{enumerate}
\item $m_0=0$:  In this case, there are exactly $d$ zero
    modes and exactly $d$ SSSs.  There are now two
    possibilities:
    \begin{enumerate}
    \item All $d$ \SSSs have a nonzero overlap with the
        affine bond elongations, and the elastic matrix
        \eref{eq:elas-matrix} then has $d$ positive and
        $d(d+1)/2- d = d(d-1)/2$ zero eigenvalues,
        which correspond to zero-energy elastic
        deformations, now referred to as Guest modes
        \cite{GuestHut2003}. This case corresponds to
        what we call an isostatic periodic lattice. As
        we shall see, this is the situation for the
        square, twisted kagome, and topological kagome
        lattices [\sref{sec:top-lattice}] in which
        there are two positive eigenvalues and one
        zero-energy elastic mode.
    \item Fewer than $d$ \SSSs have a nonzero overlap
        with affine bond elongations, and as a result
        there are fewer than $d$ positive eigenvalues
        to the elastic matrix and more than $d(d-1)/2$
        zero-energy elastic distortions.  The zigzagged
        square lattice to be discussed in
        \sref{sec:other-lattices} provides an example
        of this behavior. Finite periodic approximates
        to infinite-unit cell systems, such as packed
        spheres at the jamming transition
        \cite{GoodrichNag2012,GoodrichNag2014} and
        randomized quasi-crystalline Penrose tilings
        [\sref{sec:other-lattices}]
        \cite{StenullLub2014} exhibit smaller and
        smaller overlap as the order of the approximate
        increases, leading to shear moduli that vanish
        as $1/n$ as $n\rightarrow \infty$.
    \end{enumerate}
\item $m_0 >0$, and there are $d+m_0$ \SSSsa, that may or
    may not have an overlap with affine bond stretches.  If
    more than $d$ overlap, the additional \SSSs stabilize
    elastic response relative that of the lattice with
    $m_0= 0$, and we are presented with the curious
    situation in which additional zero modes increase
    elastic rigidity. The usual cause of this effect is the
    appearance of sample traversing straight filaments that
    support macroscopic stress, but they also, as we have
    seen, introduce additional infinitesimal zero modes.
    The untwisted kagome lattice, with three $\qv=0$ states
    of self stress produced by parallel filaments along
    three independent directions, is elastically stable
    with all eigenvalues of the elastic matrix positive. In
    spite of extra mechanisms it is possible for only some
    or even none of the \SSSs to overlap with affine bond
    elongations, in which case the elastic energy can even
    fall to zero. The latter situation occurs in
    unrandomized periodic approximates to Penrose tilings
    \cite{StenullLub2014} in which there are of order
    $\sqrt{n}$ zero modes and states of self-stress but in
    which all elastic moduli vanish
    [\sref{sec:other-lattices}].
\end{enumerate}

\subsubsection{Stiffness matrix of $2d$ periodic lattices with Guest modes}

Two-dimensional periodic lattices with one or two $\qv=0$ \SSSs
and two $\qv=0$ zero modes have two and one Guest modes,
respectively.  Fully gapped isostatic lattices have two such
\SSSs, but so do lattices that have additional zero modes at
non-zero $\qv$. When there is only one Guest mode,  the elastic
matrix of \eref{eq:elas-matrix}, which acts on the strain
vector $\stm_V$, has one zero eigenvalue with normalized
eigenvector $\ve_0=(\vs_{0,1},\vs_{0,2},\vs_{03})$ (i.e., the
strain of the Guest mode with amplitude $U_G$ is $U_G \ve_0$)
and two positive eigenvalues, $K_1$ and $K_2$, with respective
associated eigenvectors $\ve_1$, and $\ve_2$ so that
\begin{equation}
\mbb{K}_{ij}= K_1 \vs_{1,i}\vs_{1,j} + K_2  \vs_{2,i}\vs_{2,j}
\label{eq:Guest-K}
\end{equation}
The long-wavelength stiffness matrix $\ma{K}$ is determined by
$\mbb{K}$, and as \ref{App:Guest} shows, its determinant
depends only on $K_1$, $K_2$, and the Guest-mode eigenvector
$\ve_0$:
\begin{equation}
\det \Kv(\qv)=\frac{1}{4} K_1 K_2 \,(\vs_{02} q_x^2 - 2 \vs_{03} q_x q_y
+\vs_{01} q_y^2 )^2.
\label{eq:GuestK}
\end{equation}
Since $\Kv = k \Qv \cdot \Qv^\dag$, this implies that
\begin{equation}
\det \Cv = (\det \Qv)^* =  (c/2) \sqrt{K_1 K_2/k}\,\, (\vs_{02} q_x^2 - 2 \vs_{03} q_x q_y
+\vs_{01} q_y^2 ) ,
\label{Eq:detCv}
\end{equation}
where $c$ is some unit amplitude complex number. This simple
form implies that in the $q\rightarrow 0$ limit the zeroes of
$\det\Cav(\qv)$, which occur at \cite{square}
\begin{equation}
q_y = \frac{q_x}{\vs_{01}}\left(\vs_{03} \pm \sqrt{\vs_{03}^2
-\vs_{01}\vs_{02}} \right) ,
\end{equation}
depend only on $\ve_0$ and not on the elastic moduli $K_1$ and
$K_2$. The negative of the quantity under the square root is
proportional to the determinant of the strain of the Guest
mode:
\begin{equation}
\det \stm_G = \st_{xx}\st_{yy}-\st_{xy}^2 = U_G^2 (\vs_{01}\vs_{02}-\vs_{03}^2) .
\end{equation}
Thus, to order $q^2$, there is a zero mode with real values of
$q_x$ and $q_y$ if $\det \stm_G <0$ and complex values when
$\det \stm_G >0$. The complex values of $\qv$ correspond to
decaying or growing surface modes. The zeros at real values of
$\qv$ can either become complex when higher-order terms in the
$q$-expansion are included, in which case they again correspond
to surface modes albeit with an inverse decay length quadratic
rather than linear in the surface wavenumber, or they can
remain real.  In the latter case, they can occur at specific
values of $\qv$, in which case they are isolated.  These
points, which have a topologically protected winding number are
called Weyl points, and they have received considerable
attention recently in relation to Dirac semimetals and
topological insulators
\cite{km05b,Aji2012,WeiChao2012,WeiWang2012,AjiHe2013,ZhuAji2013,PhillipsAji2014,SosenkoWei2014,WeiChao2014}
and in photonic materials
\cite{LuJoan2012,LuJoan2013,LuJoan2014}. The distorted square
lattices of \fref{fig:square-kagome}(b) has isolated Weyl
points \cite{square}.  It is also possible to have Weyl lines
rather than isolated points.  These occur in distorted versions
of the $3d$ pyrochlore lattice \cite{StenullLub2014b} and
gyroid photonic crystals \cite{LuJoan2013}.

\section{Examples of periodic lattices\label{sec:examples}}
%\subsection{Discussion}
In this section, we will study three simple lattices: the
square, kagome, and twisted kagome lattices depicted in
\fref{fig:square-kagome}.  They provide specific examples of
the phenomena discussed in the preceding section. Section
\ref{ssec:square-lattice} and \ref{ssec:kagome} rely heavily of
reference \cite{Souslov2009} and section
\ref{ssec:twisted-kagome} on reference \cite{SunLub2012}.

\subsection{Square Lattice\label{ssec:square-lattice}}
\subsubsection{The nearest-neighbor lattice}
\begin{figure}
\centering
\includegraphics{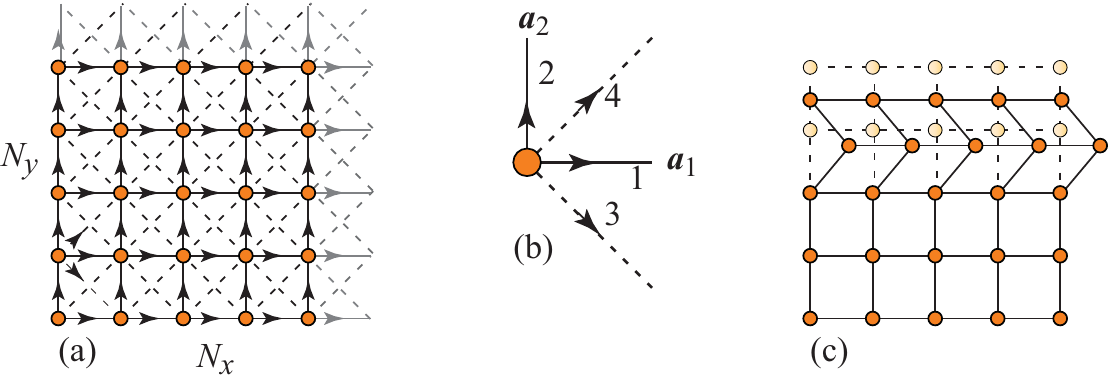}
\caption{(a) An $N_x=5$ by $N_y=5$ square lattice with $NN$ (full lines) and $NNN$
(dashed lines) bonds showing bond directions for the calculation of self stress. The gray bonds
at the right and upper boundaries are the $N_x+N_y = 10$ $NN$ and $2(N_x+N_y-1)$
$NNN$ bonds that must be cut to from a $N_x \times N_y$ lattice under periodic boundary
conditions to produce the $N_x \times N_y$ lattice under free boundary conditions.
(b) A unit cell showing its two Bravais lattice vectors $\av_1$ and $\av_2$ and
it two $NN$ ($1$ and $2$) and two $NNN$ ($3$ and $4$) bonds. (c) A finite mechanism
of the finite square lattice with $NN$ bonds only.  This is also an infinitesimal mechanism in the
periodic lattice.}
\label{fig:square1}
\end{figure}

\begin{figure}
\centering
\includegraphics{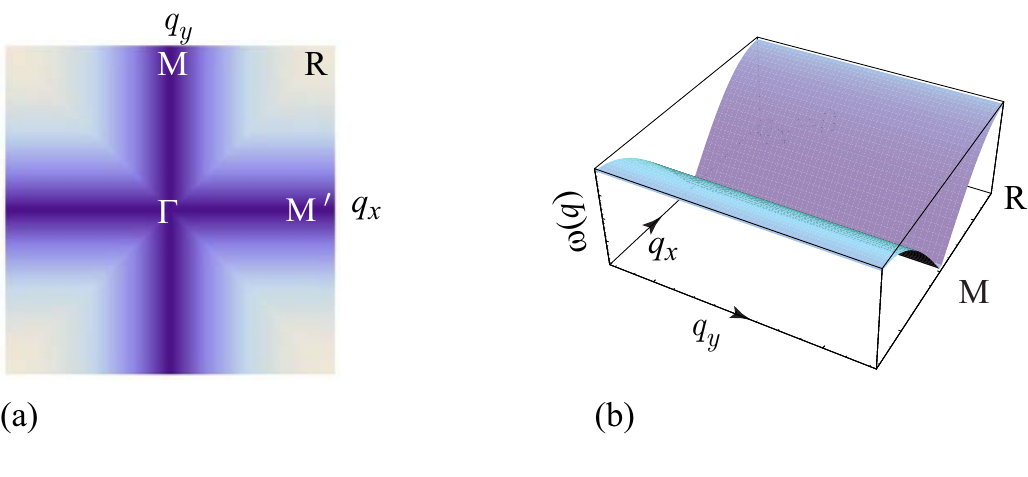}
\caption{(a) Density plot of the lowest-frequency mode of the $NN$ square
lattice showing lines of zero modes running from the Brillouin zone center
at $\Gamma$ to the midpoints $M$ and $M'$ of the zone edge. (b) $3d$ plot of the single
mode $\omega_x(\qv)$, which is independent of $q_y$ and equal to zero for $q_x=0$.
\emph{After reference \cite{Souslov2009}}}
\label{fig:NNsq-spectrum}
\end{figure}

The periodic square lattice (\fref{fig:square1}) with only
nearest neighbor bonds, along $\bv_1 = a(1,0)$ and
$\bv_2=a(0,1)$, and one site and two bonds per unit cell is the
simplest example of a periodic Maxwell lattice. Many of its
properties follow without calculation from the observations of
the previous sections.  It consists of straight filaments in
orthogonal directions, each of which develops an \SSS when
placed under periodic boundary conditions. Thus an $N_x\times
N_y$ lattice has $N_S = N_x+N_y$ states of self stress and
\begin{equation}
N_0 = N_x + N_y
\label{eq:square-N0}
\end{equation}
zero modes, which, apart from the trivial translation modes,
are infinitesimal floppy modes in which any row or column is
rigidly displaced as shown in \fref{fig:square1}(c).  These
rigid displacements form a basis for the null space of
$\ma{C}$, and their Fourier transforms do as well.  If
filaments parallel to the $x$-axis are rigidly displaced, then
$q_x=0$, and the Fourier transform of a set of displaced
filaments parallel to the $x$-axis are indexed by a wavevector
$\qv_1=(0,q_y)$. Similarly rigidly displaced filaments parallel
to the $y$-axis are indexed by $\qv_2=(q_x,0)$. There are $N_x$
independent values of $q_x$ and $N_y$ of $q_y$. Each wavevector
represents a zero mode, and since there are $N_x$ independent
values of $q_x$ and $N_y$ independent values of $q_y$, these
account for all of the $N_x+N_y$ zero modes, linear
combinations of which account for rigid translations and
rotations. Thus, the bulk phonon spectrum for the periodic
square Maxwell lattice has two lines of zero modes running from
the center $\Gamma$ of the Brillouin Zone to the midpoint $M$
and $M'$ at the zone boundaries as shown in
\fref{fig:NNsq-spectrum}.

An $N_x\times N_y$ lattice with free boundary conditions can be
created by cutting and removing $N_x+N_y$ bonds from an
$N_x\times N_y$ lattice under PBCs. Thus, the cut lattice has
only $2N_x N_y - (N_x+N_y)$ bonds [\fref{fig:square1}]. In the
cutting process, however, all $N_x+ N_y$ \SSSs are lost, and
the zero mode count of the free lattice is the same as that of
the periodic lattice.  To infinitesimal order in the
displacement, the modes themselves are identical in the two
cases. In the free lattice, however, the modes are nonlinear as
shown in \fref{fig:square1}(c). The bulk zero modes, which are
seen under periodic boundary conditions, exhaust the \Ith
count. There are no additional zero modes at the surface of the
free lattice.  It is clear that any distortion of one of the
surfaces parallel to the $x$- or $y$-axis of the free lattice
will be transmitted across the entire sample by the rigid
filaments, which support the \SSS under periodic boundary
conditions. So, zero modes are not localized near the surface;
surface distortions have infinite penetration depth and thus do
not constitute surface modes.

There are exactly two \SSSs at $\qv=0$, and they clearly
overlap affine bond elongations, which are equal to $
a\st_{xx}$ for bonds parallel to the $x$-axis and to
$a\st_{yy}$ for bonds parallel to the $y$ axis. Thus the
elastic energy density is simply
\begin{equation}
f_{\text{el}}=\tfrac{1}{2}k\, (\st_{xx}^2 + \st_{yy}^2) .
\end{equation}
There are two independent lattice distortions, $\st_{xx}$ and
$\st_{yy}$, that cost energy and one $\st_{xy}$ that does not
in agreement with the analysis of \sref{ssec:elastic2}.

\subsubsection{The next-nearest-neighbor lattice}
\begin{figure}
\centering
\includegraphics{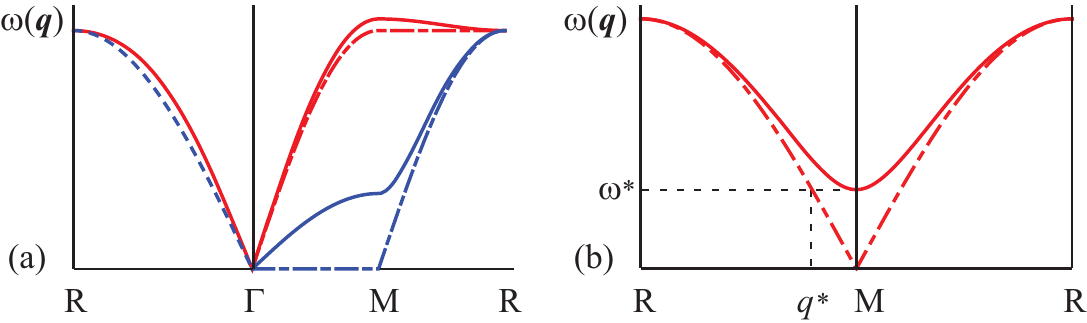}
\caption{(a) Comparison of the square-lattice phonon frequencies along symmetry directions in
the BZ with $k'=0$ (dashed lines) and $k'>0$ (full lines) The blue curves depict
$\omega_1(k',q)$, the lower and the red curves, $\omega_2(k',q)$, the higher
of the two modes.  The single blue dashed line
from $R$ to $\Gamma$ represents the curves $\omega_1(0,q)$, $\omega_1(k1,q)$,
and $\omega_2(0,q)$, all of which are the same. (b) Frequencies for $k'=0$ and $k'>0$
along the BZ edge from $R$ to $M$ and back to $R$. When $k'=0$, the frequency grows linearly with $q$ away from
$M$. When $k'>0$, there is a gap defining the characteristic
frequency $\omega^* = \sqrt{k'}$ and a length scale
when $l^* \sim 1/\omega^*$. }
\label{fig:k'>0sqdis}
\end{figure}

Introducing $NNN$ bonds,  $\bv_3 = a(1,-1)$ and $\bv_4
=a(1,1)$, increases the bond number without changing the number
of sites. If $NNN$ bonds are added one at a time, initially no
additional \SSSs are created, and each additional bond
decreases the zero-mode count by one. As additional bonds are
added, eventually additional states of self stress are created,
for example in isolated cells with two $NNN$ bonds in
configurations like that of \fref{fig:selftstress1}(d).
Consider for simplicity the case with $N_x=N_y$.  If a filament
of equivalent contiguous $NNN$ bonds (pointing either up or
down) traverses the sample, a new \SSS along that line is
created. Thus, if there are no other $NNN$ bonds, the change in
bond number and number of \SSSs relative to those in the state
with no $NNN$ bonds are, respectively, $\Delta N_B = N_x$ and
$\Delta N_S = 1$, leading to a decrease in the number of zero
modes of $\Delta N_0= -N_x +1$ for a total of $N_0 = N_x +1$.
If each unit cell has an upward pointing $NNN$ bond and not a
downward pointing one, $\Delta N_B = N_x^2$. In addition, there
is one \SSS for every $\qv$ except $\qv=0$ for which there are
three for a total of $N_S = N_x^2+2$ leading to $N_0=2$, i.e.,
leaving only the required two zero modes of uniform
translation. Each added down pointing bond increases $N_B$ and
$N_S$ by one leaving $N_0$ fixed at two. Thus, if all $NNN$
bonds are present, there are no zero modes beyond the two at
$\qv=0$, and the spectrum is gapped everywhere except at
$\qv=0$ as shown in \fref{fig:k'>0sqdis}(a). The gap at points
$M$ provides a characteristic frequency,
\begin{equation}
\omega^* = 2 \sqrt{\frac{k'}{m}} ,
\label{eq:omega*}
\end{equation}
and a comparison of the $k'=0$ linear dispersion with the
$k'\neq 0$ dispersion along the zone edge near $M$
[\fref{fig:k'>0sqdis}(b)] yields a characteristic length
\begin{equation}
\ell^* = \frac{a}{2} \sqrt{\frac{k}{k'}} ,
\label{eq:l*}
\end{equation}
which  diverges as $k' \rightarrow 0$.  The approach of
$\omega^*$ and $(\ell^*)^{-1}$ to zero as the Maxwell limit,
$k'\rightarrow 0$, is similar to the behavior of these
quantities near jamming where $\omega^* \sim (\Delta z)$ and
$l^* \sim (\Delta z)^{-1}$ \cite{SilbertNag2005}.  This result
is not altogether surprising, given that weakening the spring
constants of $NNN$ bonds to zero yields the Maxwell limit in
which there is no characteristic frequency or inverse length.
An effective medium analysis \cite{MaoLub2010} of the case in
which $NNN$ bonds are randomly added with probability $p =
(z-z_c)/4$ leads to exactly the same results as jamming.

\subsubsection{Equilibrium and dynamical matrices}
We now investigate how these results follow from explicit
calculations of the equilibrium and compatibility matrices. We
set $\av_1 = a(1,0)$ and $\av_2=a(0,1)$ and  designate the $NN$
bonds $1$ and $2$ to be to the right and above each lattice
site, respectively, and $NNN$ bonds $3$ and $4$ to be along
$\av_1-\av_2$ and $\av_1+\av_2$, respectively, as shown in
\fref{fig:square1}(b). Following the rules outlined in section
\ref{ssec:equil-comp}, the force at site $\la$ is $\fv(\Rv_\la)
= \tv_1(\Rv_\la) -\tv_1(\Rv_\la -\av_1) + \tv_2(\Rv_\la)-
\tv_2(\Rv_\la - \av_2)$, and
\begin{equation}
\Qv_{NN} (\qv)= -
    \begin{pmatrix}
    1-e^{-iq_x a} & 0 \\
    0 & 1-e^{-iq_y a}
    \end{pmatrix} .
\end{equation}
Thus if $q_x = 0$ and $q_y \neq 0$, the one-dimensional null
space of $\mathbf{Q}(\qv)$ is spanned by the unit vector
$\hat{\tv}_x(q_y)= (1,0)$; and if $q_y=0$ and $q_x \neq 0$, it
is spanned by the vector $\hat{\tv}_y(q_x) = (0,1)$.  In other
words, for each $q_y\neq 0$, there is a \SSS for each value of
$q_y$ with independent tensions in bonds parallel to the
$x$-axis that have the same value for every bond in a given
filament and similarly for $q_x\neq 0$. When both $q_x$ and
$q_y=0$, there are two \SSSs.  Thus, there is one state \SSS
for each value of $q_x$ and one for each value of $q_y$ for a
total of $N_S=N_x+ N_y$ \SSSs. The null pace of
$\Cav(\qv)=\Qv_{NN}^\dag (\qv)$, which consists of the set of
zero modes, is similar with one zero mode per $q_x$ and $q_y$
for a total of $N_0 = N_S$ zero modes as required by \Itha. It
consists of rigid displacements of individual rods as already
discussed.

The force at site $\la$ arising from $NNN$ bonds is
$\fv^{NNN}(\Rv_\la)= \tv_3(\Rv_\la)-\tv_3(\Rv_\la - \bv_3)
+\tv_4(\Rv_\la) -  \tv_4(\Rv_\la - \bv_4)$, and the $NNN$
equilibrium matrix is
\begin{equation}
\Qv_{NNN}(\qv) = -\frac{1}{\sqrt{2}}
    \begin{pmatrix}
    1-e^{-i(q_x - q_y)a} & 1-e^{-i(q_x + q_y)a} \\
    -1+e^{-i(q_x - q_y)a} & 1-e^{-i(q_x + q_y)a}
    \end{pmatrix} .
\end{equation}
When both $NN$ and $NNN$ bonds are present, there are four
bonds per unit cell, and the full equilibrium matrix is a the $
2 \times 4$ matrix
\begin{equation}
\Qv (\qv)=
    \begin{pmatrix}
    \Qv_{NN}(\qv) & \Qv_{NNN}(\qv)
    \end{pmatrix} .
\end{equation}
At $\qv=0$, all entries in $\Qv$ are zero, and all $\qv=0$ bond
vectors are in its null space.  Thus, the elastic energy is
simply the expected affine result,
\begin{subequations}
\begin{align}
f & =  \tfrac{1}{2 V}\sum_{\alpha=1}^4 k_\alpha (\bhv_{\alpha, i} \st_{ij} \bv_{\alpha,j})^2 \\
& = \tfrac{1}{2} k (\st_{xx}^2+ \st_{yy}^2)
+ \tfrac{1}{4} k' [(\st_{xx}+\st_{yy} + 2 \st_{xy})^2+
(\st_{xx}+\st_{yy} - 2 \st_{xy})^2] \\
& = \frac{1}{2}K_{11} (\st_{xx}^2 + \st_{yy}^2) + K_{12} \st_{xx}\st_{yy}
+ 2 K_{44} \st_{xy}^2 ,
\end{align}
\end{subequations}
where $K_{11} = k+k'$, and $K_{12} =K_{44} = k'$. As expected,
shear moduli vanish when $k'=0$.

The $NN$ and $NNN$ dynamical matrices are
\begin{subequations}
\begin{eqnarray}
\Dv^{NN} (\qv) & = & 4 \frac{k}{m}
    \begin{pmatrix}
    \sin^2 (q_x a/2) & 0 \\
    0 & \sin^2 (q_y a/2)
    \end{pmatrix} ,\\
\Dv^{NNN}(\qv)& =& 2 \frac{k'}{m}
    \begin{pmatrix}
    1- \cos q_x a \cos q_y a & \sin q_x a \sin q_y a \\
    \sin q_x a \sin q_y a & 1- \cos q_x a \cos q_y a
    \end{pmatrix} .
\end{eqnarray}
\end{subequations}
The spectrum arising from the sum of these dynamical matrices
is shown in \fref{fig:k'>0sqdis}.  When $k'=0$, only $\Dv^{NN}$
contributes, and there are two independent one-dimensional
modes with
\begin{equation}
\omega_{x,y} = 2\omega_0|\sin (q_{x,y}a/2)| ,
\end{equation}
where $\omega_0 = \sqrt{k/m}$. For every $q_y$, $\omega_x(\qv)$
goes to zero with $q_x$ as $c q_x$, where $c=\omega_0 a$ is the
longitudinal (compressive) sound velocity, and reaches a
maximum of $2 \omega_0$ at the point $M$ on the zone edge [$\qv
= (\pi/a,0)$] and similarly for $\omega_y(\qv)$ as shown in
\fref{fig:k'>0sqdis}. These one-dimensional spectra produce
\cite{Kittel1971} (pages 205-209) a density of states
\begin{equation}
\rho(\omega) = \frac{1}{2 \pi \omega_0 a} \left(1-\frac{\omega^2}{4 \omega_0^2}\right)^{-1/2} ,
\end{equation}
which approaches a constant as $\omega \rightarrow 0$ and
diverges as $\omega \rightarrow 2 \omega_0$ at the zone edge.

\begin{figure}
\centering
\includegraphics{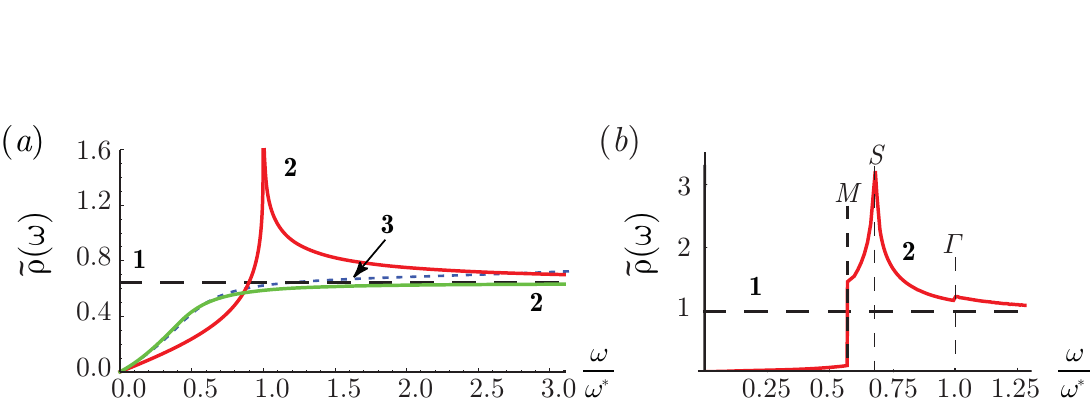}
\caption{Densities of states $\tilde{\rho}(\omega)=\rho(\omega)/\rho(0)$ for
the (a) square and (b) kagome lattices.  The dashed black line (labeled $\bf 1$) is the flat, one-dimensional
$\tilde{\rho}(\omega)$ at $\omega\ll\omega_0$ when $k'=0$, and the full red line (labeled $\bf 2$) is
$\tilde{\rho}(\omega)$ for $k'>0$ showing linear-in-$\omega$ Debye behaviour at
$\omega \ll \omega^*$, van Hove singularities near $\omega = \omega^*$ and constant
behaviour at $\omega >\omega^*$.  Lines $\bf 3$ and $\bf 4$ in (a) are effective medium
results for different probabilities of occupation of NNN bonds showing the washing-out
of the van Hove singularity and the smooth transition from Debye to one-dimensional behavior.
\emph{Adapted from references \cite{Souslov2009} and \cite{MaoLub2010}}}
\label{fig:DOS-sq-kag}
\end{figure}

When $k'>0$, modes [with $\qv = q(\cos\theta, \sin \theta)$]
exhibit a $\cos (4 \theta)$ angular modulation at low frequency
and one-dimensional $k'=0$ behavior at larger $\qv$. When $0<k'
\ll k$, $D_{ij}(\qv)$ is well approximated as a diagonal matrix
with $m D_{xx}(\qv) = k q_x^2 + 4 k' \sin^2(q_y/2)$ with
associated eigenfrequency $\omega_x(\qv) \sim
\sqrt{D_{xx}(\qv)}$. These expressions immediately define the
characteristic frequency of \eref{eq:omega*} at the point
$M=(0,\pi)$ on the BZ edge. The first term in $D_{xx}(\qv)$
represents the long-wavelength one-dimensional $NN$-modes that
are present when $k'=0$, whereas the second represents the
effects of $NNN$ coupling. When $q_x=0$, the only length scale
in the problem is the unit lattice spacing, and no divergent
length scale can be extracted from $D_{xx}( 0 , q_y) $. When
the first term is large compared to the second, $D_{xx}(\qv)$
reduces to its form for the Maxwell $k'=0$ limit, and we can
extract a length by comparing these two terms. The shortest
length we can extract is that of \eref{eq:l*}, which comes from
comparing $k q_x^2$ to $D_{xx}(\qv)$ at point $M$ on the zone
edge as depicted in \fref{fig:k'>0sqdis} If $q_y<\pi$, the
Maxwell limit is reached when $q_x
> q^*$. A similar analysis applies to $D_{yy}(\qv)$ when $q_y >
q^*$. If a square of length $l$ is cut from the bulk, the
wavenumbers of its excitations will be greater than $\pi/l$,
and for $ql^*>1$, all modes within the box will be effectively
those of the lattice without $NNN$ bonds.  This construction is
equivalent to the cutting argument of Wyart et al.
\cite{WyartWit2005b,Wyart2005}.

The characteristic length of \eref{eq:l*} is identical to the
length at which the frequency of the compressional mode
$\omega_x ( q_x\sim 1/l) = \sqrt{k}/l^* ~\sim
\sqrt{K_{11}}/l^*$ becomes equal to $\omega^*$. A meaningful
length from the transverse mode $\omega_x(0, q_y)$ cannot be
extracted in a similar fashion. The full phonon spectrum
[\fref{fig:k'>0sqdis}(a)] exhibits acoustic phonons identical
to those of a standard square lattice at $q\ll 1$ and a saddle
point at the point $M$. Thus, the low-frequency density of
states shown in \fref{fig:DOS-sq-kag}(a) is Debye-like:
$\rho(\omega) \sim \omega/\sqrt{k k'}$ with a denominator that,
because of the anisotropy of the square lattice, is
proportional to the geometric mean of longitudinal and
transverse sound velocities rather than to a single velocity.
In addition $\rho(\omega)$ exhibits a logarithmic van Hove
singularity at $\omega^*$ and approaches the one-dimensional
limit $(1/\pi)/\sqrt{k}$ at $\omega^* \ll \omega \ll
2\sqrt{k}$.  The frequency $\omega^*$ [\eref{eq:omega*}] is
recovered by equating the low-frequency Debye form at
$\omega^*$ to the high-frequency isostatic form of the density
of states.

\subsection{Kagome lattice\label{ssec:kagome}}
\begin{figure}
\centering
\includegraphics{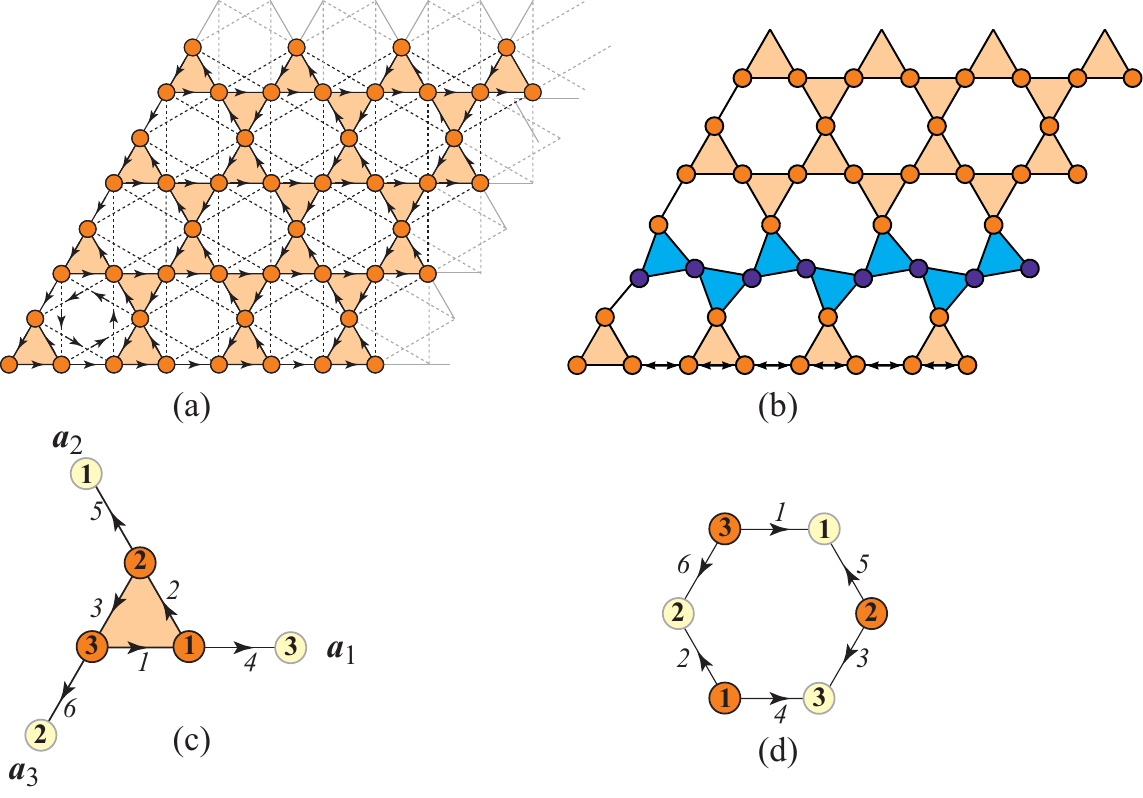}
\caption{(a) An $N_x=5$ by $N_y = 5$ kagome lattice showing $NN$ (full lines) and
$NNN$ (dotted lines) bonds.  The gray bonds along the right and upper edges
are the $2(N_x+N_y-1)$ $NN$ and $4(N_x+N_y-1)$ $NNN$ bonds that must be cut from a
$N_x \times N_y$ lattice under periodic boundary conditions to produce the
the free $N_x \times N_y$ lattice. (b) Representation of a zero modes.
(c) and (d) two different symmetric versions of the kagome unit cells showing labeling of sites and $NN$ bonds.
The vectors $\av_1$ and $\av_2$, $\av_1$ and $-\av_3$, or any other similar pair can serve
as basis vectors for the triangular Bravais lattice.}
\label{fig:kagome1}
\end{figure}

\begin{figure}
\centering
\includegraphics{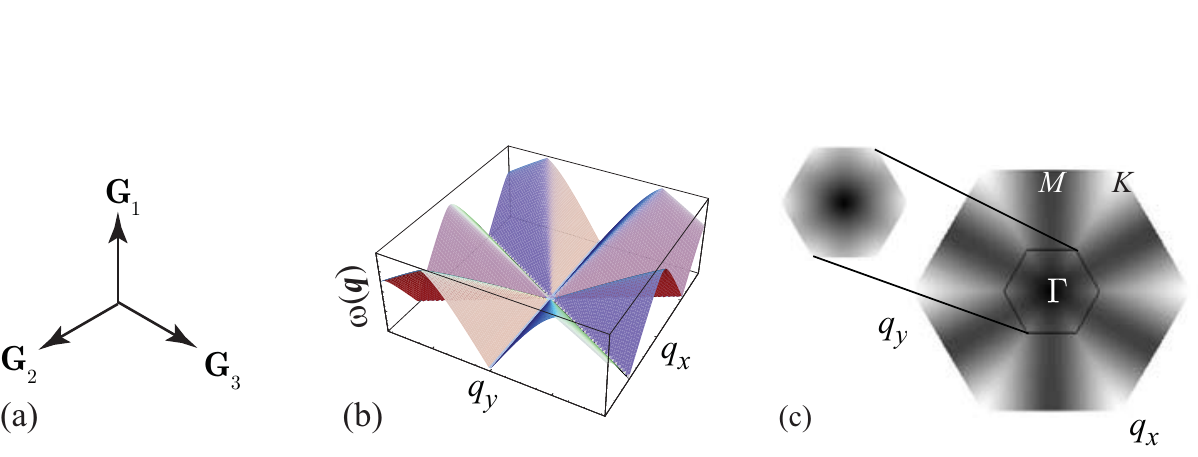}
\caption{(a) Shortest reciprocal lattice vectors, related by $3$-fold rotations, of the kagome lattice that satisfy
$\Gv_j\cdot \av_j=0$.
(b) Dispersion of lowest-frequency mode, showing ``knife-edges" along the
$\Gamma-M$ in the BZ. (c) Density plot of lowest-frequency mode with $k'/k = 0.02$.
Note the isotropic behavior near $\qv=0$. \emph{Adapted from reference \cite{Souslov2009}}}
\label{fig:kagome-disperion1}
\end{figure}

The $NN$ kagome lattice consists of three grids of straight
parallel filaments intersecting at lattice sites as shown in
\fref{fig:kagome1}. This figure also shows two different unit
cells reflecting the $3$-fold symmetry of the lattice. For the
moment, we focus on lattices with $N_x=N_y$ cells on a side as
shown in the figure. \ref{app:compat} derives the compatibility
matrix for a generalized kagome lattice, of which the simple
kagome lattice considered here is a special case. The
equilibrium and dynamical matrices are straightforwardly
calculated from it, as are the phonon spectrum and zero modes.
As in the square lattice, each of the kagome-lattice filaments
supports a \SSS under periodic boundary conditions (care must
be taken to join equivalent sites at the boundaries to create a
single filament for bonds slanting to the left from the
bottom), and the expectation is that a periodic lattice will
have $3N_x$ states of self stress, and this is indeed the case.
There is one state of self stress for each wavevector $\qv = q
\Gv_j/G_0$ along the symmetry equivalent lines from $\Gamma$ to
$M$ in the BZ [\fref{fig:kagome-disperion1}(a)] parallel to the
three reciprocal lattice vectors $\Gv_j$, where $G_0 =|\Gv_j|=
4 \pi/(\sqrt{3} a)$. Since there are $N_x$ values of $\qv$
along each of these directions, there are a total of $3N_x$
\SSSsa, from which \SSSs for individual filaments can be
constructed.

\begin{figure}
\centering
\includegraphics{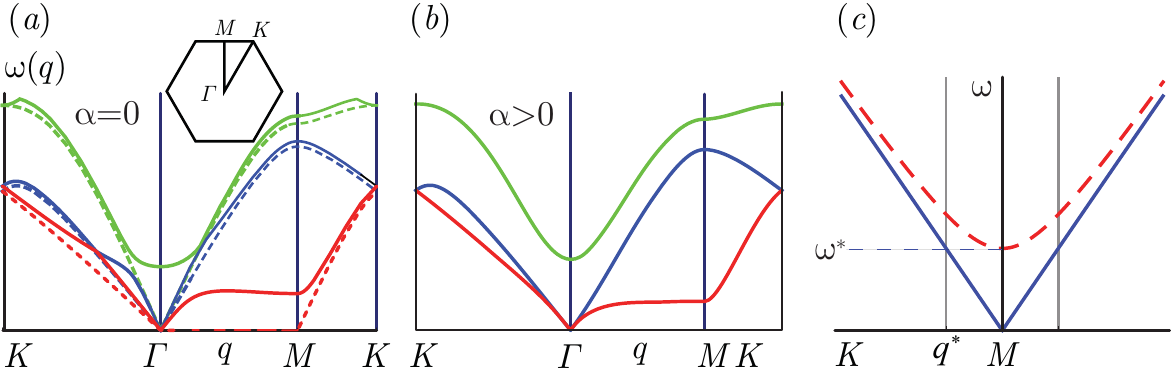}
\caption{(a) Phonon spectrum for the three lowest modes of the undistorted
kagome lattice. Dashed lines depict frequencies at $k'=0$
and full lines at $k'>0$. The inset shows the Brillouin zone
with symmetry points $\Gamma$, $M$, and $K$. (b) Phonon spectrum
of the twisted kagome lattice with $\alpha
>0$ and $k'=0$.  (c) Phonon dispersion
along the zone edges from $K$ to $M$ in schematic form for both the
undistorted lattice at $k'>0$ and for the twisted lattice with $\alpha \neq 0$,
showing the characteristic
frequency $\omega^*=2 \sqrt{k'/m}$ ($\omega_{\alpha}
\sim \sqrt{k/m} |\sin \alpha |$) and length $l^*=(q^*)^{-1}=(a/2)(k'/k)^{1/2}$
($l_{\alpha} \sim 1/|\sin \alpha|$) for the
untwisted (twisted) kagome lattice. \emph{Adapted from references \cite{Souslov2009} and
\cite{SunLub2012}.}}
\label{fig:kagome-disperion2}
\end{figure}

The $3N_x$ \SSSs require an equal number of zero modes, which,
as in the square lattice, occur along lines in reciprocal space
that have no component parallel to one of the three grids,
i.e., along the lines $\Gamma M$ in reciprocal space as shown
in \fref{fig:kagome-disperion1}(b). The zero modes for a
filament parallel to the $x$-axis consist of displacements of
all $1$-sites and $3$ sites  by $s (\cos \pi/6, \sin \pi/6)$
and $s(\cos \pi/6, -\sin\pi/6)$, respectively, for
infinitesimal $s$.  This corresponds to rigid rotations of
triangles about site $2$ as shown in \fref{fig:kagome1}(b). An
alternative description of the mode is that the entire filament
is displaced a distance $s \cos \pi/6$ to the right, and sites
$1$ and $3$ are, respectively, displaced upward and downward a
distance $s\sin\pi/6$ producing the zero-modes structure of
\fref{fig:selftstress2}(b). The spectrum of the
lowest-frequency modes has a linear dispersion with $\omega=cq$
($c=\sqrt{3} k a/8$) in the direction perpendicular to the
$\Gamma-M$ zero modes (\fref{fig:kagome-disperion2}(c)).

As in the square lattice, adding $NNN$ bonds gaps the spectrum
leading to a characteristic frequency $\omega^* \sim \sqrt{k'}$
and associated length scale $\ell^* \sim 1/\sqrt{k'}$
calculated from the dispersion along the line $M$ to $K$ at the
zone edge (\fref{fig:kagome-disperion2}). Other characteristic
frequencies can be calculated from the lowest frequency optical
modes or from the frequency at which the low-frequency acoustic
phonon modes crosses over to a nearly flat dispersion.

A total of $4 N_x -1$ bonds must be cut to liberate a $N_x
\times N_x$-unit-cell free lattice from its periodic parent.
There are no states of self-stress in the free lattice, so
there must be as many zero modes as bonds that are cut.  This
is more zero modes than the $3N_x$ in the periodic system.  The
difference between the two numbers arises from the joining of
lines slanting to the left under periodic conditions.  The
number of horizontal and right slanting filaments is the same
in both cases. However, under PBCs, there are $N_x$ distinct
left-slanting filaments; under free BCs, there are $N_x$ such
lines terminating at the bottom and $N_x-1$ terminating at the
right side of the lattice.

Because of the three sets of filaments aligned along $\av_n$,
at $\qv=0$, there are now three rather than two \SSSsa,
characterized by bond vectors
$\hat{\mathbf{t}}_1=(1,0,0,1,0,0)/\sqrt{2}$,
$\hat{\mathbf{t}}_2=(0,1,0,0,1,0)/\sqrt{2}$, and
$\hat{\mathbf{t}}_1=(0,0,1,0,0,1)/\sqrt{2}$, each of which has
a nonzero overlap with the vector of affine distortions. This
gives enough \SSSs to fully stabilize the elastic energy of the
$NN$ kagome lattice with nonzero Lam\'{e} coefficients
\cite{HutchinsonFle2013}. Addition of $NNN$ bonds increases
these coefficients.
\begin{equation}
\lambda = \mu = \frac{\sqrt{3}}{8}( k +3 k')
\label{eq:kagome-elas}
\end{equation}
The response is affine even though three sites per unit cell
introduces the possibility of their undergoing nonaffine
displacement to lower energy.  However, the geometry of this
lattice is special, and response is necessarily affine.

\subsection{Twisted kagome lattice\label{ssec:twisted-kagome}}

The twisted kagome lattice is  constructed from the finite zero
modes by oppositely rotating triangles along all of the
filaments of the untwisted lattice through an angle $\alpha$ as
shown in \fref{fig:twisted-kagome1}. There are a continuum of
lattices indexed by the angle $\alpha$ that bond $1$ makes with
the $x$-axis. As we shall see, this lattice has properties that
at first blush seem surprising but that in fact are simple
consequences of the \Ith.

\subsubsection{Bulk and elastic properties}
The straight filaments of the untwisted kagome lattice become
``zigzagged" and lose their ability to sustain \SSSs. The
result is that there are only the two $\qv=0$ states of self
stress required by the \Ith and the existence of two zero modes
of translation, and there are no zero modes other than those at
$\qv=0$.  Thus the simple rotation of triangles to create the
twisted lattice from the the untwisted ones gaps all but the
$\qv=0$ bulk phonons, just as does adding $NNN$ bonds to the
untwisted lattice (\fref{fig:kagome-disperion2}). The untwisted
spectrum is approached continuously as $\alpha \rightarrow 0$
leading to a characteristic frequency (measured by the gap at
the symmetry point $M$) and associated length scale,
\begin{equation}
\omega_\alpha \sim |\sin \alpha|, \qquad l_\alpha \sim \frac{1}{|\sin \alpha|} .
\end{equation}
that, respectively, vanish and diverge as $\alpha \rightarrow
0$.

\begin{figure}
\centering
\includegraphics{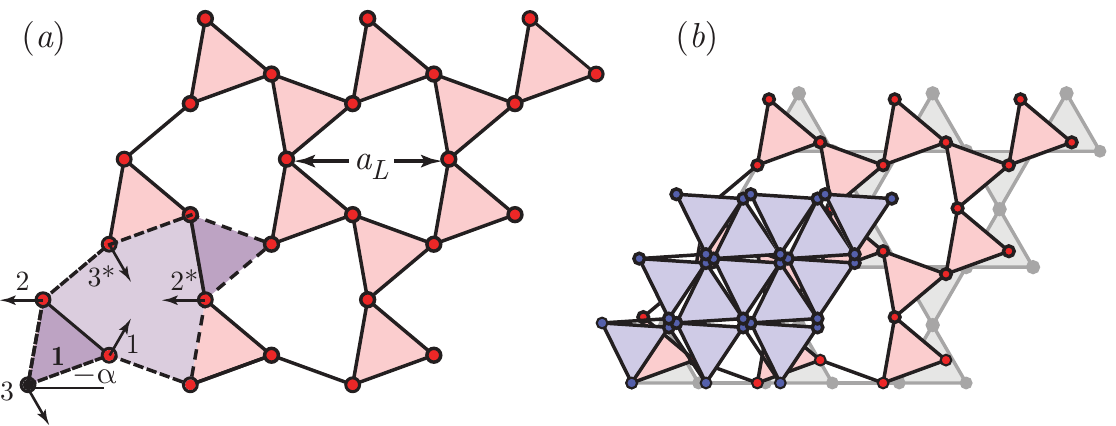}
\caption{(a) Twisted kagome lattice showing the displacements of sites in the
two unit cells shown in \fref{fig:kagome1} and rotation of triangles
through $\pm \alpha$. Bond $\bf 1$ (connecting sites $3$ and $1$) in this figure is rotated through an angle $-\alpha$.
The lattice spacing for bonds of length $a/2$ is
$a_L = a \cos \alpha$. (b) Lattices with different values of $\alpha$ superposed showing
how changing $\alpha$ reduces the lattice area. \emph{From reference \cite{SunLub2012}}}
\label{fig:twisted-kagome1}
\end{figure}

As \fref{fig:twisted-kagome1} shows, twisting the lattice
uniformly compresses it.  If bond lengths are fixed at $a/2$,
the Bravais lattice vectors are reduced in length from $a$ to
$a_L=a \cos \alpha$, and the volume of each unit cell from
$(\sqrt{3}/2) a^2$ to $(\sqrt{3}/2)a^2\cos^2 \alpha$. Thus
angle changes modify the area of the lattice without changing
any bond length of a $NN$ lattice, implying that the bulk
modulus $B$ of these lattices vanishes for all $\alpha \neq 0$.
Observe that the twisted lattice has the peculiar property that
it expands or contracts isotropically at no energy cost.  If it
expands in one direction, it will also do so in the opposite
direction. Elastic materials with this property have a negative
Poisson ratio \cite{Landau1986}; they are \emph{auxetic}. The
twisted kagome lattice has the most negative allowed Poisson
ratio of $-1$, which it retains for all strain, and it
sometimes called \emph{maximally auxetic} \cite{SunLub2012}. In
addition, the unit cell contracts isotropically.  Such lattices
are termed \emph{equiauxetic} \cite{MitschkeGue2013}. There are
not many naturally occurring materials that have this property,
but artificial ones can be created \cite{Lakes1987}.

As indicated above, there are two $\qv=0$ states of self-stress
that have the potential to create non-vanishing elastic moduli.
The long-wavelength elasticity is necessarily isotropic, so if
the bulk modulus is zero, the only option if for there to be an
isotropic shear modulus or for none of the elastic moduli to be
nonzero.  The two states of self-stress overlap with affine
strain, the shear modulus $\mu = \sqrt{3} k/8$ is nonzero (and
curiously independent of $\alpha$ and identical to that of the
untwisted kagome lattice [\eref{eq:kagome-elas}], and the
elastic energy density is
\begin{equation}
f_{\text{el}} = \frac{1}{2}\mu \,\tilde{\st}_{ij} \tilde{\st}_{ij}
\end{equation}
where $\tilde{\st}_{ij} = \st_{ij} - \frac{1}{2}\st_{kk}$ is
the symmetric, traceless shear strain tensor.

The strain is related to the metric tensor $g_{ij}$ via
$\st_{ij} = (g_{ij} - \delta_{ij})/2$.  The traceless part of
the strain,  $\tilde{\st}_{ij}=(1/2)(g_{ij} -
\frac{1}{2}\delta_{ij} g_{kk})$, which is zero for $g_{ij} =
\delta_{ij}$, is invariant, and thus remains equal to zero,
under conformal transformations that take the metric tensor
from its reference form $\delta_{ij}$ to $h(\xv )\delta_{ij}$
for any continuous function $h(\xv)$ of position $\xv$. The
zero modes of the theory thus correspond simply to conformal
transformations, which in two dimensions are best represented
by the complex position and displacement variables $z=x+iy$ and
$w(z) = u_x(z) + i u_y ( z)$.  All conformal transformations
are described by an analytic displacement field $w(z)$.  Since
by Cauchy's theorem, analytic functions in the interior of a
domain are determined entirely by their values on the domain's
boundary (the ``holographic" property \cite{Susskind1995}), the
zero modes of a given sample are simply those analytic
functions that satisfy its boundary conditions. For example, a
disc with fixed edges ($\uv=0$) has no zero modes because the
only analytic function satisfying this FBC is the trivial one
$w(z)=0$; but a disc with free edges (stress and thus strain
equal to zero) has one zero mode for each of the analytic
functions $w(z) = a_n z^n$ for integers $n \geq 0$. The
boundary conditions $\lim_{x\to \infty} \uv (x,y) = 0$ and
$\uv(x,y) = \uv(x+L, y)$ on a semi-infinite cylinder with axis
along $x$ are satisfied by the function $w(z) = \rme^{\rmi q_x
z}=\rme^{\rmi q_x x} e^{-q_x y}$ when $q_x = 2n\pi/L$, where
$n$ is an integer. This solution is identical to that for
classical Rayleigh waves \cite{Landau1986} on the same
cylinder. Like the Rayleigh theory, the conformal theory puts
no restriction on the value of $n$ (or equivalently $q_x$).
Both theories break down, however, at $q_x =q_c \approx \min
(l_\alpha^{-1}, a^{-1})$ beyond which the full lattice theory,
which yields a complex value of $q_y=q_y'+i
q_y^{\prime\prime}$, is needed.

\subsubsection{Surface modes\label{ssec:surface-m}}
As we have seen free two-dimensional lattices of $N$ sites cut
from a periodic Maxwell lattice necessarily have of order
$\sqrt{N}$ zero modes because of order $\sqrt{N}$ bonds must be
cut, and any sample-spanning states of self stress are lost
under the cut. In the untwisted kagome lattice, these modes are
identical to the bulk zero modes calculated under PBCs. In the
twisted lattice, whose cut lattice must have the same number of
zero modes as the untwisted lattice, there are no bulk zero
modes (except at $\qv=0$), and as a result the zero modes must
be localized at surfaces. In the long-wavelength limit, these
modes must reduce to the zero-frequency Rayleigh waves of an
isotropic elastic continuum with vanishing bulk modulus with
decay length $l_s \equiv \kappa^{-1}$ equal to the inverse
surface wavenumber $q$.  At shorter wavelength, $l_s$ is
determined by the length $l_\alpha$ associated with the twisted
phonon gap.

\begin{figure}
\centering
\includegraphics{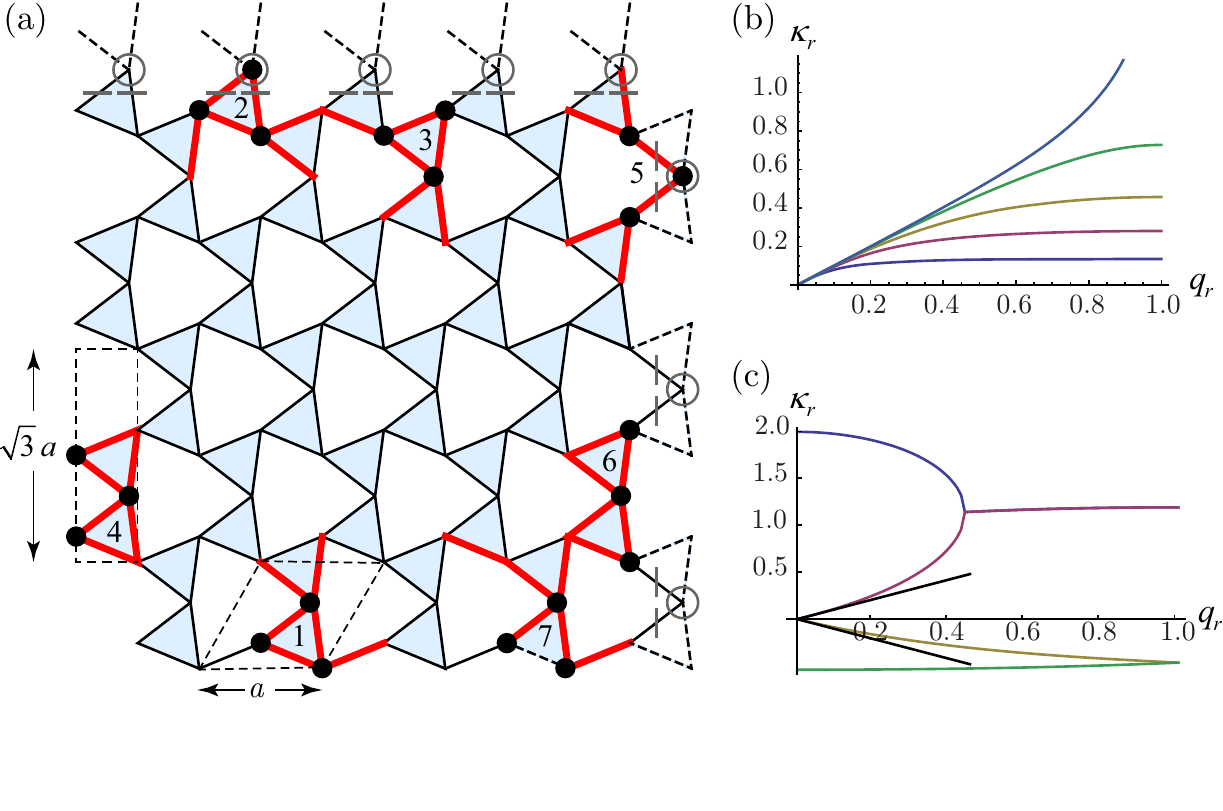}
\caption{(color online) (a) A free twisted kagome lattice with
free horizontal and vertical boundaries.  Sites and bonds of unit cells that match
the surface are shown in black and red, respectively, and the geometric form of these unit
cells are indicated by dashed quadrilaterals. Cutting the two (four) dashed bonds per cell on the
horizontal (vertical) boundary produces a lattice with bottom (left) and top (right) boundary
cells $1$ ($4$) and $2$ ($5$). Alternatively removing these bonds along with one circled site and
the two bonds crossed by grey lines changes the top (right) boundary cell from $2$ ($5$) to $3$ ($6$).
The number of zero modes per site is $1$ ($2$) per site on both the bottom (right)
and top (left) boundaries. Cell $7$ has one additional bond cut from it. (b) The reduced inverse
decay length $\kappa_r = \kappa/(G_r/2)$ of the horizontal boundaries
as a function of $q_r = q/(G_r/2)$, where $G_r =2 \pi/a$ for
$\alpha = \pi/20, \pi/10, 3 \pi/20, \pi/5, \pi/4$ in order from bottom to top.  All curves
follow $\kappa_r = q_r$ near $q_r = 0$.  The curve at $\alpha = \pi/4$ diverges at $q_r = 1.0$.
(c) $\kappa_r$ as a function of $q_r$ with $G_r = 2 \pi/\sqrt{3}$ for $\alpha = \pi/8$.  The positive
curves are for the left boundary and the negative ones are minus $\kappa_r$ for the right boundary. The
two positive curves merge at $q_r \approx 0.4$.  There are still two distinct decays beyond this point
with the same real part and different imaginary parts.  The straight grey lines are the
elastic limit $\kappa_r = q_r$.
}
\label{twisted-kagome-surface}
\end{figure}

\Fref{twisted-kagome-surface} depicts a finite rectangular
sample with free horizontal surfaces parallel to $\av_1$ and
vertical surfaces parallel to $\av_2 - \av_3$ along with unit
cells constructed so that all sites and bonds on a surface lie
in periodically repeated continuous cells.  It also shows which
bonds (or bonds and sites) must be removed to liberate the
finite lattice from the one under PBCs.  The removal of two
bonds or four bonds and one site per unit cell liberate the
horizontal surfaces. In either case, the number of zero modes
per cell is $n_0 = 2 \Delta n - \Delta n_b = 2$, where $\Delta
n=0$, $\Delta n_b = -2$ in the first case and $\Delta n=-1$,
$\Delta n_b = -4$ in the second.  Similar arguments yield $n_0
= 4$ for vertical surfaces.  As might be expected, the modes
are distributed equally between opposite surfaces, i.e., there
is one zero mode per unit cell on horizontal surfaces and two
modes per unit cell on vertical surfaces. Equivalently, there
is one (two) mode per surface wavenumber $-\pi/G_s <q<\pi/G_s$,
where $G_s$ is the magnitude of the surface reciprocal lattice
vector, $2\pi/a$ for horizontal and $2\pi/(\sqrt{3}a)$ for
vertical surfaces.

The amplitude of the surface modes decay as
$\exp[-\tilde{\kappa} s]$ with distance $s$ away from the
surface all of the way to the opposite free surface, where
$\tilde{\kappa}$ is in general complex indicating oscillations
along with decay. $\kappa \equiv \text{Re}\, \tilde{\kappa} =
l_s^{-1}$ is the inverse decay length. In the case of
horizontal surfaces, the decay length is the same on opposite
surfaces. \Fref{twisted-kagome-surface}(b) plots the single
$\kappa(q,\alpha)$ for different values of $\alpha$.  In the
case of vertical surfaces, the two decay lengths for the left
surface differ from those of the right one.
\Fref{twisted-kagome-surface}(c) plots these for $\alpha =
\pi/8$.  In all cases, one $\kappa(q,\alpha)$ reduces to
$\kappa(q,\alpha)=q$ in the long-wavelength limit as required
by the continuum theory.

Surface zero modes are by definition in the null space of the
compatibility matrix $\ma{C}$.  Systems with parallel free
surfaces with PBCs along their direction of alignment can be
viewed as a series of layers $n=1,...,N$, and $\ma{C}$ can be
decomposed as
\begin{equation}
\ma{C} =
    \begin{pmatrix}
        \Cv_{11} & \Cv_{12} & 0 & \dots & 0 & 0 \\
        0 & \Cv_{11} & \Cv_{12} & \dots & 0 & 0 \\
        \hdotsfor{6} \\
        0 & 0 & 0 & \dots & \Cv_{11}& \Cv_{N-1,N} \\
        0 & 0 & 0 & \dots & 0 & \Cv_{NN}
    \end{pmatrix} ,
\label{eq:matrix-C}
\end{equation}
where all $\Cv_{ij}$'s depend on the surface wavenumber $q$ and
$\alpha$. $\Cv_{11}$ is the $6\times6$ matrix connecting bonds
and sites within in a single unit cell and $\Cv_{12}$ is the
$6\times 6$ matrix connecting bonds in one unit cell to sites
in unit cells one layer deeper in the sample. The same unit
cells are used throughout the sample. The opposite surface may
terminate with only a partial version of these cells, and the
exact forms of $ \Cv_{N-1,N}$ and $\Cv_{NN}$ depend on how the
surface is terminated. Consider, for example, horizontal
surfaces, the bottom of which is characterized by the unit cell
labeled $1$ in \fref{twisted-kagome-surface}(a) and the top of
which is characterized by unit cell $2$.  For modes localized
at the bottom surface, unit cell $1$ is used through out the
sample. The termination cell $2$ at the opposite surface has
all three sites but only four bonds of unit cell $1$. The
missing bonds ($5$ and $6$) are not affected by cell $N-1$.
Therefore, $\Cv_{N,N-1} = \Cv_{12}$ is a $6\times 6$ matrix,
and $\Cv_{NN}$ is a $4\times 6$ matrix. If the top surface is
terminated by unit cell $3$, which has only $2$ sites and $2$
bonds of cell $1$, $\Cv_{N-1,N}$ is a $6 \times 4$ matrix and
$\Cv_{N,N}$ is a $2 \times 4$ matrix. Thus displacements
$\ma{U} = (\uv_1, ... ,\uv_N)$ in the null space of $\ma{C}$
satisfy
\begin{equation}
\Cv_{11} \uv_n + \Cv_{12} \uv_{n+1} = 0 ,
\end{equation}
for $n=1, ... N-2$.  These equations are solved by $\uv_{n+1} =
\lambda\uv_n$ and
\begin{equation}
\det (\Cv_{11} + \lambda \Cv_{12}) = 0
\label{eq:detC-s}
\end{equation}
subject to the boundary conditions that $\Cv_{11}\uv_{N-1} +
\Cv_{N-1,N} \uv_N = 0$ and $\Cv_{NN} \uv_N = 0$.

The inverse penetration depth is  determined by $\lambda$ via
$\exp[-\kappa a_{\perp}] = \lambda$, where $a_{\perp}= a
\sqrt{3}/2$ is the distance between unit cells in the direction
perpendicular to the surface. In case of termination with
either unit cell $2$ or unit cell $3$, $\uv_{N}$ must equal
$\lambda \uv_{N-1}$ to solve the first boundary condition.
Though required by the known existence of the zero mode, it is
a remarkable fact that the projection of $\uv_1$, which is in
the null space of $\Cv_{11} + \lambda \Cv_{12}$, onto the space
of displacements of the last layer is also in the null space of
$\Cv_{NN}$, and as advertised earlier, the surface mode decays
exponentially from one free surface to the next.  A similar
analysis applies to the vertical surface for which $a_{\perp}=
a/2$. Of course, any linear combination of the exponentially
decaying modes on the two sides of a strip is a zero mode if
both are individually. In the usual situation in which the
Rayleigh waves have a nonzero frequency, the eigenstates of a
finite strip are symmetric and anti-symmetric combinations of
states on the two surfaces that interact across the strip
yielding a peristaltic mode with $\omega \sim q$ and a bending
mode with $\omega\sim q^2$. It should be noted that $\kappa$
for modes that are not localized within the surface unit cell
can also be calculated from $\det C(\qv)$, which is a function
of $\exp[\rmi \qv\cdot \av_i]$, $i = 1,...,3$, by setting $\qv=
q_{\perp} \hat{\nv}_\text{in} + \qv_{||}$ where $\qv_{||}$ is
the component of $\qv$ parallel to the surface and
$\hat{\nv}_\text{in}$ is the unit inward normal to the surface,
setting $q_{\perp} = i \kappa$, and solving for $\kappa$ in
$\det C(i \kappa, \qv_{||})=0$.  This approach does not
directly provide eigenvectors satisfying boundary conditions.

A surface consisting of unit cells $7$ in
\fref{twisted-kagome-surface}(a) differs from surfaces composed
of other unit cells shown in that figure in that it is missing
a bond on the surface: it is obtained from cell $1$ by cutting
the dashed downward pointing bond, and as a result, this
surface has two, not one zero mode per cell.  The calculation
just outlined indeed produces two zero modes per $q$, one of
which is localized completely on the first row because
$\Cv_{11}$ has a non-empty null space. Similarly, $\kappa$
diverges as shown in \fref{twisted-kagome-surface}(b) at $q =
\pi/G_s$ for $\alpha=\pi/4$ because $\Cv_{11}(\pi/G_s,\pi/4)$
has a non-empty null space.

\subsection{Other lattices\label{sec:other-lattices}}

\begin{figure}
\centering
\includegraphics{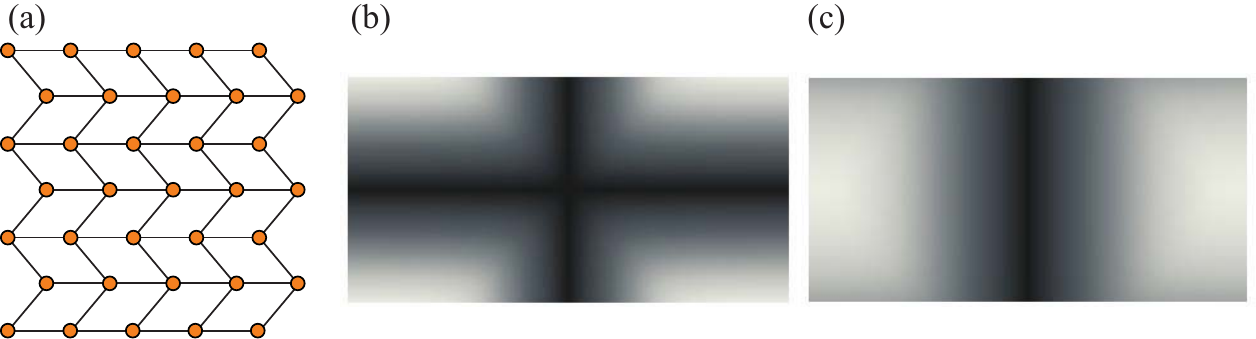}
\caption{(a) The zigzagged square lattice with two sites per unit cell and
rectangular unit cells and Brillouin zone.  (b) and (c) density plots of the two
lowest-frequency bands, showing lines of zero modes along $q_y=0$ and $q_x=0$
in (a) and along $q_x=0$ in (b).  The total number of zero modes in the two bands
is the same as in the undistorted square lattice.}
\label{fig:zigzag-square}
\end{figure}

So far, we have focussed on the three simplest examples of
two-dimensional periodic Maxwell lattices and the free lattices
cut from them.  There are many others that can be constructed
from these without changing the local coordination numbers or
the lengths of any bonds and whose properties can easily be
understood in the context of the \Itha.  One of the simplest of
such lattices is the ``zigzagged" square lattice with two sites
per unit cell \cite{SouslovLub2014}, shown in
\fref{fig:zigzag-square}, in which every other row is displaced
to the right while allowing the requisite compression along the
vertical direction. This lattice retains the straight
horizontal filaments of the original square lattices but loses
those in the vertical direction.  It does not, however, lose
any vertical \SSSs because the \SSSs from individual straight
filaments are converted to ones like those of
\fref{fig:selftstress2}(e) on pairs of zigzagged filaments of
which there are total of $N_x$ under PBCs . These \SSSs must be
accompanied by zero modes in the spectrum, which show up as
horizontal and vertical lines of zeros in the lowest-frequency
mode and a vertical line of zeros in the second-lowest
frequency mode as shown in figures \ref{fig:zigzag-square}(b)
and (c). The vertical \SSSs have no overlap with affine strain,
and the lattice does not resist vertical compression.  The
result is that the elastic energy density is simply
$f_{\text{el}} = k u_{xx}^2/2$.

\begin{figure}
\centering
\includegraphics{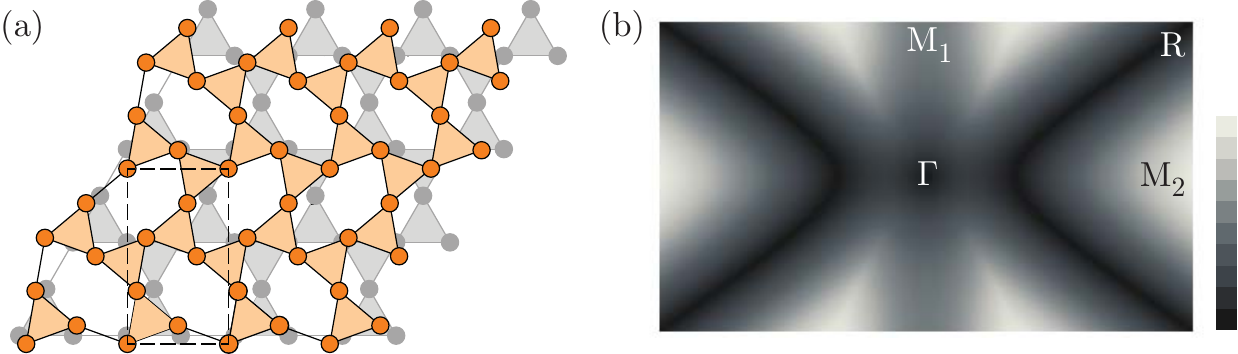}
\caption{(a) An example of one of the many lattices that can be constructed
from the kagome lattice without changing any bond lengths overlayed on an
undistorted kagome lattice. This one is
a stacked lattice with alternating rows of oppositely tilted hexagons.
(b) Density map of the lowest-frequency band showing the gapped spectrum
near the origin and two curved lines of zero modes passing between two corners.
\emph{Adapted from reference \cite{SunLub2012}}}
\label{fig:kagome-variation}
\end{figure}

The kagome lattice offers more interesting variations
\cite{SouslovLub2014,SunLub2012}.
\Fref{fig:kagome-variation}(a) shows one such variation with
intriguing properties.  It consists of alternating rows of
oppositely tilted distorted hexagons.  It has rectangular
symmetry with $6$ sites per unit cell.  As
\fref{fig:kagome-variation}(b) shows, it has an unusual
spectrum:  its modes are fully gapped (except at $\qv=0$) near
the origin but exhibit curved lines of zero modes at large
$\qv$.  It has zero-frequency surface modes for surface
wavenumbers in the gapped region but not in the ungapped
region. Curiously, the elastic energy density is identical to
that of the twisted kagome lattice even though it has a lower
symmetry.

\begin{figure}
\centering{\includegraphics{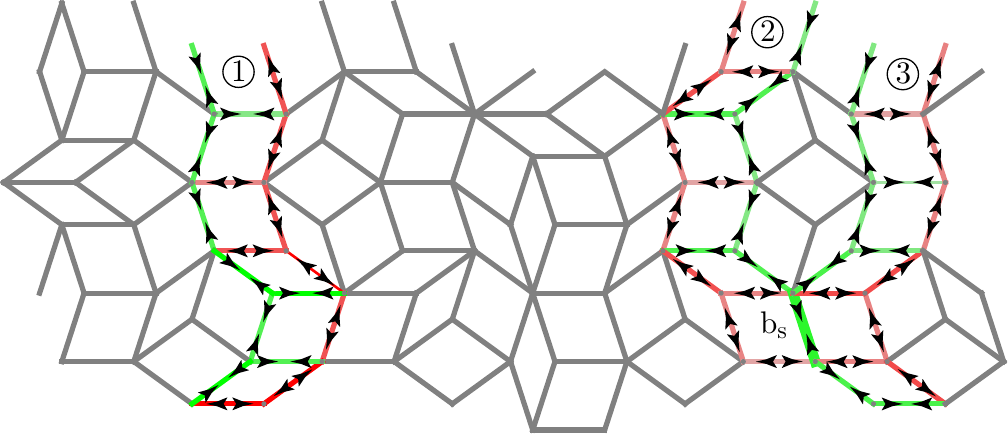}}
\caption{(color online) Unit cell of the second periodic approximant of the Penrose tiling \
showing \SSSs (circled $1$, $2$, and $3$).  In all states, stress is localized
on vertical ladders with different signs of stress on opposite sides as in \fref{fig:selftstress2}(c).
States $2$ and $3$ share
the bond marked $\text{b}_s$, and are not orthogonal.  They can be orthogonalized to produce states mostly,
but not completely localized on the two ladders. \emph{From reference \cite{StenullLub2014}}}
\label{fig:Penrose}%
\end{figure}

The square and kagome lattice are the two-dimensional Maxwell
lattices with the smallest unit cells.  As discussed above, the
kagome lattice can be distorted in various ways to produce
larger-unit-cell Maxwell lattices, but there are many other
lattices including random ones \cite{Henley1991} that can be
created. An intriguing set of periodic Maxwell lattices
\cite{StenullLub2014} are those arising from rational
approximates to Penrose rhombohedral tilings \cite{Penrose74}
that approach a quasi-crystalline lattice with $5$-fold
symmetry
\cite{LevineSte1984,SteinhardtOst1987,DiVincenzoSte1991}. A
unit cell for the second periodic approximate with $80$ sites
and $160$ bonds per cell is shown in \fref{fig:Penrose}. These
periodic lattices, which can be constructed via projection from
a five-dimensional cubic lattice \cite{deBruijn1981}, all have
an average coordination of exactly four even though the
coordination of local sites varies from three to as high as
ten.  The size of the unit cell increases rapidly from $N=30$
in the first approximate to $N=25840$ in the eighth
approximate. Each of the approximates is a legitimate periodic
lattices with a full set of phonon branches with dispersions
depending on lattice wavenumber $\qv$.  They can also be
interpreted as a single-cell system under toroidal PBCs that
approach the infinite-cell quasicrystalline limit.  In this
interpretation, which we pursue here, only the $\qv=0$ part of
the spectrum is of physical interest as is the case for
periodic approximates to randomly packed spheres at jamming.
These lattices have a number of interesting properties: (1)
their undistorted versions have of order $\sqrt{N}$ \SSSs and
zero modes, but none of the \SSSs overlap with affine strain,
and all elastic moduli are zero, much as in rigidity
percolation.  (2) Randomizing site positions removes all but
the two required \SSSs and zero modes.  The two \SSSs overlap
with affine strain to produce two independent positive
eigenvalues of the elasticity matrix, one corresponding in the
large $N$ limit to the bulk modulus with associated eigenvector
of pure compression and one to a shear modulus with associated
shear eigenvector.  The bulk modulus increases with $N$
reaching a saturation value at $N\approx 10^4$.  The shear
modulus on the other hand approaches zero as $1/N$.  The latter
behavior is required by  $5$-fold-symmetric quasicrystalline
limit whose elasticity must be isotropic with both shear moduli
equal. Because there are only two \SSSsa, one shear modulus
must be zero if the bulk modulus is nonzero, and the other must
approach zero with $N$. This is essentially the same behavior
observed in randomly packed spheres at the jamming threshold
with $z=2d$ \cite{GoodrichNag2012,GoodrichNag2014}.

\section{Topological phonons\label{sec:topological}}

Twisting the kagome lattice gaps all zero modes of the
untwisted lattice except those at $\qv=0$.  This gapping is
reminiscent of that of the electron spectrum in systems like
polyacetylene \cite{ssh,NiemiSem1986}, quantum Hall materials
\cite{halperin82,haldane88}  and topological insulators
\cite{km05b,bhz06,mb07,fkm07,HasanKane2010,QiZhang2011}, which
is associated with the appearance of topologically protected
zero modes at free boundaries and at boundaries separating two
topological classes. An interesting and natural question is
then whether or not the phonon spectra of Maxwell lattices can
be gapped in different ways to produce distinct topological
classes with protected boundary modes. The answer is yes they
can be, and this section, which is mostly based on reference
\cite{KaneLub2014}, will explore both how they can be
constructed and the nature of their interface states.

\subsection{A one-dimensional Model  \label{sec:one-d-model}}
\begin{figure}
\centering
\includegraphics{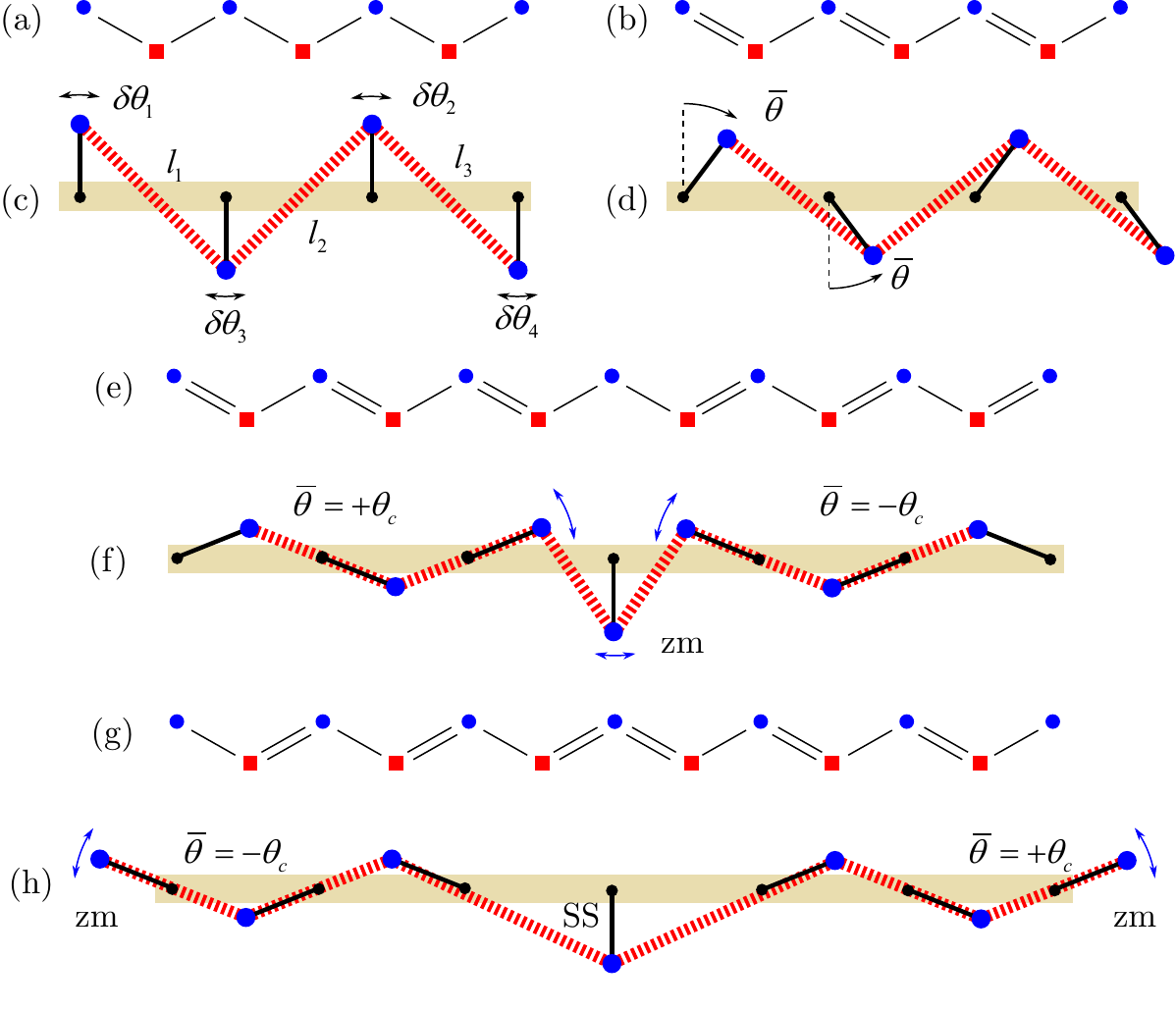}
\caption{(a) and (b) depict
the SSH model of polyacetylene, with A and B sublattices indicated by blue circles
and red squares, respectively.
(a) describes the gapless state with all bonds identical, while (b) describes the
gapped AB dimerized state, with double (single) bonds on the AB (BA) bonds. The BA
dimerized state with single and double bonds interchanged is not shown.
(c) and (d) show the 1D isostatic lattice model in which masses, represented by the
larger blue dots, are connected by springs in red and are constrained
to rotate about fixed pivot points, represented by small black dots.
(c) is the gapless
high-symmetry state with $\bar\theta=0$, and (d) is the gapped lower-symmetry phase
with  $\bar\theta >0$.   (c) and (d) are equivalent to (a) and (b) if we identify
the masses (springs) with the A (B) sublattice sites.
(e) shows a domain wall in polyacetylene connecting the AB and BA dimerized states.
There is a topologically protected zero-energy state associated with the A sublattice at the defect. (f) shows
the equivalent state for the mechanical model with a topologically protected
zero-frequency mode at the domain wall  connecting a $\btheta=+\theta_c$ lattice with
a $\btheta=-\theta_c$ lattice. (g) shows a domain wall connecting the BA and AB dimerized states,
which has a zero energy state associated with the B sublattice.
(h) shows the equivalent isostatic state with a state of self-stress (SS) at the domain wall and zero modes at
the end so that Index count $N_0-N_S=1$ is satisfied. \emph{Adapted from reference \cite{KaneLub2014}}}
\label{fig:1d-model}
\end{figure}

We begin with a one-dimensional model whose phonon spectrum
(including both positive and negative frequencies) is identical
to that of the one-dimensional Su-Schrieffer-Heeger model for
polyacetylene \cite{ssh,NiemiSem1986} schematically depicted in
figures \ref{fig:1d-model}(a) and (b). Our model, depicted in
figures \fref{fig:1d-model}(c) and (d), consists of rigid bars
of length $r$ that can rotate freely about fixed positions on a
one-dimensional periodic lattice. The ends of neighboring bars
are connected by harmonic springs whose lengths are adjusted so
that the equilibrium configuration is one in which alternate
rods make an angle $\bar{\theta}$ with the upward or downward
normals. Bars tilt towards the right if $\btheta>0$ and to the
left if $\btheta<0$. Each rod $s$ has one degree of freedom
$\theta_s =\bar{\theta}-\delta \theta_s$, and each spring
provides one constraint. Under periodic boundary conditions,
the number $N$ of rods equals the number $N_B$ of springs.  In
the linearized limit, the compatibility matrix with components
$C_{\beta s}$ connects the stretch in spring $\beta$ with
rotations of rod $s$: $\delta l_\beta = C_{\beta s} \delta
\theta_s$, and the equilibrium matrix $Q_{s\beta}$ relates
torques on rod $s$ to tensions $t_\beta$ in spring $\beta$:
$\tau_s = -Q_{s\beta} t_\beta$. With appropriate sign
convention for the torque, $C_{\beta s} = Q^T_{\beta s}$. In a
state of self stress, springs are under tension, but there are
no torques on the rods. The \Ith thus applies directly to this
system.  Under periodic boundary conditions $N_B= N$ and $N_0 =
N_S$.  Cutting one bond creates a free lattice with $N_0 = N_S
+1$.  Thus, if there are no \SSSsa, the free system must have
one zero mode, which can either be a mode in the bulk spectrum
or a surface mode on one of the boundaries. But what determines
whether it is localized on the right or left boundary?  As we
shall now, it is the topological class.

We proceed now to a more detailed analysis of the our model.
The components of the compatibility matrix at rest angle
$\btheta$ are
\begin{equation}
C_{s\beta}(\bar{\theta}) = c_1(\btheta) \delta_{s,\beta} - c_2(\btheta)\delta_{\beta,s+1} ,
\end{equation}
where
\begin{equation}
c_{1(2)} = \frac{(a \pm 2 r \sin \btheta )r \cos \btheta}{\sqrt{a^2 + 4 r^2 \cos^2 \btheta}} .
\end{equation}
Thus $|c_1|>|c_2|$ for all $0<\btheta<\pi$, and $|c_1|<|c_2|$
for all $-\pi<\btheta<0$. The energy of the system (contained
entirely in the stretching of the springs) is then
\begin{equation}
E =\tfrac{1}{2} k \sum_{\beta} (\delta l_\beta)^2 =
\tfrac{1}{2} k \sum_s (c_1 \delta \theta_s - c_2 \delta \theta_{s+1})^2.
\end{equation}
The Fourier transform of $C_{s\beta}$ is
\begin{equation}
C(q) = c_1 -  \rme^{\rmi q a} c_2 ,
\label{eq:C(q)}
\end{equation}
and bulk phonon modes have frequency
\begin{equation}
\omega(q) = \pm|C(q)| = \pm \sqrt{(c_1 -c_2)^2 + 4 c_1 c_2 \sin^2 (qa/2)}
\end{equation}
(for unit mass). When $\btheta = 0$ (vertical rods), $c_1 =
c_2$, the energy becomes invariant with respect of $\delta
\theta_s \rightarrow \delta \theta_s +\delta$ for every $s$,
and there is necessarily a bulk zero mode at $q=0$ - this in
spite of the fact that the bases of the rods are anchored,
breaking translational invariance. For other values of
$\btheta$, the phonon spectrum is fully gapped.

In a finite system, the compatibility matrix can be expressed
in the form of \eref{eq:matrix-C} with $\Cv_{11} = c_1$ and
$\Cv_{12} = -c_2$.  The decay length of the surface state is
determined by $c_1 - \lambda c_2 = 0$ or $\lambda = c_1/c_2 =
e^{-\kappa a}\,\,\,(e^{\kappa a})$ for states localized on the
left (right) boundary.  Thus, the one zero mode is on the left
if $|c_1|<|c_2|$ and on the right for $|c_1|>|c_2|$.  The left
zero mode is particularly easy to see at the critical angle
$\theta_c = -\sin^{-1} [a/(2 r)]$ at which $c_1 = 0$ as shown
in \fref{fig:1d-model}(s). The equilibrium matrix $\Qv=\Cv^T$
for a finite system has no zero modes, and there are no \SSSs
as required by the Index count.  This follows because $\Cv_{11}
t_1 = 0$ for any $\tv$ in the null space of $\Qv$. Then the
equation $-\Cv_{12} t_n + \Cv_{11} t_{n+1}=0$ sets all $t_n$
for $n>1$ equal to zero. Under periodic boundary conditions,
there must be one localized state of self stress for each
localize zero mode.

\begin{figure}
\centering
\includegraphics{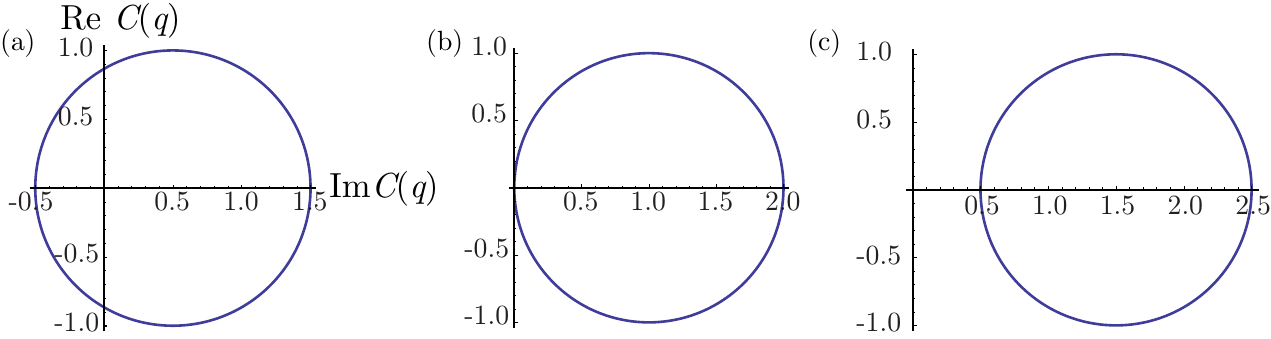}
\caption{Contour plots in the complex plane of $C(q)$ for complete circuit
of $q$ from $q=0$ to $q= 2\pi/a$.  (a) $c_2=1.0$ and $c_1=0.5<c_2$; the circuit contains the origin
and the topological charge is $n=1$. (b) $c_1=c_2 = 1.0$; this is the transition case in
which the contour just touches the origin, and $n=0$. (c) $c_1=1.5>c_2 = 1.0$; the
circuit does not contain the origin and $n=0$. Note the circuits begin with $q=0$ on the far left and
circulate counterclockwise.}
\label{fig:SSH-contour}
\end{figure}

But what does this have to do with topology? The compatibility
matrix $C(q)\equiv |C(q)|^{i \phi}$ (or more generally its
determinant) maps points in the Brillouin zone ($-\pi/a < q\leq
\pi/a$) to a path in the complex plane.  Since it depends on
$\rme^{\rmi q a}$, the path will be a closed and return to its
starting point as $q$ advances between equivalent points in the
zone [$q\rightarrow q + (2 \pi/a)$]. These curves are
characterized by an integer winding number:
\begin{equation}
n = \frac{1}{2 \pi} \int_0^{2 \pi} d\phi =
\frac{1}{2 \pi \rmi}\int_0^{2\pi/a} dq \frac{\rmd}{\rmd q}\Im \ln \det C(q) ,
\label{eq:top-index1}
\end{equation}
which for the $C(q)$ of \eref{eq:C(q)} is either $+1$ or $0$.
Clearly $n=0$ if the path in the complex plane does not enclose
the origin, and $n=1$ if it does. The first case occurs
whenever $|c_1|>|c_2|$ and the second whenever $|c_1|<|c_2|$ as
shown in \fref{fig:SSH-contour}. When $|c_1|=|c_2|$, the
boundary of the curve passes through the origin.  The winding
number is thus a topological invariant equal to $1$ for all
$-\pi<\btheta<0$ and to $0$ for all $0<\btheta<\pi$. The only
way it can change values is by passing through the critical
angles $\btheta=0$ or $\btheta = \pi$ (bars lie along the
horizontal axis). When $n=0$, the zero mode is on the right and
if $n=1$, the zero modes is on the left. The connection between
the topological invariant and the existence of zero modes is
easy to see in this case. A zero surface mode exists if there
is a solution to $c(\lambda)=c_1 - c_2 \lambda=0$ with
$|\lambda|<1$.  The compatibility matrix is simply
$C(q)=c(\lambda =\rme^{\rmi q a})$, and along the closed path
it describes in the complex plane, $|\lambda|=1$. In the
interior of the path $|\lambda|<1$.  Thus, if the path encloses
the origin, which is the point at which $c(\lambda)=0 $, the
solution for $\lambda$ will have a magnitude less than one. A
complementary perspective, based on the Cauchy argument
principle, is discussed in \ref{App:gauge},

Here we considered the winding number associated with $C(q)$.
We could equally well have considered that associated with the
equilibrium matrix $Q(q) = C^*(q)$.  Because $\rme^{-\rmi qa}$
winds clockwise rather than counter clockwise with increasing
$a$, its winding number is either $-1$ or $0$, with $-1$
corresponding to $|c_1|<|c_2|$.  Thus the surface state is at
the left when the winding number of $Q$ is $-1$.  We will use a
generalization of the $Q$-winding number in our
characterization of topological classes of kagome lattices in
what follows.

As is well known for the SSH model \cite{ssh,jackiw76}, an
interface between the two dimerizations binds a zero mode, as
indicated.   Similarly, the interface separating the two signs
of $\btheta$ does as well.  This is most easily seen for
$\bar\theta = \pm \theta_c$ where the springs are colinear with
the bars, so that $c_1$ or $c_2 = 0$. \Fref{fig:1d-model}(f)
shows a domain wall between $+\theta_c$ and $-\theta_c$, in
which the center two sites share a localized zero mode.
\Fref{fig:1d-model}(h) shows an interface between $-\theta_c$
and $+\theta_c$ with a state of self-stress localized to the
middle three bonds, in addition to floppy modes localized at
either end. As long as there is a bulk gap, the zero modes
cannot disappear when $\btheta$ deviates from $\pm \theta_c$.
The zero modes remain exponentially localized, with a
localization length, $l = a/\ln|c_1/c_2|$, that diverges when
$\btheta\rightarrow 0$.

\subsection{Topological lattices \label{sec:top-lattice}}

We have just seen the intimate connection between topological
properties of the compatibility or equilibrium matrices and
zero-modes at boundaries in a one-dimensional system.  Here we
discuss how these ideas can be extended to higher-dimensional
Maxwell lattices. We will for the most part only quote results,
and not provide details of how they were obtained.  The latter
can be found in reference \cite{KaneLub2014}. There are two
significant differences between these lattices and those in one
dimension. First, because $\Qv(\qv)$ (or $\Cv(\qv)$) depends on
the vector wavenumber $\qv$ with two rather than one
independent component, topological characterization will
require a vector rather than a scalar winding number. Second
there are boundary surface modes imposed by the \Ith that are
required whether or not lattices have any topologically
properties.  We will thus need to divide the surface mode count
into those parts imposed by the \Ith and those parts implied by
topological considerations. The result is that topology cannot
change the global count of the \Itha, but it can move zero
modes from one boundary to another and give rise to
topologically protected zero modes at interfaces between two
different topological classes.

\subsubsection{Topological and total mode
count\label{ssec:top-tot-count}}

To unify the treatment of zero modes arising from mismatch of
sites and bonds and those that arise in locations where there
is no local mismatch, reference \cite{KaneLub2014} generalized
the \Ith so that it determines a zero mode count
$\nu^S=N_0^S-N_S^S$ in a subsystem $S$ of a larger system. This
is well defined provided the boundary of $S$ is deep in a
gapped phase where zero modes are absent. This count has the
two separate contributions just discussed:
\begin{equation}
\nu^S = \nu_L^S + \nu_T^S ,
\label{eq:surfacenu}
\end{equation}
where $\nu_L^S$ is a local count of sites and bonds in $S$ and
$\nu_T^S$ is a topological count, which depends on the
topological structure of the gapped phases in the boundary of
$S$.  The topological contribution has a form similar to that
of the one-dimensional system. For the periodic lattices we are
considering, it is best expressed as a count per unit cell on
an edge indexed by a reciprocal lattice vector $\Gv$ (i.e.,
$\Gv$ is normal to the surface with a magnitude of $2
\pi/a_\perp$, where $a_{\perp}$ is the distance between lines
of cells identical to the line of surface cells),
\begin{equation}
\tnu_T^S =\nu^S/N_{\text{cell}}=\Gv\cdot \Rv_T /(2 \pi) ,
\label{eq:tnuT}
\end{equation}
where $N_{\text{cell}}$ is the number of surface unit cells and
$\Rv_T$, a generalization of the one-dimensional winding
number, is a lattice vector
\begin{equation}
\Rv_T = \sum_i n_i \av_i ,
\label{eq:RvT}
\end{equation}
where $\av_i$ are the primitive translation vectors and
\begin{equation}
n_i = \frac{1}{2\pi \rmi}\oint_{C_i} d\qv\cdot {\rm Tr}[\Qv(\qv)^{-1}
\nabla_{\qv} \Qv(\qv) ].
= \frac{1}{2\pi }\oint_{C_i} d \qv\cdot \nabla_\qv \phi (\qv) ,
\label{eq:ni}
\end{equation}
where $\phi(\qv)$ is the phase of $\det \Qv(\qv)$ ($\Qv(\qv) =
|\Qv(\qv)|^{\rmi \phi(\qv)}$).  Here $C_i$ is a cycle of the BZ
connecting $\qv$ and $\qv + \Bv_i$, where $\Bv_i$ is a
primitive reciprocal vector satisfying $\av_i \cdot \Bv_j =
2\pi \delta_{ij}$ ($\Bv_1=-\Gv_2$ and $\Bv_2=\Gv_1$ in
\fref{fig:kagome-disperion1}a). The $n_i$ are winding numbers
of the phase of $ \det \Qv(\qv)$ around the cycles of the BZ,
where $\Qv(\qv)$ is the equilibrium matrix in a Bloch basis.
This winding number is independent of path, and thus
independent of $\qv$ so long as the spectrum is gapped
everywhere except the origin. The zero-mode at the origin of
$\qv$ is not topologically protected (i.e, it can be gapped by
a weak potential breaking translational symmetry), so it does
not cause any problems.  It is possible, however, for there to
be topologically protected gapless points.  These would be
point zeros around which the phase of $\det \Qv({\bf k})$
advances by $2\pi$. These lead to topologically protected bulk
modes that form the analog of a ``Dirac semimetal" in
electronic systems like graphene
\cite{km05b,Aji2012,WeiChao2012,WeiWang2012,AjiHe2013,ZhuAji2013,PhillipsAji2014,SosenkoWei2014,WeiChao2014}.
These singularities could be of interest, but they do not occur
in lattices derived from the kagome lattice we study below.
They do, however, occur in modified square lattices
\cite{square} of the type shown in \fref{fig:square-kagome}
considered in reference \cite{GuestHut2003} and in the
pyrocholore lattice \cite{StenullLub2014b}.

In general, the winding number is not gauge invariant and
depends on how the sites and bonds are assigned to unit cells.
It is, however, possible to adopt a ``standard unit cell" with
basis vectors $\rv_{\mu(\beta)}$ for the $n_s$ sites ($n_b =d
n_s$ bond centers) per cell for which the ``dipole" moment of
the site and bond charges, $\Rv_{\text{stan}} = d \sum_\mu
\rv_\mu - \sum_\beta \rv_\beta$,  is equal to zero.   The two
unit cells of \fref{fig:kagome1} satisfy this zero dipole
condition even after being distorted to those of the twisted
lattice. $\Qv(\qv)$ is defined using Bloch basis states
$|\qv,a=\mu,\beta\rangle \propto \sum_{l}\exp \rmi
\qv\cdot(\Rv_l + \rv_a)|\Rv_l + \rv_a \rangle$, where $\Rv_l$
is a Bravais lattice vector.   In this gauge, $\Rv_T$ is
uniquely defined, and the zero-mode count is given by equations
\eref{eq:surfacenu} to \eref{eq:RvT}.

Because $\Rv_{\text{stan}}$ is zero, we could equally well use
a basis in which all $\rv_\mu$ and $\rv_\beta$ are simply zero.
This is in fact the basis we use for all of the calculations
presented here.  It should be noted that it is not always
possible to find a symmetric ``standard" unit cell with a
vanishing dipole moment defined in terms of charges at sites
and bond centers.  Indeed, there is no such cell for the SSH
model or for 3D pyrochlore lattices.  Not having such a
standard cell is not a problem, however.  The number and the
nature of surface modes do not depend on the choice of a
``standard" or reference unit cell and are unambiguous.  The
easiest choice is usually to set the positions of all of the
sites and bonds in the unit cell to be zero. As discussed in
\ref{App:gauge},  $\det\Cv$ for different unit cell choices,
such as those used to calculate the zero surface modes in
\sref{ssec:surface-m}, will have different phase factors, which
account for the differences in the topological integral of
\eref{eq:ni} for different choices of unit cells..

The local count, $\nu^S_L$, depends on the details of the
termination at the surface and can be determined by evaluating
the macroscopic ``surface charge" that arises when charges $+d$
($-1$) are placed on the sites (bonds) in a manner analogous to
the ``pebble game" \cite{JacobsTho1995}. This can be found by
defining a bulk unit cell with basis vectors $\tilde{\rv}_a$
that accommodates the surface with no leftover sites or bonds
(see figures \ref{twisted-kagome-surface} and
\ref{fig:parameter}) as discussed in \sref{ssec:surface-m}.
This unit cell depends on the surface termination and, in
general, will be different from the ``standard" unit cell
(\fref{fig:kagome1}) used for the calculation of $\nu^S_T$. The
local count is then the surface polarization charge given by
the dipole moment $\Rv_L$  per unit cell:
\begin{equation}
\tilde{\nu}_L \equiv \nu^S_L/N_{\text{cell}} = \Gv\cdot \Rv_L /2\pi,
\label{eq:tnuL}
\end{equation}
where
\begin{equation}
\Rv_L = d \sum_{\text{sites}\, \mu} \tilde{\rv}_\mu - \sum_{\text{bonds}\,\beta} \tilde{\rv}_\beta.
\label{eq:RL}
\end{equation}
The total zero-mode count on the surface then follows from
equations (\ref{eq:surfacenu}), (\ref{eq:tnuT}), and
(\ref{eq:tnuL}). The polarization of the standard cell is zero
so that $\Rv_L = \Rv_L - \Rv_{\text{stan}}$ can be calculated
from the displacements $\tilde{\rv}_a - \rv_a$ of the surface
cell sites and bonds relative to those of the standard cell.

\subsubsection{Constructing topological
lattices\label{ssec:cons-topo}}

\begin{figure}
\centering
\includegraphics{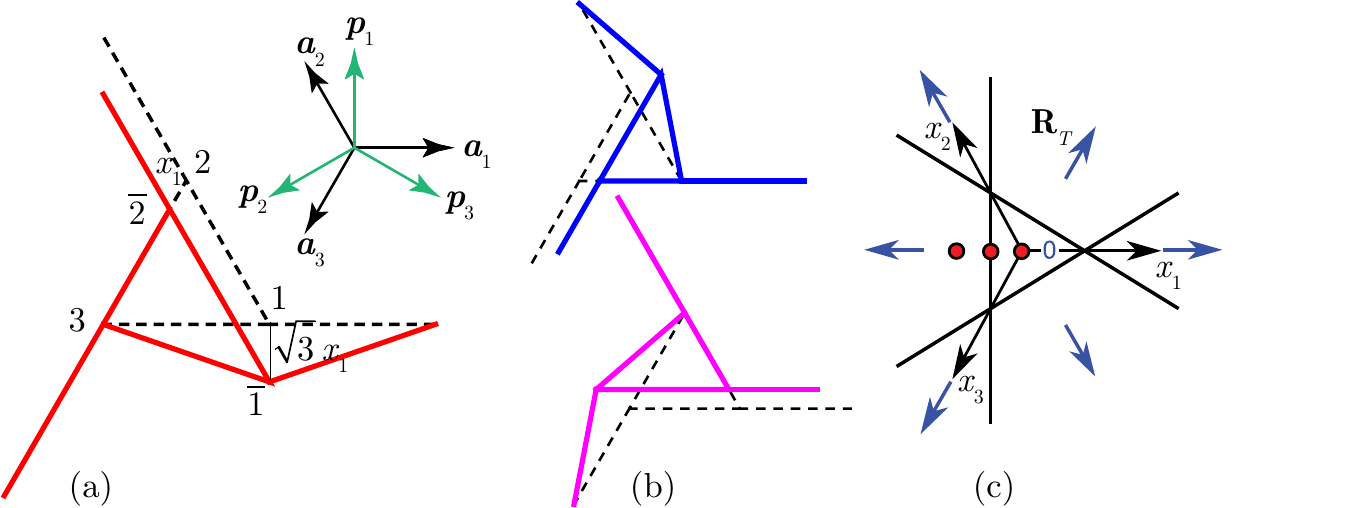}
\caption{(a) and (b) depict the geometry used to derive equations (\ref{eq:top-parameters1}) and
(\ref{eq:top-parameters2}).  In (a), site $1$ of the symmetric unit cell of \fref{fig:kagome1}(c) is displaced
downward by $\sqrt{3}\, x_3\, \pv_1$, perpendicular to the lattice vector $\av_1$, and
site $2$ is displaced along $\av_3$  by $x_1 \av_3$.  The result is that the filaments along
$\av_2$ and $\av_3$ remain straight whereas those along $\av_1$ are zigzagged.  (b) depicts similar
displacements that zigzag only filaments along $\av_2$ or $\av_3$.  (c) shows the ternary phase diagram
for fixed $x_1+x_2+x_3 >0$.  The space enclosed by the central triangle corresponds to the
to the state with $\Rv_T=0$.  The direction of $\Rv_T$ is the six sectors surrounding the
central triangle are indicated by the blue arrows. The red dots from right to left correspond to the
three lattices shown in the opposite order in \fref{fig:top-lat-dis}}
\label{fig:parameter}
\end{figure}

\begin{figure}
\centering
\includegraphics{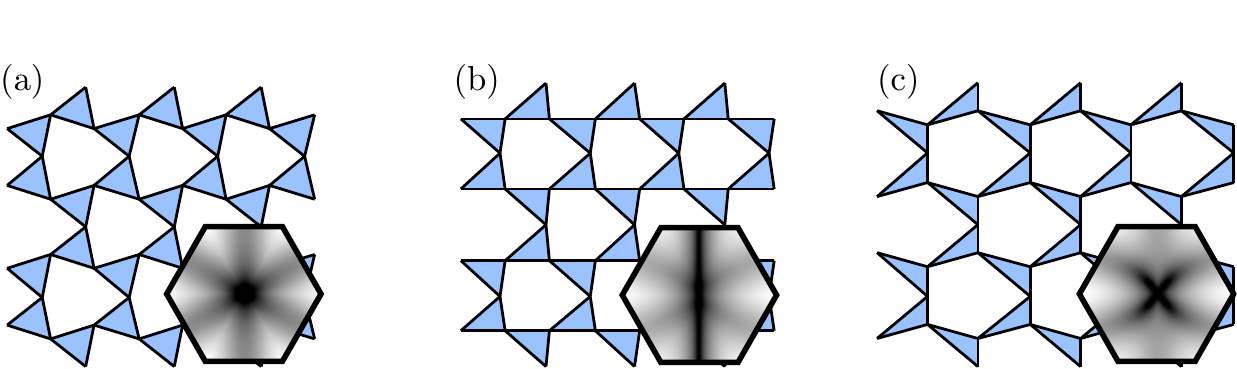}
\caption{Representations of lattices and the density maps of  their associated
lowest frequency modes for (a) a twisted kagome lattice with
$X=(0.1,0.1,0.1;0)$ and $\Rv_T=0$ (right-most red dot in \fref{fig:parameter}(c)), (b) a critical lattice with
$X=(0,0.1,0.1;0)$ (central red dot), and (c) a topological lattice with
$X=(-0.1,0.1,0.1;0)$ and $\Rv_T = -\av_1$ (left-most red dot).  Note the isotropically gapped
spectrum (except for the origin) of (a), the line of zero modes in (b), and the soft-mode cross in
(c). \emph{Adapted from reference \cite{KaneLub2014}}}
\label{fig:top-lat-dis}
\end{figure}

The kagome lattice has three sites per unit cell, and
displacing these sites  while maintaining the size and shape of
the unit cell creates different lattices, each of which can be
smoothly connected with the other across domain walls.  The
twisted kagome lattice, with bond length increased by $1/\cos
\alpha$ to connect smoothly with the untwisted lattice, is an
example of this operation.  The most general such lattice is
described by four parameters (one site in the the unit cell can
always be fixed). The most useful parametrization is one in
which the ``straightness" of the three sets of filaments are
controlled individually.  Such a set is depicted in
\fref{fig:parameter}. Displacing site $1$ of the unit cell by
$-\sqrt{3} x_1 \pv_1$, where $\pv_1$ is the vector of length
$a$ perpendicular to lattice vector $\av_1$, ``zigzags" the
filaments parallel to $\av_1$. Though this operation leaves
filaments parallel of $\av_3$ straight, it zigzags the
filaments parallel to $\av_2$. The latter can be straightened
by the simple operation of displacing site $2$ along $\av_3$ by
$x_1 \av_3$, as shown in \fref{fig:parameter}(a). This process
is repeated to zigzag filaments parallel of $\av_2$ and $\av_3$
as shown in figures \ref{fig:parameter} (b) and (c). Finally
the $1-2-3$ triangle can be isotropically expanded by
displacing the three sites the same amount along directions
$\pv_1$, $\pv_2$ and $\pv_3$.  The basis vectors for the unit
cell are then
\begin{equation}
\rv_\mu = \rv_\mu^0 - \sqrt{3} x_\mu \pv_\mu + x_{\mu-1} \av_{\mu+1}
+ (z/\sqrt{3}) \pv_{\mu-1} ,
\label{eq:top-parameters1}
\end{equation}
where $\rv_\mu^0$ (e.g., $\rv_1=\av_1/2$, $\rv_2 = - \av_3/2$,
$\rv_3 = 0$) are the basis vectors of the untwisted unit cell
and it is understood that all subscripts are evaluated mod $3$.
The bond vectors are then
\begin{equation}
\bv_\beta = \dv_{\beta+1} - \dv_\beta =\tfrac{1}{2} \av_
\beta -(-1)^{\rm{Int((\beta-1)/3)}}[(x_{\beta+1}-x_{\beta-1} - z) \av_\beta
+\sqrt{3} x_\beta \pv_\beta ] ,
\label{eq:top-parameters2}
\end{equation}
where ${\rm Int}(x)$ is the integer part of $x$ and it is
understood that $\av_{\beta+3} = \av_\beta$ and $\pv_{\beta+3}
= \pv_\beta$. The expressions for $\bv_\beta$ for $\beta =
4,5,6$ are obtained from those for $\beta=1,2,3$ via the
relation $\bv_{\beta+3} = \av_\beta - \bv_{\beta}$. The
untwisted kagome lattice corresponds to
$X=(x_1,x_2,x_3;z)=(0,0,0;0)$ and the twisted kagome with twist
angle $\alpha = \tan^{-1}(2 \sqrt{3} x)$ to $X=(x,x,x;0)$. For
a lattice with straight filaments along $\av_1$ only,
$X=(0,x_2,x_3;z)$ and similarly for straight filaments along
$\av_2$ and $\av_3$. These lattices have \SSSs and associated
zero modes along a single direction in the Brillouin Zone.
Moving off $x_1=0$ gaps the spectrum as shown in
\fref{fig:top-lat-dis}.

The topological polarization in terms of $X$ is
\begin{equation}
\Rv_T = \frac{1}{2} \sum_{p=1}^3 \av_p \,\text{sign}( x_p ).
\end{equation}
This leads to the ternary phase diagram shown in
\fref{fig:parameter}(c) as a function of $x_1$, $x_2$, and
$x_3$ for fixed $x_1+x_2+x_3$ and $z=0$. It has eight octants
corresponding to the eight possible sign combinations of
$(x_1,x_2,x_3)$. The $(+,+,+)$ and $(-,-,-)$ octants correspond
to the class of the twisted kagome lattice. The remaining 6
octants are states that are topologically equivalent, but
related to each other by $C_6$ rotations.
\Fref{fig:top-lat-dis} shows representative structures for the
$\Rv_T=0$ phase (\fref{fig:top-lat-dis}(a)), the ${\bf R}_T \ne
0$ phase (\fref{fig:top-lat-dis}(c)), and the transition
between them (\fref{fig:top-lat-dis}(b)).  The insets show
density plots of the lowest frequency mode, which highlight the
gapless point due to the acoustic mode in
\fref{fig:top-lat-dis}(a) and the gapless line due to states of
self stress in \fref{fig:top-lat-dis}(b). In
\fref{fig:top-lat-dis}(c), the gap vanishes only at the origin,
but there is a low-frequency cross that arises because acoustic
modes vary quadratically rather than linearly with $\qv$ along
its axes.  This behavior will be discussed in the next section.

\begin{figure}
\centering
\includegraphics{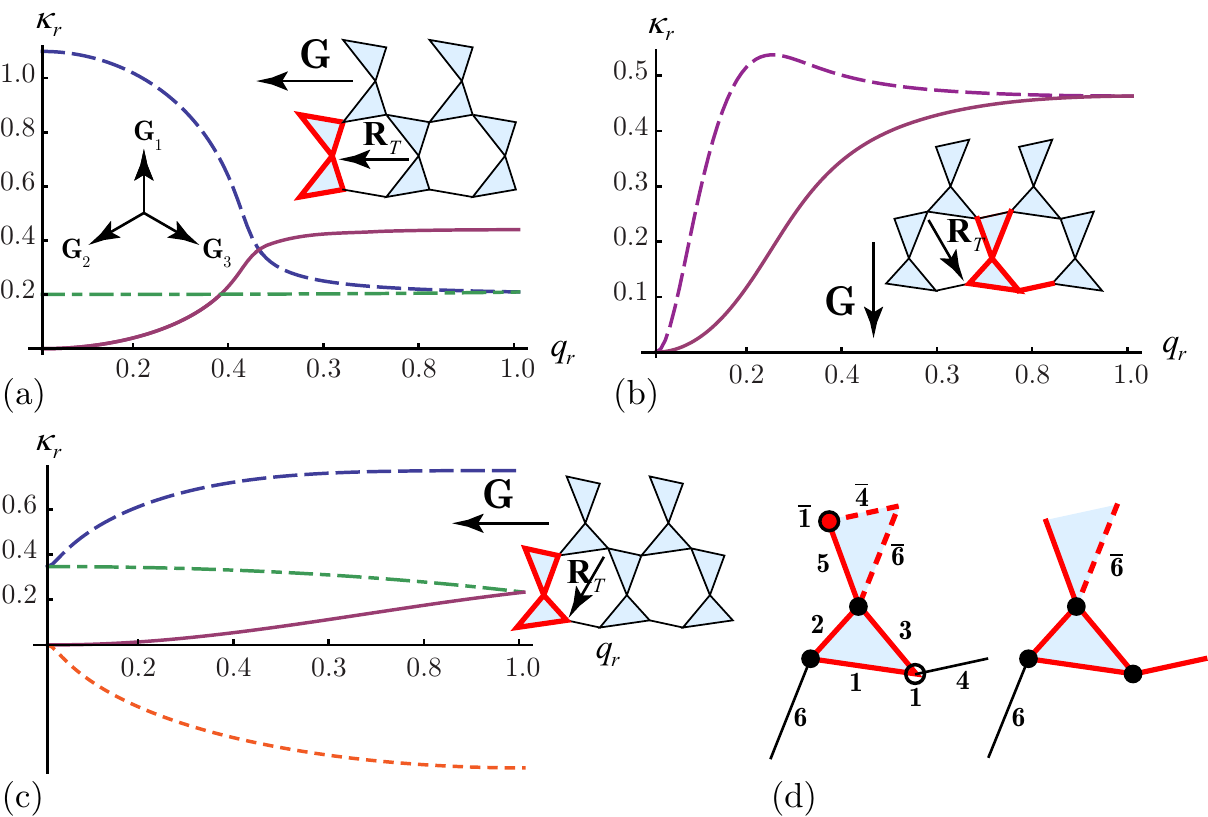}
\caption{Real part of the reduces inverse penetration depth $\kappa_r =2 \kappa/G_S$
for various orientations of surfaces and $\Rv_T$
as a function of reduced surface wavenumber $q_r = 2 q/
G_S$, where $G_S$ is the magnitude of the surface reciprocal lattice vector. In
(a) the surface lattice vector $\Gv=\Gv_2-\Gv_3$, $\Rv_T = -\av_1$,
$\Gv\cdot \Rv_T/(2\pi) = 2$, and
there are four zero surface modes.  The opposite surface with $-\Gv$ has no
zero modes. The full magenta curve is doubly degenerate with opposite imaginary
parts, and the two dashed curves are not degenerate. (b) $\Gv=-\Gv_1$, $\Rv_T = \av_3$ and $\Gv\cdot \Rv_T/(2 \pi) = 1$.
There are two surface zero modes. (c)
$\Gv= \Gv_2 - \Gv_3$, $\Rv_T = -\av_2$, $\Gv\cdot \Rv_T/(2 \pi) = 1$, and there are three surface zero modes.
The full magenta curves in (a) to (c) correspond to acoustic surface states, and in all cases $\kappa_r$ approaches
$0$ as $q_r^2$ as $q_r \rightarrow 0$.
(d) shows how to construct the dipole moments $\Rv_L$ for the surface unit cells in (a) to (c).  On the left, black bonds
$\bf{4}$ and $\bf{6}$ of the symmetric unit cell are, respectively, displaced by $\av_2$ and $-\av_3$
while the circled site $1$ is displaced by $\av_2$ to produce the surface unit cell
of (a) and (c) with $\Rv_L= 2 \av_2 -(\av_2-\av_3) = -\av_1$ and $\tnu_L^S=2$.
On the right, the black bond $\bf{6}$ is displaced
through $-\av_3$ to produce the surface unit cell of (b) with $\Rv_L=\av_3$ ad $\tnu_L=1$. }
\label{fig:topological-surface}
\end{figure}

\begin{table}
\caption{\label{table1} Reciprocal lattice vectors $\Gv$
indexing surfaces, dipole moment $\Rv_L$, and local count
$\tilde{\nu}_L$ for the seven surfaces cells of
\fref{twisted-kagome-surface}(a).  The reciprocal lattice
vectors $\Gv_1$, $\Gv_2$, and $\Gv_3$ are depicted in
\fref{fig:kagome-disperion1} and the Bravais lattice vectors
$\av_1$, $\av_2$, and $\av_3$ in \fref{fig:kagome1}.}
\begin{ruledtabular}
\begin{tabular}{@{}c|ccccccc}%\br
cell & 1 & 2 & 3 & 4 & 5 & 6 & 7 \\
%\mr
$\Gv$ & $-\Gv_1$ & $\Gv_1$ & $\Gv_1$ & $\Gv_2 - \Gv_3$ & $\Gv_3 - \Gv_2$ & $\Gv_3 - \Gv_2$ & $- \Gv_1$ \\
$\Rv_L$ & $\av_3$ & $\av_2$ & $-\av_3$ & $- \av_1$ & $\av_1$ & $\av_1$ & $\av_3- \av_2$  \\
$\tilde{\nu}_L$ &  $1$ & $1$ & $1$ & $2$ & $2$ & $2$ & $2$  \\
%\br
\end{tabular}
\end{ruledtabular}
\end{table}

We are now in a position to analyze zero interface modes for
different $X$.  Surfaces in the twisted-kagome octants are in
the same class of those discussed in \sref{ssec:surface-m} and
in \fref{twisted-kagome-surface}.  The bonds and sites to be
displaced to convert the standard unit cell to a surface one
are depicted for two cells in
\fref{fig:topological-surface}(d). Table \ref{table1} lists the
surface polarization vector, the surface reciprocal lattice
vector $\Gv$, and the reduced surface index $\tnu_L$ (which
because there are no \SSSs equals the number of zero modes per
surface cell) for the seven surface cells shown in
\fref{twisted-kagome-surface}. As required, the count
corresponds to the number obtained by direct evaluation.

\begin{figure}
\centering
\includegraphics{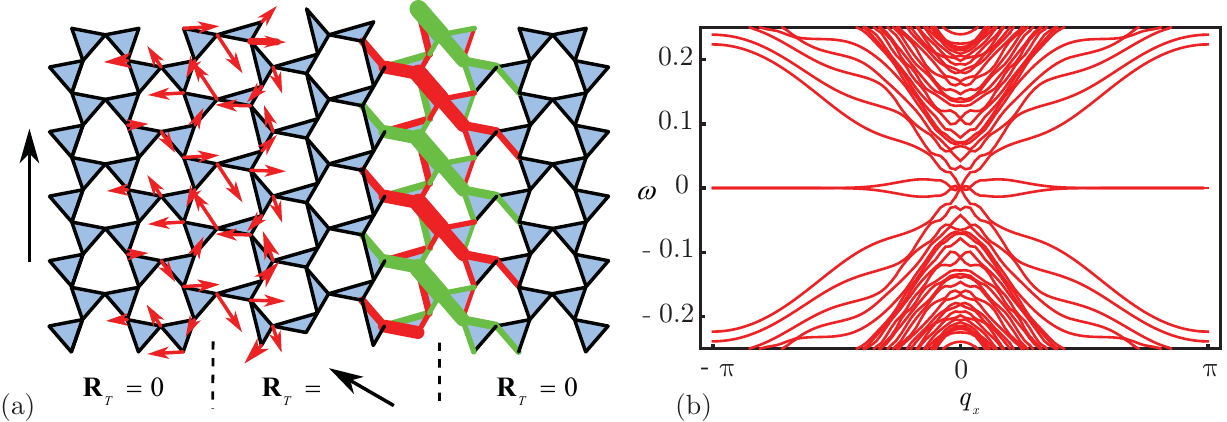}
\caption{Zero modes at domain walls.  (a) shows a
lattice with periodic boundary conditions and two domain walls,
the left one between $(.1,.1,.1;0)$ and $(.1,.1,-.1;0)$ with zero modes
and the right one between
$(.1,.1,-.1;0)$ and $(.1,.1,+.1;0)$ with states of self stress.
The zero mode eigenvectors at $q_x=\pi$ are indicated
for the floppy mode (arrows) and the state of self-stress
(red (+) and green (-) thickened bonds).  (b) shows the vibrational
mode dispersion as a function of $q_x$. \emph{From reference \cite{KaneLub2014}}}
\label{fig:topological-domain}
\end{figure}

Converting  $\Rv_T=0$ to $\Rv_T \neq 0$ modifies the surface
count by the term $\Rv_T \cdot \Gv/(2 \pi )$ in $\tnu$. Since
$\Gv$ has opposite signs on opposite parallel surfaces, the
effect is to move zero surface modes between the two surfaces
while keeping the total count fixed as required.
\Fref{fig:topological-surface} plots the real part of the
surface penetration wavenumber, $\kappa$, of three
representative surfaces and orientations of $\Rv_T$ relative to
the surface normals.  In \fref{fig:topological-surface}(a),
$\Rv_T$ is parallel to $\Gv$ of a vertical surface, and there
are four rather than the two zero modes of the non-topological
surface of \fref{twisted-kagome-surface} with surface cell $4$.
Thus on the opposite surface (with surface cells $5$ an $6$ in
\fref{twisted-kagome-surface}), there are no zero modes.
Similarly in \fref{fig:topological-surface}(b) with $\Rv_T$ at
$30^\circ$ to $\Gv$ of a horizontal surface, there are two zero
surface modes whereas the same surface (cell $1$ in the
$\Rv_T=0$ lattice of \fref{twisted-kagome-surface}) has only
one.  Finally \fref{fig:topological-surface}(c) with $\Rv_T$ at
$60^\circ$ to $\Gv$ of a vertical surface, there are three
rather than two surface modes per unit cell. A striking feature
is that the curves versus the surface wavenumber $q$ of
$\kappa$ for  acoustic modes when $\Rv_T \neq 0$ is that their
approach to $0$ as $q\rightarrow 0$ is quadratic rather than
linear in $q$. This is a consequence of the modes with $q^2$
dispersion shown in \fref{fig:top-lat-dis}.

\subsubsection{Continuum limit}

\begin{figure}
\centering
\includegraphics{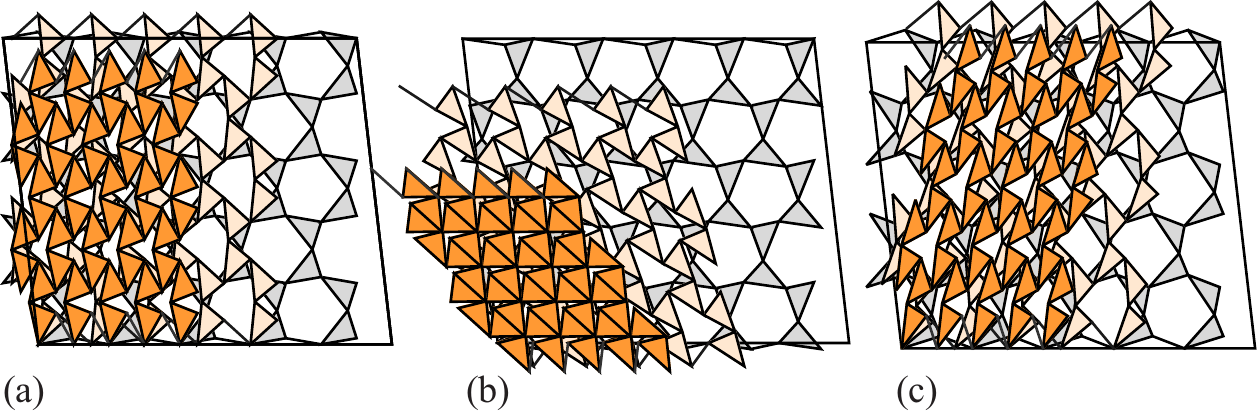}
\caption{Evolution
 of the nonlinear Guest mode in  topological
lattices in response to uniform compression along the $x$-axis.
(a) $X=\{-0.05,0.05,0.05\}$:  Small compression along $x$ produces area change
along with pure shear with a positive Poisson ratio in agreement with \eref{eq:elastic}.
At large compression, there is contraction along both $x$ and $y$
reflecting a nonlinear negative Poisson ratio.  (b) $X=\{0.05,-0.05,0.05\}$: Compression
along $x$ now produces a simple shear component with length contraction
in all directions (negative Poisson ratio).  (c) $X=\{-0.1,0.025,0.075\}$:
Compression along $x$ now produces some simple shear, but the Poisson ratio
($-u_{yy}/u_{xx}$) remains positive for all compressions.}
\label{Fig:PositivePoisson}
\end{figure}

As we have seen, there are two related long-wavelength
properties of the topological lattices that differ from their
non-topological counterparts: Their spectrum has peculiar
low-frequency lobes [\fref{fig:top-lat-dis}(c)], and
penetration wavenumbers of their acoustic modes approach zero
with $q$ as $q^2$ rather than as $q$.  These properties must be
reflected in the in the form of the long-wavelength elastic
energy, and indeed they are.  For simplicity we focus on states
with $X=(x_1,x_2,x_2;0)$, where $x_2>0$ is fixed and $x_1$ is
allowed to vary. The elastic energy density $f$ can be written
\begin{equation}
f=\tfrac{1}{2}K[(u_{xx}-r_1 u_{yy})^2 + 4 r_4 u_{xy}^2] ,
\label{eq:elastic}
\end{equation}
where $r_{1}\propto x_1$ for small $x_1$, while $r_4>0$ and $K$
are positive and smoothly varying near $x_1=0$. Thus, the
$\Rv_T=0$ and $\Rv_T \neq 0$ sectors are distinguished by the
sign of $r_1$. The Guest mode \cite{GuestHut2003}, for which
$f=0$, corresponds to shape distortions with $u_{xx} = r_1
u_{yy}$ and $u_{xy}=0$. When $r_1>0$, $u_{xx}$ and $u_{yy}$
have the same sign, and the distortion has a negative Poisson
ratio \cite{Lakes1987}, expanding or contracting in orthogonal
directions (a feature shared by the twisted kagome lattice
\cite{SunLub2012}); when $r_1<0$, $u_{xx}$ and $u_{yy}$ have
the same sign, and the distortion has the more usual positive
Poisson ratio. Finally when $r_1=0$, uniaxial compressions
along $y$ costs no energy. Note that this energy consists of
two independent positive definite quadratic forms as it must in
a periodic isostatic lattice with two \SSSsa.
\Fref{Fig:PositivePoisson} shows the effects of evolution of
the nonlinear Guest mode in response to compression along the
$x$-axis of various topological lattices.

Expanding $\det \Cv$ for small $\qv$ provides useful
information about the bulk- and surface-mode structure. To
order $q^3$,
\begin{equation}
\det \Cv = A[q_x^2+ r_1 q_y^2 + i \gamma (q_x^3 - 3q_x q_y^2)]
+O(q^4),
\label{eq:detQT}
\end{equation}
where $A>0$ and  $\gamma> 0$ for small $x_1$. The quadratic
part of the equation follows from \eref{Eq:detCv} and the
elastic matrix, \eref{eq:elas-matrix}, with $K_{xxxx}= K$,
$K_{yyyy} = r_1^2 K$, and $K_{xyxy}= r_4 K$, associated with
the free energy of \eref{eq:elastic}. Long-wavelength zero
modes are solutions of $\det \Cv = 0$.   The quadratic term,
which corresponds to the elastic theory reveals an important
difference between the bulk acoustic modes of $\Rv_T= 0$ and
$\Rv_T \neq 0$. In the former case, $r_1>0$, $\det \Cv=0$ only
at $\qv=0$. For $r_1<0$, though, to order $q^2$, $\det \Cv = 0$
for $q_x = \pm \sqrt{|r_1|} q_y$, so the elastic theory
predicts lines of gapless bulk modes. The degeneracy is lifted
by the $q^3$ term, leading to a $q^2$ dispersion along those
lines, which can be seen by the cross in the density map of
\fref{fig:top-lat-dis}(c).

As we have seen, $\det \Cv (\qv\rightarrow 0)$ vanishes for
complex wavenumbers associated with zero-frequency Rayleigh
surface waves. Writing $\qv = q_{\perp} \hat{\nv} + q_{||} \hat
z \times \hat \nv$ for a surface whose outward normal $\hat
\nv$ makes an angle $\theta$ with $\hat x$, there is an
$\omega=0$ Rayleigh wave with penetration depth $\kappa^{-1} =|
{\rm Im}\,q_{\perp}|^{-1}$ if ${\rm Im}\, q_{\perp} <0$. To
order $q_{||}^2$ there are two solutions,
\begin{equation}
q_{\perp}^\pm = \rmi \kappa
\frac{\sin \theta \pm \rmi \sqrt{r_1}\cos \theta}{\cos \theta \mp \rmi \sqrt{r_1} \sin \theta} q_{||}
+ \frac{\sqrt{r_1} ( 3 + r_1) \gamma}{2(\cos \theta \pm \rmi \sqrt{r_1} \sin \theta )^3}
q_{||}^2 .
\end{equation}
When $r_1>0$, the linear term is always finite and nonzero, and
${\rm Im}\ q_{\perp}^\pm$ have opposite signs, indicating that
there can be acoustic surface zero modes on all surfaces. These
are the classical Rayleigh waves predicted by the elastic
theory \cite{Landau1986}, with penetration depth
$O(q_{||}^{-1})$. When $r_1<0$, the linear term in $q_{||}$ is
real and $\kappa_r = \Im\, q_{\perp}^{\pm} \propto q_{||}^2$.
The number of long wavelength surface zero modes depends on the
angle of the surface.  When $|\theta| <\theta_c = \cot^{-1}
\sqrt{|r_1|}$, $\Im\,q_{\perp}^{\pm}$ are both positive, and
there are no acoustic surface zero modes.   The opposite
surface, $|\theta-\pi|<\theta_c$, has two acoustic surface
modes.  For $\theta_c < \theta < \pi - \theta_c$,
$\Im\,q_{\perp}^{\pm}$ have opposite sign, so there is one
mode. This is consistent with the results shown in figures
\ref{twisted-kagome-surface} and \ref{fig:topological-surface}.
In the former figure, $r_1>0$, and there are acoustic zero
modes on every surface. In the latter, $\Rv_T$ is chosen so
that the coordinate system can always be rotated so that its
direction corresponds to the $-x$ direction.  Thus, in
\fref{twisted-kagome-surface}(a), the surface normal $\Gv$ and
$\Rv_T$ are parallel, implying by the above considerations that
$\theta = \pi$ and that there should be two acoustic zero
surface modes.  In (b), $\theta = 5\pi/6$, and there are also
two zero acoustic modes whereas in (c), $\theta = 3\pi/3$, and
there is only one acoustic mode.  This is consistent with the
above long-wavelength analysis if $\pi/6 < \theta_c < \pi/3$.
Finally, for surfaces such as those with cells $1$, $2$, $3$,
and $7$ in \fref{twisted-kagome-surface}, $\theta = \pm\pi/2$
for systems with $\Rv_T$ in the $\pm x$ direction, and there
should be one acoustic zero mode on each surface as there are,

\section{Review and future directions\label{sec:review}}

In this review, we have presented a pedagogical introduction to
Maxwell frames, in free versions of which $N_B = dN - d(d+1)/2$
and in the periodic versions of which $N_B = dN$, and to
related frames on the verge of mechanical collapse. We made
extensive use of the \Ith \cite{Calladine1978}, which relates
the number of zero modes and \SSSs of a frame to its site
number $N$ and bond number $N_B$, and of a relation between
elastic energy and \SSSs \cite{Pellegrino1993,Goodrich2014} to
frame our discussion of the elasticity and vibrational spectrum
of these frames. We concentrated on Maxwell lattices, whose
sites can be collected into unit cells whose origins lie on
sites in a Bravais lattice, and we paid particular attention to
the surface zero-modes that necessarily arise when periodic
Maxwell lattices are cut to create free surfaces.

All of the physical examples we studied were two-dimensional,
both because important concepts are most easily explored in two
rather than higher dimensions and because there is very little
work of the type we discuss on higher dimensional systems.
There is nonetheless, interesting work to be done on
three-dimensional systems.  The most obvious frames to
investigate are variants of the pyrochlore lattice, a
generalization of the two-dimensional kagome lattice composed
of corner sharing tetrahedra arranged so that lines of bonds
form straight filaments.  Under periodic boundary conditions,
this lattice is a Maxwell lattice with $N_B = 6N$. Preliminary
work \cite{StenullLub2014b} indicates that this lattice can be
distorted in much the same way as the kagome lattices to gap
the spectrum and produce topologically distinct states with
protected interfacial zero modes. Three-dimensional Maxwell
lattices other than the simple cubic lattice include the many
zeolite lattices, which like the pyrochlore lattice are based
on corner sharing tetrahedra and which are in a sense $3D$
generalizations of the distorted kagome lattices like the
twisted lattice of \fref{fig:square-kagome}(d) or the more
complex layered lattice of \fref{fig:kagome-variation}.  These
lattices all have a ``flexibility window" that, like the
twisted kagome lattice allow for easy change in volume
\cite{Sartbaeva2006}. It is likely that judicious
rearrangements of lattice sites in these lattice will yield
different topological states.

We considered here only linearized elasticity and vibrational
structure. Maxwell frames obviously have interesting nonlinear
properties, which in the end are responsible, for example, for
the ability of the kagome lattice to undergo large area change
upon twisting.  References \cite{ChenVit2014,VitelliChe2014}
explored the nonlinear properties of the one-dimensional model
\cite{KaneLub2014} discussed in \sref{sec:one-d-model} and
found that under appropriate boundary conditions, the surface
zero-modes become elevated to zero-energy nonlinear topological
modes that propagate freely throughout the bulk. The system is
a nonlinear mechanical conductor whose carriers are nonlinear
solitary waves not captured by the linearized theory.  An
obvious question is whether similar behavior will be found in
appropriately designed two- or three-dimensional frames.

As we have seen Maxwell lattices exhibit a surprisingly rich
variety of vibrational responses.  Ideal lattice-structures are
generally the exception rather than the rule, and one can ask
what effect do defects like dislocations have on the linear and
nonlinear vibrational spectrum of Maxwell lattices. Reference
\cite{PauloseVit2014} studied just this question in dislocated
topological kagome lattices and found that zero modes can be
localized at dislocations as a result of the interplay between
between the topological dipole $\Rv_T$ of the lattice and the
topological Burgers vector of the dislocations. Thus zero modes
can be localized at a point by dislocations and along a a grain
boundary separating two topologically distinct phases.
Localized modes can also be created by enclosing a region of
topological type B in a lattice of topological type A. There is
certainly the potential for interesting and perhaps useful
generalizations of these ideas.

{\bf Acknowledgments} We are grateful for illuminating
discussions with Simon Guest, who brought reference
\cite{Calladine1978} to our attention, members of the
soft-matter theory group at the University of Pennsylvania
including Randall Kamien, Andrea Liu, Carl Goodrich, and Daniel
Sussman, and Bryan Chen and Vincenzo Vitelli of the University
of Leiden. This work was supported in part by NSF under grants
DMR-1104707 and DMR-1120901 (TCL), DMR-1207026 (AS),
DMR-0906175 (CLK), by a Simons Investigator award to CLK, the
Georgia Institute of Technology (AS).

\appendix
\section{States of self stress and the elastic
energy\label{app:elas-SSS}}

In this appendix, we present a derivation of
\eref{eq:elastic-self-stress}.  Our starting point is
\eref{eq:Vel} for the stretch energy in terms of the $N_B$
dimensional bond-elongation vector $\ma{E}$, which we break up
into an affine part $\ma{E}_{\text{aff}}$ with components given
by \eref{eq:affine-stretch} and a part $\ma{C}\ma{U} = \ma{Q}^T
\ma{U}$ describing additional elongation that relaxes in
response to the macroscopic strain imposed by
$\ma{E}_{\text{aff}}$.  Multiplying $\ma{Q}^T$ on the left by
the decomposition of the $N_B$-dimensional into the projection
operators $\ma{P}^Q_{r}$ and $\ma{P}^Q_s$, onto $\oc (\ma{Q})$
and $\ker (\ma{Q})$, respectively, and on the right by the
decomposition of the $dN$-dimensional unit matrix into
projections operators $\ma{P}^C_r$ and $\ma{P}^C_z$ onto $\oc
(\ma{C})$ and $\ker (\ma{C})$, respectively, $\ma{Q}^T$ can be
decomposed into $\ma{Q}^T =\ma{Q}^T_{rr} +
\ma{Q}^T_{sr}+\ma{Q}^T_{rz} +\ma{Q}^T_{sz}$, where
$\ma{Q}^T_{sr}=\ma{P}^Q_{s}\ma{Q}^T\ma{P}^C_r$, and so on. When
$\ma{Q}^T$ multiplies any vector to its right, the components
of that vector in $\ker(\ma{C})$ are annihilated, and only the
components in $\oc (\ma{C})$ survive.  Thus, $\ma{Q}^T \ma{U} =
(\ma{Q}^T_{rr} + \ma{Q}^T_{sr})\ma{U}_r$. When this quantity is
multiplied on the left by any $N_B$-dimensional vector
$\ma{W}$, the term involving $\ma{Q}^T_{sr}\ma{U}_r$ vanishes
because it is orthogonal to $\oc (\ma{Q})$ and it gives zero
for any vector in $\ker (\ma{Q})$.  Thus, in $V_{\text{el}}$,
$\ma{Q}^T \ma{U}$ can be replaced by $\ma{Q}^T_{rr} \ma{U}_r$.
Similarly, $\ma{U}^T \ma{Q}$ can be replaced by $\ma{U}^T_r
\ma{Q}_{rr}$, and the elastic energy as a function of $\ma{U}$
and $\ma{E}_{\text{aff}}$ is
\begin{equation}
V f_{\text{el}} = \tfrac{1}{2} (\ma{E}_{\text{aff}}^T
+\ma{U}^T_r \ma{Q}_{rr}) \ma{k} (\ma{E}_{\text{aff}}
+\ma{Q}^T_{rr} \ma{U}_r) .
\end{equation}
Minimizing over the $dN - N_0$ independent components of
$\ma{U}_r$ (or $\ma{U}^T_r$) yields
\begin{equation}
V\frac{\partial f_{\text{el}}}{\partial \ma{U}^T_r} =
\ma{Q}_{rr} \ma{k} (\ma{E}_{\text{aff}}
+\ma{Q}^T_{rr} \ma{U}_r)
\end{equation}
or, because both $\ma{Q}_{rr}$ and $\ma{k}_{rr}$ are
invertible,
\begin{equation}
\ma{E}_{\text{aff},r}
+\ma{Q}^T_{rr} \ma{U}_r =
-(\ma{k}_{rr})^{-1} \ma{k}_{rs} \ma{E}_{\text{aff},s}
\end{equation}
and
\begin{equation}
V f_{\text{el}}= \tfrac{1}{2} \ma{E}_{\text{aff},s}^T
(\ma{k}_{ss} - \ma{k}_{sr} (\ma{k}_{rr})^{-1} \ma{k}_{rs}) \ma{E}_{\text{aff},s}
=\tfrac{1}{2} \ma{E}_{\text{aff},s}^T[(\ma{k}^{-1})_{ss}]^{-1}  \ma{E}_{\text{aff},s} ,
\end{equation}
where $(\ma{k}^{-1})_{ss}$ is the projection of $\ma{k}^{-1}$
onto the null space of $\ma{Q}$. This is
\eref{eq:elastic-self-stress}.  The final relation can be
derived as follows:  Let $\ma{p}$ the inverse of $\ma{k}$ and
decompose the two matrices into their projections onto the
range and nullspace of $\ma{Q}$,
\begin{equation}
\ma{k} =
    \begin{pmatrix}
        \ma{k}_{ss} & \ma{k}_{sr} \\
        \ma{k}_{rs} & \ma{k}_{rr}
    \end{pmatrix}  \qquad
\ma{p}=
     \begin{pmatrix}
        \ma{p}_{ss} & \ma{p}_{sr} \\
        \ma{p}_{rs} & \ma{p}_{rr} .
    \end{pmatrix}
\end{equation}
Then $\ma{k}\, \ma{p} = \ma{I}$ implies that $\ma{k}_{ss}
\ma{p}_{ss} + \ma{k}_{sr} \ma{p}_{rs} = \ma{I}_{ss}$ and
$\ma{k}_{rs} \ma{p}_{ss} + \ma{k}_{rr} \ma{p}_{rs} = 0$.  Thus,
$\ma{k}_{ss} +\ma{k}_{sr} \ma{p}_{rs} \ma{p}_{ss}^{-1}
=\ma{p}_{ss}^{-1}$ and $\ma{k}_{rs} = - \ma{k}_{rr} \ma{p}_{rs}
\ma{p}_{ss}^{-1}$, and finally $\ma{k}_{ss} - \ma{k}_{sr}
\ma{k}_{rr}^{-1} \ma{k}_{rs} = \ma{p}_{ss}^{-1} =
[(\ma{k}^{-1})_{ss}]^{-1}$ as required.

\section{Zeroes of $\det\ma{K}(\qv)$ and Guest Modes.}
\label{App:Guest}

In this appendix we derive \eref{eq:GuestK} relating the
determinant of the long-wavelength dynamical matrix and the
eigenvectors of the elastic Guest mode. We begin by
representing $\st_{ij}$ as $\frac{1}{2}(q_i u_j + q_j u_i)$.
The the strain vector becomes
\begin{equation}
\stm_V =
\begin{pmatrix}
i q_x u_x \\
i q_y u_y \\
\tfrac{i}{2}(q_x u_y + q_y u_x)
\end{pmatrix} \equiv i
\begin{pmatrix}
q_x & 0 \\
0 & q_y \\
\tfrac{1}{2} q_y & \tfrac{1}{2} q_x
\end{pmatrix}
\begin{pmatrix}
u_x \\
u_y
\end{pmatrix} \equiv
\Ss_{a i} u_i ,
\end{equation}
where the summation convention is understood, $a=1,2,3$, and $i
= x,y$.  The elastic energy density for a fixed $\qv$ is thus
\begin{equation}
f_{\text{el}} = \tfrac{1}{2} u_i \Ss^T_{ia} \mbb{K}_{ab} \Ss_{bj} u_j=
\tfrac{1}{2} u_i(-\qv) K_{ij}(\qv) u_j(\qv) ,
\end{equation}
where $\ma{K}_{ij} = \Ss^T_{ia} K_{ab} \Ss_{bj}$. Then using
\eref{eq:Guest-K},
\begin{equation}
\ma{K}_{ij} = \Ss^T_{ia} \left(\sum_{p=1,2}K_p \vs_{pa} \vs_{pb}\right) \Ss_{bj}
\equiv \sum_{p=1,2} K_p \tSS_{pi}\tSS_{pj},
\end{equation}
where $\tSS_{pj} = \vs_{pb}\Ss_{bj}$, or
\begin{eqnarray}
\tSS_{11} & = & \ve_1 \cdot (q_x,0,\tfrac{1}{2}q_y) = q_x \vs_{11} + \tfrac{1}{2} q_y\vs_{13} \nonumber\\
\tSS_{22} & = & \ve_2 \cdot (0,q_y,\tfrac{1}{2}q_x )= q_y \vs_{22} + \tfrac{1}{2} q_x \vs_{23} \nonumber\\
\tSS_{12} & = & \ve_1 \cdot (0,q_y,\tfrac{1}{2}q_x )= q_y \vs_{12} + \tfrac{1}{2} q_x \vs_{13} \nonumber\\
\tSS_{21} & = & \ve_2 \cdot (q_x,0,\tfrac{1}{2}q_y) = q_x \vs_{21} + \tfrac{1}{2} q_y \vs_{23} .
\end{eqnarray}
Thus,
\begin{equation}
\ma{K}_{ij} =
\begin{pmatrix}
K_1 \tSS_{11}^2 + K_2 \tSS_{21}^2 & K_1 \tSS_{11} \tSS_{12} + K_2 \tSS_{21}\tSS_{22} \\
K_1 \tSS_{12} \tSS_{11} + K_2 \tSS_{21}\tSS_{22} & K_1 \tSS_{12}^2 + K_2 \tSS_{22}^2
\end{pmatrix}
\end{equation}
and
\begin{eqnarray}
\det\ma{K} & = & (K_1 \tSS_{11}^2 + K_2 \tSS_{21}^2)(K_1 \tSS_{12}^2+K_2 \tSS_{22}^2)
-(K_1 \tSS_{11}\tSS_{12} + K_2 \tSS_{21}\tSS_{22} )^2 \nonumber\\
& = & K_1 K_2 (\tSS_{11}\tSS_{22}-\tSS_{12}\tSS_{21})^2 \nonumber\\
&\equiv & K_1 K_2 (\det\tSS )^2 ,
\label{eq:det-ma(K)}
\end{eqnarray}
where
\begin{equation}
\det \tSS = \Ss_{1i}\Ss_{2j} (\vs_{1i}\vs_{2j} - \vs_{2i}\vs_{1j} ) .
\end{equation}
The three-component vectors $\ve_1$, $\ve_2$, and $\ve_0$ form
an orthonormal triad such that $(\vs_{1i}\vs_{2j} -
\vs_{2i}\vs_{1j} ) =\epsilon_{ijk}\vs_{0k}$, where
$\epsilon_{ijk}$ is the anti-symmetric Levi-Civita symbol, and
\begin{equation}
\det \tSS= \Ss_{1i}\Ss_{2j}\epsilon_{ijk} \vs_{0k}=
-\frac{1}{2} (\vs_{02} q_x^2 - 2 \vs_{03} q_x q_y + \vs_{01} q_y^2) .
\end{equation}
Using this equation and \eref{eq:det-ma(K)} yields
\eref{eq:GuestK}.

\section{Compatibility matrix for kagome
lattices\label{app:compat}}

The kagome lattice such a central role in this review.  Here we
derive and display the $\qv$-dependent compatibility matrix for
a general lattice for the symmetric unit cell of
\fref{fig:kagome1}(c), which identifies and labels the three
sites and six bonds in the cell.  The labeling of the sites and
bonds does not change when the cell is distorted to produce the
twisted kagome of topological lattices. The equations relating
the six bond elongations in the unit cell to sites
displacements in that and nearest neighbor cells are
\begin{align}
e_1 &= (\uv_1(\Rv)-\uv_3(\Rv))\cdot \bhv_1 \qquad &e_2  = (\uv_2(\Rv)-\uv_1(\Rv))\cdot \bhv_2 \nonumber \\
e_3 &  = (\uv_3(\Rv)-\uv_2(\Rv))\cdot \bhv_3 \nonumber
& e_4   = (\uv_3(\Rv + \av_1) - \uv_1(\Rv))\cdot \bhv_4  \nonumber \\
e_5 & = (\uv_1(\Rv + \av_1) - \uv_2(\Rv))\cdot \bhv_5 \qquad
& e_6  = (\uv_2(\Rv + \av_1) - \uv_3(\Rv))\cdot \bhv_6 ,
\label{eq:A-e-ueqs}
\end{align}
where $\Rv$ is the position of the unit cell, and $\av_i$ are
the lattice vectors shown in \fref{fig:kagome1}(c) and the
resultant compatibility matrix (in a gauge in which the site
and bond basis vectors $\rv_{\mu}$ and $\rv_\beta$ are set to
zero) is
\begin{equation}
\Cav_\text{
sym}(\qv)  =
\begin{pmatrix}
\bh_{1x} & \bh_{1y} & 0 & 0 & - \bh_{1x} & - \bh_{1y}\\
-\bh_{2x} & -\bh_{2y}& \bh_{2x} & \bh_{2y} & 0 & 0 \\
0 & 0 & -\bh_{3x} & -\bh_{3y} & \bh_{3x} & \bh_{3y} \\
-\bh_{4x} & -\bh_{4y} & 0 & 0 & \rme^{\rmi \qv\cdot \av_1} \bh_{4x} & \rme^{\rmi \qv\cdot \av_1} \bh_{4y} \\
\rme^{\rmi \qv\cdot \av_2} \bh_{5x} & \rme^{\rmi \qv\cdot \av_2} \bh_{5y} & -\bh_{5x} & - \bh_{5y} & 0 & 0\\
0 & 0 & \rme^{\rmi \qv\cdot \av_3} \bh_{6x} & \rme^{\rmi \qv\cdot \av_3} \bh_{6y} & -\bh_{6x} & - \bh_{6y}
\end{pmatrix}
\label{eq:C-kagome}
\end{equation}

\section{Gauge choice and surface zero modes\label{App:gauge}}

This appendix presents details about properties of $\det \Cv$
in different choices of the unit cell (different gauges) and
how switching from the symmetric unit cell (i.e., those shown
in \fref{fig:kagome1}) to ones that match surface boundaries
increases the count of topological count.

We begin with the observation that $\det \Cv_{\text{Sym}}$ is
the compatibility matrix in the symmetric gauge
[\eref{eq:C-kagome}] of the kagome lattice scan be expressed in
terms of four parameters $\g_0, ..., \g_3 $ (each a function of
the unit vectors $\bhv_\beta$) as
\begin{equation}
\det \Cv_{\text{Sym}}= g_1 S_2 S_3 + \g_2 S_3 S_1 + \g_3 S_1 S_2
+ g_0 S_1 S_2 S_3 ,
\label{eq:AppdetC}
\end{equation}
where $S_n = w_n -1$ with $w_n = e^{i \qv \cdot \av_n}$.  Note
that there is no constant term or terms linear in $w_n$.  This
is a reflection of the fact that the translation invariance
requires the stiffness matrix $\Kv= k\Cv^T \Cv$
\eref{eq:stiffness2}  have two eigenvalues that vanish as $q$
in the long-wavelength $q\rightarrow 0$ limit and thus that
$\det \Kv$ vanish as $q^4$, or equivalently that $\det \Cv$
(for any gauge) vanish as $q^2$. With the help of the identity
$w_1 w_2 w_3 = 1$, \eref{eq:AppdetC} expressed in terms of the
$w_n$s becomes
\begin{equation}
\det \Cv_{\text{Sym}}= a + \sum_{n=1}^3 b_n w_n +
\sum_{n=1}^3 c_n w_n^{-1} ,
\end{equation}
where $a = \g_1 + \g_2+\g_3$, $b_n = \g_0 +\g_n - a$, and $c_n
= \g_n - \g_0$, or setting $w_1 = z_x$, $w_2 = z_x^{-1/2} z_y$,
and $w_3 = z_x^{-1/2} z_y^{-1}$, where $z_x = e^{i q_x}$ and
$z_y = e^{i \sqrt{3} q_y /2}$,
\begin{equation}
\det \Cv_{\text{Sym}} = (a+b_1 z_x + c_1 z_x^{-1})+
(b_2 z_x^{-1/2} +c_3 z_x^{1/2}) z_y +
(b_3 z_x^{-1/2} + c_3 z_x^{1/2})z_y^{-1} .
\end{equation}
This equation can be used without further modification to study
the topological number associated with $\det \Cv_{\text{Sym}}$
for $\qv$-vectors an paths between equivalent points in the BZ
parallel to the $x$ and $y$ axes.

The Cauchy argument principle provides a relative count of the
number of zeros and poles of any meromorphic function $F(z)$ of
a complex number $z$ in the interior of a contour $\cal{C}$:
\begin{equation}
\frac{1}{2 \pi i} \oint_{\cal{C}}\frac{F'(z)}{F(z)}dz =
\frac{1}{2 \pi i} \oint_{\cal{C}}\frac{d \ln F(z)}{dz} dz = n-p ,
\end{equation}
where $n$ is the number of zeros and $p$ the number of poles
counted with their order (e.g., if $F(z) = z^{-2}$, $p=2$) in
the region bounded by $\cal{C}$.  If the contour $\cal{C}$ is
the unit circle, the the magnitude at $z$ at any zero of $F(z)$
is less than unity.  Thus, the argument principle applied to
$\det \Cv$ with $\cal{C}$ the unit circle will count the number
of zeros with $|z|<1$, and thus the number of decaying surface
modes, minus the number of poles.

Consider first the case for $\qv$ along the $y$-direction (i.e.
perpendicular to a surface parallel to the $x$-axis). $z_y$
remains unchanged under the transformation $q_y \rightarrow q_y
+4 \pi/\sqrt{3} a$ or equivalently under $\qv \rightarrow
\qv+\Gv_1$, where $\Gv_1$ is defined in \fref{fig:kagome1},
reflecting the fact that the width of the symmetric unit cell
along $y$ is $a\sqrt{3}/2$. $\det \Cv_{\text{Sym}}$ is equal to
$z_y^{-1}$ times a quadratic polynomial in $z_y$ and therefore
has one pole and either $n = 0,1$, or $2$ zeros. Thus by the
argument principle,
\begin{equation}
\frac{1}{2 \pi i} \oint \frac{d}{dz_y} \ln \det \Cv_{\text{Sym}}=
n-1 \equiv m_y,
\label{eq:Apptopint}
\end{equation}
where $m_y = -1,0,1$.  In the non-topological lattices, $m_y =
0$ indicating that as a function of $q_y$ there is one zero in
$\det \Cv_{\text{Sym}}(z_y)$ within the unit circle. In the
topological lattice shown in \fref{fig:top-lat-dis} (c), $\Rv_T
= -\av_1$, and $m_y=1$ (for the configuration shown in
\fref{fig:top-lat-dis}(b))  indicating two zeros in $\det
\Cv_{\text{Sym}}$ within the unit circle.  Direct evaluation of
the the integral in \eref{eq:Apptopint} confirms this result.

For $\qv$ parallel to $x$ [perpendicular to surface parallel to
$y$-axis as in \fref{fig:topological-surface}(a) or (c)], $\det
\Cv_{\text{Sym}}$ is a function of $z\equiv z_x^{1/2}$ and is
equal to $z^{-2}$ times a fourth order polynomial in $z$, and
it has two poles and four zeros in $z$. The topological
integral of $\ln \det \Cv_{\text{Sym}}$ around the unit circle
in $z$ (i.e., for $q_x$ from $0$ to $4 \pi$) can, therefore,
take on values of $m_x=-2, -1, ...,2$. For the non-topological
lattice, $m_x = 0$ for all orientation of $\Rv_T$ relative to
the $x$-axis, indicating two zeros within the unit circle.  For
the topological lattices, $m_x = 2$ and $m_x=1$, respectively,
for configuration \fref{fig:topological-surface} (a) with
$\Rv_T = -\av_1$ and configuration
\fref{fig:topological-surface}(a) with $\Rv_T = \av_3$.

Transformations from the symmetric unit cell to cells
compatible with specific surfaces constitute gauge
transformations in which the phase of $\det\Cv$ changes by a
factor of the local polarization charge of \eref{eq:RL}, $-
\qv\cdot\Rv_L = -d \sum \tilde{\rv}_\mu + \sum_{\beta}
\tilde{\rv}_\beta$:
\begin{equation}
\det \Cv_{\text{Sur}}=e^{-i \qv \cdot \Rv_L} \det \Cv_{\text{Sym}} .
\end{equation}
The phase factor can be determined directly from
\eref{eq:A-e-ueqs} relating bond stretches to site
displacements.   For example, the left-hand (LH) surface cell
of \fref{fig:topological-surface}(d) requires the displacement
of site $1$  through $\av_2$ and bonds $4$ and $6$ through
$\av_2$ and $-\av_3$, respectively.  A bond translation through
a lattice vector $\av$ changes the arguments of the
displacements of sites at the both ends of the translated bond
by $\av$. For example, a displacement of bond $4$ through
$\av_2$ with no site displacement would change the equation for
the stretch of bond $4$ from that in \eref{eq:A-e-ueqs} to $e_4
= (\uv_3(\Rv+\av_1 + \av_2)-\uv_1(\Rv
+\av_2))\cdot\bhv_4\cdot$. On the other hand, a translation of
a site requires the change in the arguments of the displacement
of all occurrences of that site by the negative of the
translation.  Thus the equation for bond $4$, which includes a
displacement of site $1$ whose argument must be augmented by
the negative of that displacement, i.e., by $-\av_2$ in the
case of the LH surface cell of
\fref{fig:topological-surface}(d), becomes $e_4 =
(\uv_3(\Rv+\av_1)-\uv_1(\Rv))\cdot\bhv_4\cdot$ as can be
verified by calculating the expression for $e_4$ direction from
the surface cell.

Thus for the LH cell of \fref{fig:topological-surface}(d),
$\Rv_L= -\av_1$, and $e^{-i \qv\cdot\Rv_L} = e^{i q_x}=z^2$,
and $\det \Cv_{\text{Sur}}=z^2 \det \Cv_{\text{Sym}}$ is simply
a fourth-order polynomial in $z=e^{i q_x/2}$. The two poles in
$\det \Cv_{\text{Sym}}$ have been removed, and the topological
integral around the unit circle of $z$ counts only the zeros of
$\det \Cv_{\text{Sym}}$ within the unit circle, four in the
case of \fref{fig:topological-surface}(a) and three in the case
of \fref{fig:top-lat-dis}(c).

The right-hand (RH) surface cell of
\fref{fig:topological-surface}(d), on the other hand, requires
only the translation of bond $6$ through $-\av_3$, and $\det
\Cv_{\text{Sur}}=z_x z_y \det \Cv_{\text{Sym}}$, which has no
poles in $z_y$, is simply a quadratic polynomial in $z_y$.
Again, the topological integral around the unit circle of $z_y$
counts only zeros of $\det \Cv_{\text{Sur}}$.

It is clear that $\Cv_\text{Sur}$ for a given surface is
identical with $e^{i \qv\cdot \Gv_s}$ replaced by $\lambda$,
where $\Gv$ is the reciprocal lattice vector associated that
surface, is identical to $\Cv_\text{Sur}$ calculate from the
surface cells directly as discussed in \sref{ssec:surface-m},
\eref{eq:matrix-C} to \eref{eq:detC-s}.  The latter equations
are set up from the beginning so that there are no negative
powers of $\lambda$ in $\Cv_{\text{Sur}}$, so the integral
around the unit circle of $\det \Cv_{\text{Sur}}$ give the
count of the number of zero modes of that surface.

%\newpage

%\bibliographystyle{unsrt}
%\bibliography{RPP}

\end{document}